\documentclass[12pt]{Latex/PhDthesisPSnPDF}
\pdfoutput=1
\usepackage{amsmath, amsthm, amssymb, amsfonts}

\usepackage[numbers]{natbib}

\title{Infinite Derivative Gravity:\\ A Ghost and Singularity-free Theory}

\author{Aindri\'{u} Conroy}

\collegeordept{Physics \\ Department of Physics}
\university{Lancaster University}
\crest{\includegraphics[width=4cm]{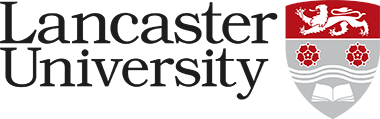}}

\renewcommand{\submittedtext}{A thesis submitted to Lancaster University for the degree of }
\degree{Doctor of Philosophy in the Faculty of Science and Technology}
\degreedate{April 2017}
\supervisor{\emph{\\Supervised by Dr Anupam Mazumdar}}

\begin{document}
\newcommand{\nn}{\nonumber}
\newcommand{\bg}{\begin{equation}}
\newcommand{\en}{\end{equation}}
\renewcommand\[{\begin{equation}}
\renewcommand\]{\end{equation}}
\newcommand{\al}{\alpha}
\newcommand{\bt}{\beta}
\newcommand{\ga}{\gamma}
\newcommand{\da}{\delta}
\newcommand{\n}{\nabla}
\newcommand{\ba}{\begin{eqnarray}}
\newcommand{\ea}{\end{eqnarray}}
\newcommand{\LF}{\left(}
\newcommand{\RF}{\right)}
\newcommand{\LT}{\left[}
\newcommand{\RT}{\right]}
\newcommand{\Ld}{\left.}
\newcommand{\Rd}{\right.}
\newcommand{\cO}{{\cal O}}
\newcommand{\cF}{{\cal F}}
\newcommand{\Ra}{\Rightarrow}
\newcommand{\bl}{\begin{align}}
\newcommand{\el}{\end{align}}
\newcommand{\fl}{}
\newtheorem{theorem}{Theorem}[section]
\theoremstyle{definition}
\newtheorem{definition}{Definition}[section]

\renewcommand\baselinestretch{1.5}
\baselineskip=18pt plus1pt


\maketitle  








\begin{abstracts}        
The objective of this thesis is to present a viable extension of general relativity free from cosmological singularities. A viable cosmology, in this sense, is one that is free from ghosts, tachyons or exotic matter, while staying true to the theoretical foundations of General Relativity such as general covariance, as well as observed phenomenon such as the accelerated expansion of the universe and inflationary behaviour at later times. To this end, an infinite derivative extension of relativity is introduced, with the gravitational action derived and the non-linear field equations calculated, before being linearised around both Minkowski space and de Sitter space. The theory is then constrained so as to avoid \emph{ghosts} and \emph{tachyons} by appealing to the modified propagator, which is also derived. Finally, the Raychaudhuri Equation is employed in order to describe the ghost-free, \emph{defocusing} behaviour around both  Minkowski and de Sitter spacetimes, in the linearised regime.
\end{abstracts}




\frontmatter
\begin{dedication} 

For Eli

\end{dedication}


\begin{acknowledgements}      
Special thanks go to my parents, M\'{a}ir\'{i}n and Kevin, without whose encouragement, this thesis would not have been possible; to Emily, for her unwavering support; and to Eli, for asking so many questions.
\\\\
I would also like to thank my supervisor Dr Anupam Mazumdar for his guidance, as well as my close collaborators Dr Alexey Koshelev, Dr Tirthabir Biswas and Dr Tomi Koivisto, who I've had the pleasure of working with. Further mention goes to Dr Claus Kiefer and Dr David Burton for fruitful discussions and to my fellow students in the Lancaster Cosmology group - Ernestas Pukartas, Lingfei Wang, Spyridon Talaganis, Ilia Teimouri, Saleh Qutub and James Edholm - without whom, my time in Lancaster would not have been so productive and enjoyable. 
\\\\
Finally, I'd like to extend my gratitude to STFC and Lancaster University for funding me throughout my studies.  
\\\\
The work in this thesis was supported by the STFC grant ST/K50208X/1.

\end{acknowledgements}




\begin{declaration}        

This thesis is my own work and no portion of the work referred to in this thesis has been submitted in support of an application for another degree or qualification at this or any other institute of learning.

\end{declaration}


\vspace{20mm}
\begin{quote}

\textit{"Even now there are places where a thought might grow -- "}\\
Derek Mahon.

\end{quote}


\setcounter{secnumdepth}{3} 
\setcounter{tocdepth}{3}    
\tableofcontents            


\listoffigures	



\begin{pubs}        
\newif\ifshowcitations\showcitationsfalse%
\newif\ifshowlinks\showlinksfalse%

%
%
%

\ifshowlinks%
  \usepackage[
         colorlinks=true,
         urlcolor=blue,       
         ]{hyperref}
  \newcommand*{\inspireurl}[1]{\\\href{#1}{INSPIRE-HEP entry}}
\else
  \makeatletter
  \newcommand*{\inspireurl}[1]{\@bsphack\@esphack}
  \makeatother
\fi
\ifshowcitations%
  \newcommand*{\citations}[1]{\\* #1}
\else
  \makeatletter
  \newcommand*{\citations}[1]{\@bsphack\@esphack}
  \makeatother
\fi
\renewcommand{\labelenumii}{\arabic{enumi}.\arabic{enumii}}


\underline{Chapter 2}
\begin{itemize}
\item
{\bf ``Generalized ghost-free quadratic curvature gravity''}
  \\{}T.~Biswas, A.~Conroy, A.~S.~Koshelev and A.~Mazumdar.
  \\{}\emph{ Class.\ Quant.\ Grav.\ }  {\bf 31}, 015022 (2014),  {}arXiv:1308.2319 [hep-th]
\inspireurl{http://inspirehep.net/record/1247404}
\citations{58 citations counted in INSPIRE as of 27 Mar 2017}
\end{itemize}
\underline{Chapter 3}
\begin{itemize}
\item
{\bf ``Nonlocal gravity in D dimensions: Propagators, entropy, and a bouncing cosmology''}
  \\{}A.~Conroy, A.~Mazumdar, S.~Talaganis and A.~Teimouri.
  \\
  {}\emph{Phys.\ Rev.\ D } {\bf 92}, no. 12, 124051 (2015) {}arXiv:1509.01247 [hep-th]
\inspireurl{http://inspirehep.net/record/1391767}
\citations{12 citations counted in INSPIRE as of 27 Mar 2017}

\end{itemize}
\underline{Chapter 4}
\begin{itemize}
\item
{\bf ``Defocusing of Null Rays in Infinite Derivative Gravity''}
  \\{}A.~Conroy, A.~S.~Koshelev and A.~Mazumdar.
   \\ 
  {}\emph{JCAP} {\bf 1701}, no. 01, 017 (2017) {}arXiv:1605.02080 [gr-qc]
\inspireurl{http://inspirehep.net/record/1456776}
\citations{5 citations counted in INSPIRE as of 27 Mar 2017}
\item
{\bf ``Geodesic completeness and homogeneity condition for cosmic inflation''}
  \\{}A.~Conroy, A.~S.~Koshelev and A.~Mazumdar.
  \\{}\emph{Phys.\ Rev.\ D} {\bf 90}, no. 12, 123525 (2014) {}arXiv:1408.6205 [gr-qc]
\inspireurl{http://inspirehep.net/record/1312353}
\citations{13 citations counted in INSPIRE as of 27 Mar 2017}
\end{itemize}
\underline{Abstract D}
\begin{itemize}
\item
{\bf ``Wald Entropy for Ghost-Free, Infinite Derivative Theories of Gravity''}
  \\{}A.~Conroy, A.~Mazumdar and A.~Teimouri.
  \\{}\emph{Phys.\ Rev.\ Lett.\ }  {\bf 114}, no. 20, 201101 (2015){} arXiv:1503.05568 [hep-th]
\inspireurl{http://inspirehep.net/record/1353504}
\citations{16 citations counted in INSPIRE as of 27 Mar 2017}


\end{itemize}
\end{pubs}



\mainmatter




\chapter{Introduction}
\label{chap:Intro}
Just over a century has passed since Einstein first presented his work on General Relativity (GR) to the Prussian academy, ushering in a new paradigm for modern physics. In the intervening years, Einstein's remarkable theory has withstood the enormous advancements in experimental physics and observational data - with each new discovery adding further weight to this colossus of scientific endeavour. General Relativity is not only considered one of mankind's greatest scientific discoveries but one of the most significant intellectual achievements in human history. Outside of the scientific sphere, the influence of relativity can be found in the arts -- whether it be in the work of existential playwright Luigi Pirandello, who played with traditional notions of the observer, or Pablo Picasso, whose distorted perspective was reportedly inspired by the idea of displaying a fourth dimension on canvas \cite{Miller}. That is not to say, however, that Einstein's gravity does not have its shortcomings, specifically in constructing a quantum field theory of gravity; as well
as describing a viable theory of gravity, which is devoid of singularities.

String theory (ST) remains a popular candidate in formulating a consistent quantum theory of gravity \cite{polchinski1998string}, as does Loop Quantum Gravity (LQG) \cite{Ashtekar:2012np},\cite{Nicolai:2005mc}, to name just two. Whereas string theory approaches the problem rather grandly, with the intention of unifying gravity with the fundamental forces of nature, LQG makes no such claim, with the stated aim of quantizing the gravitational field. Such an approach centres around the notion of \emph{renormalisation}, in that unwanted divergences in  the  loop Feynman diagrams may be curtailed \cite{Veltman:1975vx},\cite{DeWitt:2007mi}. A third course of action can be found in the causal set programme, which considers the continuous Lorentzian manifold of GR to be an approximation of a discrete spacetime structure \cite{Henson:2006kf}. A common thread running through these fundamentally different approaches is the presence of \emph{non-locality}, where interactions occur not at a specific spatial point but over a (finite) region of space \cite{Deser:2007jk},\cite{Woodard:2014iga}\cite{Woodard:2014wia},\cite{Deser:2013uya}. Indeed, non-locality arising from infinite derivative extensions of GR has been shown to play a pivotal role in the more classical context of resolving the cosmological singularity problem, \cite{Dimitrijevic:2013ofa},\cite{Calcagni:2013vra},\cite{Dimitrijevic:2012kb},\cite{Biswas:2005qr},\cite{Biswas:2011ar}\cite{Biswas:2010zk},\cite{Conroy:2016sac},\cite{Conroy:2014dja},\cite{Chialva:2014rla}\cite{Biswas:2012bp},\cite{Craps:2014wga}, which is our focus here.


The concept of singularities is a particularly confounding, yet intriguing, topic. Often casually referred to as a `place' where curvature `blows up', or a `hole' in the fabric of spacetime -- the concept of a singular spacetime raises thorny questions for a physicist. If a singularity is a `hole' in the fabric of spacetime, can it be  said to exist within the framework of spacetime? Could we not simply omit the singularity from our spacetime manifold? On the other hand, if the singularity does indeed exist within the spacetime, what does it mean to have a `place' within this framework where the normal physical laws that govern the universe no longer apply? The difficulty lies in a unique characteristic of general relativity in that it is formulated without stipulating the manifold and metric structure in advance. This is in contrast to other physical theories, such as special relativity, where these are clearly defined. As such, without a prescribed manifold, it is not possible to discuss the concept of `outside' the manifold. Neither can one consider the notion of a `place' where curvature may `blow up' as this `place' is undefined a priori \cite{Geroch:1968ut},\cite{Wald:GR}.  Such intuitive inconsistencies lead many to believe that singularities are not physically present in our Universe and that GR's admittance of singularities is evidence of the need to extend this powerful gravitational theory. In this way, we see the Einstein-Hilbert action of GR as a first approximation of a broader theory.

Proposals for modifying general relativity have been put forward since almost its inception. Early examples include Eddington's reformulation of GR in terms of the affine connection instead of the metric tensor or Kaluza and Klein's $5$-dimensional reimagining  \cite{Clifton:2011jh}. The latter of these proposals found that the Einstein equation in $5$ dimensions yielded the $4$-dimensional Einstein field equations along with Maxwell's equations, giving hopes for a unified theory of gravitation and electromagnetism. While mathematically elegant, the Kaluza-Klein model predicted an additional massless scalar field which is in conflict with experimental data. Despite this, the technique of introducing higher dimensions is considered to be a great influence on the development of string theory \cite{zwiebach2004first}. Much later, in 1977, Stelle proposed a fourth order extension of GR \cite{Stelle:1976gc},\cite{Stelle1978}, given by
\[
{\cal L}\sim R+f_1 R^2 +f_2 R_{\mu\nu}R^{\mu\nu}+f_3 R_{\mu\nu\sigma\lambda}R^{\mu\nu\sigma\lambda}.
\] 
Fourth order or four-derivative gravity -- so-called as each term in the resulting field equations contains four derivatives of the metric tensor -- is a somewhat natural extension of gravity, if seen as a generalisation of the Gauss-Bonnet term, which appears in Lovelock gravity \cite{LovelockGB},\cite{Lovelock1969}, and is trivial in four dimensions. We return to this point briefly in Section \ref{sec:Mod}. What is remarkable, however, is that Stelle found that such theories are perturbatively renormalizable, leading to a boon in the field of quantum gravity \cite{Schmidt:2006jt},\cite{Kiefer:2012boa},\cite{Stelle:1976gc},\cite{Hamber:2007fk}. A particular instance of fourth order gravity, known as the Starobinsky model \cite{Starobinsky:1979ty}, with
\[
{\cal L}\sim R+f_0 R^2
,\]
created further interest due to its description of successful primordial inflation. Starobinsky's initial idea was to formulate a gravitational theory that mimics the behaviour of the cosmological constant. For sufficiently large $R$ this model does precisely that through the $R^2$ term, leading to the formation of the large scale structures we see in the Universe today. The quadratic curvature term becomes less dominant as the theory moves away from the Planck scale, signalling the end of inflation. 

However, finite higher derivative theories, such as fourth-order gravty, can open the door to \emph{ghosts} -- physical excitations with negative residue in the graviton propagator. This negative residue presents itself as negative kinetic energy, leading to instabilities even at a classical level \cite{Himmetoglu:2009qi}, and a breakdown in unitarity when one considers the renormalization of the theory \cite{Stelle:1976gc},\cite{Biswas:2013kla},\cite{Rubakov:2014jja}, see Chapter \ref{chap:GF} for further details.  


Infinite derivative theories, in contrast, have the potential to describe a theory that is free of ghosts by modifying the graviton propagator via an exponent of an entire function \cite{Biswas:2005qr},\cite{Biswas:2011ar}. This exponential suppression of the propagator results in an exponential enhancement of the vertex factors of the relevant Feynman diagrams \cite{Talaganis:2016ovm}. Furthermore, the nature of this modification is such that one can always construct a modified propagator that contains no additional degrees of freedom, other than the massless graviton, so that negative residues will not propagate \cite{Biswas:2005qr},\cite{Biswas:2011ar},\cite{Biswas:2010zk},\cite{Conroy:2016sac},\cite{Conroy:2014dja},\cite{Chialva:2014rla},\cite{Biswas:2012bp},\cite{Craps:2014wga}. Infinite derivative extensions of relativity have been shown to display improved behaviour in the UV, in terms of alleviating the $1/r$ behaviour of the Newtonian potential \cite{Biswas:2011ar}, and curtailing quantum loop divergences \cite{Talaganis:2014ida},\cite{Talaganis:2015wva}. Recent progess has also been made in terms of the resolution of the black-hole singularity problem \cite{Frolov:2016pav} and a study of the dynamical degrees of freedom via Hamiltonian analysis \cite{Mazumdar:2017kxr}. Infinite derivatve extensions of relativity also allow for the formulation of non-singular cosmologies \cite{Conroy:2016sac},\cite{Conroy:2014dja}, which we cover extensively in Chapter \ref{chap:sing}, and forms the basis of the present work.  In simple terms, the \emph{objective} of this thesis is 
\begin{quote}
\emph{To present a viable extension of general relativity, which is free from cosmological singularities.}
\end{quote}
A viable cosmology, in this sense, is one that is free from ghosts, tachyons or exotic matter, while staying true to the theoretical foundations of General Relativity such as the principle of general covariance, as well as observed phenomenon such as the accelerated expansion of the universe and inflationary behaviour at later times \cite{Borde:2001nh}.

Several competing theories have been proposed as an alternative to the Big Bang model of GR. One such example is the Steady State universe. This approach is based on an extension of the cosmological principle, which imposes that the universe is homogenous and isotropic at large scales, to the \emph{Perfect} Cosmological Principle, which extends this uniformity to include time as well as space. In this sense, it conjectures that the universe has and always will exist in a state statistically similar to its current one \cite{Bondi:1948qk},\cite{Bondi:1954qj}. The steady state model, however, has suffered setbacks following the discovery of the cosmic microwave background in 1965  \cite{Weinberg:100595}, though some proponents of ``quasi-steady'' models remain \cite{Aguirre:2001ks}. 

Another popular resolution to the cosmological singularity problem is the \emph{bouncing universe} model, where the Big Bang singularity is replaced by a Big Bounce \cite{Novello:2008ra},\cite{Bozza:2005qg}. Such a cosmology issues from a scale factor that is necessarily an even function \cite{Calcagni:2013vra},\cite{Martin:2004qj}. Although the term ``Big Bounce" was not popularised until the 1980s \cite{Overduin:2006hb}, such cosmologies have a long history of interest, stretching back to the time of Willem de Sitter \cite{anderson2015cosmic}. 

Unlike bouncing models of the universe, we make no such stipulations on the nature of the cosmological scale factor a priori, preferring to confront the cosmological singularity problem by employing the Raychaudhuri equation (RE) \cite{Raych},\cite{ellis2012relativistic}, first devised in 1955 \cite{Senovilla:2014gza}. The RE is a powerful identity, which relates the geometry of spacetime to gravity, so that the behaviour of ingoing and outgoing causal geodesic congruences can be understood in a gravitational context. If these geodesics converge to a point in a finite time, they are called \emph{geodesically incomplete}, resulting in a singularity in a geometrically-flat or open cosmology \cite{Hawking:1973uf},\cite{Ellis:2003mb},\cite{Kar:2006ms},\cite{Wald:GR},\cite{Borde:1996pt},\cite{Borde:2001nh},\cite{Joshi:2013xoa}. Similarly, one can deduce the physical conditions, whereby these causal `rays' diverge, or \emph{defocus}, as a means of describing a viable non-singular cosmology. 

We will return to these points shortly, but it is perhaps instructive to first review some of the central tenets that GR relies upon - detailing what it is about GR that makes it such a special theory, before expanding on the need to modify or extend GR.
\section{General Relativity}
\label{sec:GRintro}
\emph{The Weak Equivalence Principle}\\
A key stepping stone in the formulation of GR was Einstein's Equivalence Principle, which states that, locally, inertial and gravitational mass are equivalent. Roughly speaking, this is tantamount to saying that the physics of a freely falling observer is indistinguishable from the physics of an observer in the absence of a gravitational field, which is why this principle is sometimes referred to as the \emph{universality of free fall}. In terms of Newtonian gravity, inertial mass $m_i$ is the form of mass that makes up Newton's second law of motion, i.e. $F=m_{i} a$, whereas gravitational mass $m_g$ appears in Newton's Law of Gravitation, $F=\frac{G m_{g_1}m_{g_2}}{r^2}$. The equivalence of these two forms of mass can be seen as a direct result of Galileo's leaning tower of Pisa experiment, where balls of two different masses reach the ground at the same time, in that the acceleration due to gravity is independent of the inertial or gravitational mass of the body in question. This simple insight led Einstein to formulate a theory where gravity is not described as a force but by geometry - by the curvature of spacetime \cite{Carroll:2004st},\cite{Wald:GR},\cite{Blau}.
\begin{quote}
\emph{``All uncharged, freely falling test particles follow the same trajectories once the initial position and velocity have been prescribed"} \cite{Clifton:2011jh} \\- The Weak Equivalence Principle (WEP)
\end{quote}
A fine-tuning of the Einstein Equivalence Principle led to the Weak Equivalence Principle, stated above, which has been tested rigorously over the years, beginning with the experiments of Lor{\'a}nd E{\"o}tv{\"o}s in 1908. Current experiments place the constraint on the WEP and therefore any viable relativistic theory to be
\[
\eta=2\frac{|a_1-a_2|}{|a_1+a_2|}=(0.3\pm 1.8)\times 10^{-13}
.\]
Here, beryllium and titanium were used to measure the relative difference in acceleration of the two bodies, $a_1$ and $a_2$, towards the galactic centre. This is considered to be a null result, wholly consistent with General Relativity \cite{Clifton:2011jh}.
\\\\\emph{Principle of General Covariance}\\
Another central tenet of General Relativity, which was instrumental in the formulation of GR and is perhaps more relevant to the present work, is the principle of general covariance. General covariance insists that each term making up a gravitational action will transform in a coordinate-independent way. The principle was first struck upon by Einstein when formulating the theory of special relativity, where it was proclaimed that physical laws will remain consistent in all inertial frames. Furthermore, the universal nature of the tensor transformation law offered a simple means of rendering physical equations generally covariant. That is to say that  any gravitational action expressed in terms of tensors (and covariant operators) would be a generally covariant action. Reformulating gravity in terms of tensors - with the graviton represented by a type $(2,0)$ metric tensor - allowed for a gravitational theory to be described by curvature alone. This proved to be the cornerstone of General Relativity and any valid modification or extension of GR should conform to this principle.
\\\\\emph{Gravitational Action}\\
We have now established that the central idea behind GR, as opposed to the Newtonian theory of gravitation, is that what we perceive as the force of gravity arises from the curvature of spacetime.   Mathematically, this can expressed by the gravitational action which defines the theory
\[
\label{EH}
S=\frac{1}{2}\int d^4x \sqrt{-g}\left(M_P^2 R-2\Lambda\right)
,\]
known as the Einstein-Hilbert action, where $M_P=\kappa^{-1/2}=\sqrt{\frac{\hbar c}{8\pi G}}$
is the Planck mass, with $\hbar=c=1$ (natural units); $R$ is the curvature scalar, defined in Appendix \ref{sec:AppCurv}, the determinant of the metric tensor is given by $g=\det(g_{\mu\nu})$;  and  the cosmological constant is $\Lambda$, which we take to be of mass dimension $4$ in our formalism. Variation of the action with respect to the metric tensor gives rise to the famous Einstein equation
\[
\label{EinsteinEq}
M_P^2 G_{\mu\nu}+g_{\mu\nu}\Lambda= T_{\mu\nu}
,\]
where $G_{\mu\nu}\equiv R_{\mu\nu}-\frac{1}{2}g_{\mu\nu}R$ and $T_{\mu\nu}$ are the Einstein and energy-momentum tensors respectively.
\section{Modifying General Relativity}
Despite the phenomenal success of the theory of relativity, outstanding issues remain, which suggests that the theory is incomplete. As mentioned in the introduction, one of these issues concerns the construction of a theory which marries  quantum field theory (QFT) with GR. This has been an open question in modern physics since almost the inception of QFT in the 1920s, but gained particular traction with the rise of string theory in the 1960s and 70s.
A more classical shortcoming of GR arises from its admittance of singularities, where the normal laws of physics can be said to `break down'. We now discuss how GR cannot describe a viable, non-singular cosmology, in order to motivate the need to extend the theory.
\\\\ \emph{Singularities}\\
The Cosmological Singularity Problem is the focal point of the present text, with Chapter \ref{chap:sing} devoted to the description of a stable, \emph{extended} theory of relativity devoid of an initial singularity. The requirement of extending GR in order to avoid a Big Bang singularity can be seen by referring to the Raychaudhuri equation (RE), see Section \eqref{sec:RE} for full details. The RE is a powerful identity which relates the geometry of spacetime to gravity, so that the behaviour of ingoing and outgoing causal geodesic congruences can be understood in a gravitational context. From this, one can deduce the necessary conditions whereby ingoing causal geodesics will converge to the same event in a finite time. This convergence is known as \emph{geodesic incompleteness} and a freely falling particle travelling along this geodesic will, at some finite point in time, cease to exist. We call such a spacetime \emph{singular} and the associated condition is known as the \emph{convergence condition} \cite{Geroch:1968ut}. 

Here, we merely outline the convergence conditions in GR, which are discussed in greater detail in Chapter \ref{chap:sing}, as a means of motivating the need to modify or extend the theory. From the RE, one can deduce that a spacetime will be null-geodesically incomplete if either of the following conditions are met \cite{Kar:2006ms},\cite{Borde:1996pt},\cite{Vachaspati:1998dy},
\[
\frac{d\theta}{d\lambda}+\frac{1}{2}\theta^2\leqq 0,\qquad R_{\mu\nu}k^\mu k^\nu\geq 0.
\]
Leaving aside the left hand inequality for the moment, which describes the convergence condition in terms of geometric expansion, let us focus on the right hand inequality within the framework of GR. From the Einstein-Hilbert action \eqref{EH}, we derive the Einstein equation \eqref{EinsteinEq}, while also noting that null geodesic congruences vanish when contracted with the metric tensor, according to $g_{\mu\nu}k^{\mu}k^{\nu}=0$. Thus,
\[
R_{\mu\nu}k^\mu k^\nu = \kappa T_{\mu\nu}k^\mu k^\nu 
,\]
must be positive in order to retain the null energy condition (NEC), see Appendix \eqref{NEC}, and to avoid the propagation of potentially exotic matter \cite{Capozziello:2013vna},\cite{Elder:2013gya}. Thus, in GR we are left with the choice of either accepting singularities or accepting potentially non-physical matter. As neither option is desirable, we conclude that GR must be extended in order to describe a viable non-singular cosmology.
\\\\\emph{Lovelock's Theorem}\\
An important theorem in both the formulation of GR and concerning any valid extension of the theory is \emph{Lovelock's Theorem}\cite{Clifton:2011jh},\cite{Lovelock1969}
\begin{theorem}[Lovelock's Theorem]
The only possible \emph{second-order} Euler-Lagrange expression obtainable in a four-dimensional space from a scalar density with a Lagrangian dependent on the metric tensor (i.e. ${\cal L}={\cal L}(g_{\mu\nu})$) is
\[
E^{\mu\nu}=\sqrt{-g}\left(\alpha G^{\mu\nu}+g^{\mu\nu}\lambda\right),
\]
where both $\alpha$ and $\lambda$ are constants 
\end{theorem}
 This is a remarkable result when one considers that by taking $\lambda=\Lambda$, this is precisely the Einstein equation in the presence of the cosmological constant, modified only by the constant $\alpha$. What this theorem says is that any gravitational theory in a four-dimensional Riemannian space, whose subsequent field equations are of second order or less will be defined solely by the Einstein equation. As we have seen, the Einstein Hilbert action \eqref{EH} produces the the Einstein field equations \eqref{EinsteinEq} precisely, but a more general action does exist (in four dimensions) that also reproduces the same result, and this is given by
 \[
 \label{EHgen}
 {\cal L}=\sqrt{-g}\left(\alpha R-2\Lambda\right)+\beta\sqrt{-g}\left(R^2-4R^{\mu\nu}R_{\mu\nu}+R^{\mu\nu\lambda\sigma}R_{\mu\nu\lambda\sigma}\right)+\gamma\epsilon^{\mu\nu\lambda\sigma}R^{\alpha\beta}{ }{ }_{\mu\nu}R_{\alpha\beta\lambda\sigma}
 .\]
 In four dimensions the final two terms do not contribute to the field equations. Whereas this is true for the final term in any number of dimensions, the second term is what as known as the Gauss Bonnet term and is non-trivial in theories of dimensions higher than four.

What Lovelock's theorem means for modified theories of gravity is that, if we assume that we want to describe a generally covariant, four-dimensional, metric-tensor-based theory of gravity, whilst retaining the variational principle, we have two options: 
\begin{enumerate}
\item[1.] Extend our approach into field equations that contain higher than second order derivatives and/or 
\item[2.] Allow a degree of non-locality to enter the system.\cite{Clifton:2011jh}
\end{enumerate}
\leavevmode

\subsection*{Examples of Modified Theories}
\label{sec:Mod}
\emph{Fourth Order Gravity}\\
We have already noted that the action \eqref{EHgen} is the most general action that reproduces the Einstein-field equation. A generalisation of the Gauss-Bonnet term
\[
G_{GB}=R^2-4R^{\mu\nu}R_{\mu\nu}+R^{\mu\nu\lambda\sigma}R_{\mu\nu\lambda\sigma},
\]
forms the basis for what is called \emph{Fourth Order Gravity},
\[
{\cal L}=R+f_1 R^2 +f_2 R_{\mu\nu}R^{\mu\nu}+f_3 R_{\mu\nu\sigma\lambda}R^{\mu\nu\sigma\lambda}.
\]
As stated in the introduction, Stelle observed that fourth order theories were perturbatively renormalisable, leading to a great generation of interest in quantum gravity \cite{Stelle:1976gc}. However, such theories are beset by the presence of ghosts, see Section \ref{sec:patho} for further details.
\\\\\emph{$f(R)$-gravity}\\
Perhaps the simplest generalisation of the Einstein-Hilbert action \eqref{EH} comes in the form of $f(R)$-gravity, where the curvature scalar $R$ is replaced by an arbitrary function $f$ acting on the curvature $R$,
\[
S=\frac{M_P^2}{2}\int d^4x \sqrt{-g} f(R)
.\]
By varying with respect to the metric tensor, we can then read off the $f(R)$ field equations
\[
\kappa T_{\mu\nu}=f^{\prime}(R)R_{\mu\nu}-\frac{1}{2}g_{\mu\nu}f(R)+g_{\mu\nu}\square f'(R)-\nabla_{\mu}\nabla_{\nu}f^{\prime}(R),\]
where $f^{\prime}=\partial f(R)/\partial R$, $\Box=g^{\mu\nu}\nabla_\mu\nabla_\nu$ and using $\delta f(R)=f'(R)\delta R$.
\\\\ \emph{Starobinsky Model}\\
The Starobinsky model is a particular instance of $f(R)$-gravity with
\[
f(R)=R+c_0 R^2
,\]
for some real constant $c_0$. Recall that Starobinsky's initial idea was to formulate a gravitational theory that mimics the behaviour of the cosmological constant, leading to successful primordial inflation. This model will be of particular interest when discussing the defocusing conditions of infinite derivative theory, where it is found that the Starobinsky model struggles to pair successful inflation with the avoidance of singularities. See Section \ref{sec:defocusmink}.
\section{Infinite Derivative Theory of Gravity}
\label{sec:IDGintro}
The most general infinite derivative action of gravity that is quadratic in curvature was first derived in \cite{Biswas:2005qr}, and was found to take the form
\[
\label{actionint}
S=\int d^4x\frac{\sqrt{-g}}{2}\left[M_P^2R+ \lambda R{\cal F}_1(\Box)R
+\lambda R_{\mu\nu} {\cal F}_2(\Box)R^{\mu\nu}
+\lambda C_{\mu\nu\lambda\sigma} {\cal F}_{3}(\Box)C^{\mu\nu\lambda\sigma}\right],
\]
where the form factors ${\cal F}_i(\Box)$ are given by
\[
{\cal F}_i(\Box)=\sum^{\infty}_{n=0}f_{i_n}\left(\Box/M\right)^n
\]
and $M$ is the scale of non-locality. 
In this form, we can see this as a natural generalisation of fourth order gravity to include all potential covariant operators in accordance with the principle of general covariance.
The above action has been studied extensively in terms of the modified propagator \cite{Biswas:2013kla}; Newtonian potential \cite{Edholm:2016hbt},\cite{Biswas:2011ar}; gravitational entropy \cite{Conroy:2015wfa},\cite{Conroy:2015nva}; loop quantum gravity \cite{Talaganis:2014ida},\cite{Talaganis:2015wva},\cite{Talaganis:2016ovm}; and indeed, singularity avoidance \cite{Conroy:2016sac},\cite{Conroy:2014dja},\cite{Frolov:2016pav}. We will summarise some relevant results shortly. Firstly, however, let us briefly expand on the notion of non-locality, alluded to in the introductory paragraphs.
\\\\\emph{Non-locality} \\
We stated earlier that a consequence of Lovelock's theorem is that a valid modified theory of gravity must include derivatives that are of second order or higher and/or allow a degree of non-locality. In this sense, the action \eqref{actionint} conforms to both of Lovelock's stipulations in that it is both of higher order than 2 and non-local, so that \eqref{actionint} can be understood as an effective action \cite{Biswas:2005qr},\cite{Barvinsky:2002uf}.  A theory featuring an \emph{infinite series of higher-derivative} terms, such as the infinite derivative gravity (IDG) theory introduced above, and derived in Chapter \ref{chap:2}, yields non-local interactions and a relaxation of the principle of locality, which, in simple terms, states that a particle may only be directly influenced by its immediate surroundings.

The quantum interactions of these infinite derivative terms have been studied and found to potentially alleviate divergences in the UV, by allowing for the super-renormalizability of the theory \cite{Tomboulis:1997gg},\cite{Moffat:2010bh},\cite{Talaganis:2014ida}. Non-local objects, such as strings and branes, are a fundamental  component of string theory, while the formulation of Loop Quantum Gravity is based on non-local objects, such as Wilson Loops \cite{Talaganis:2015wva}. The IDG theory, defined by the action \eqref{actionint} was inspired by the non-locality that arises from  exponential kinetic corrections, common in string theory, see \cite{Biswas:2005qr},\cite{zwiebach2004first}. In terms of the Feynman diagrams, non-local interactions result in an exponential enhancement of the vertex operator, meaning that interaction does not take place at this point, as in a local theory \cite{Talaganis:2014ida, Stelle:1976gc, Goroff:1985th}. 
Note also, that while a series of infinite derivatives is a common feature of non-local theories, it is not true to say that this is a defining characteristic. For example, massive gravity theories which modify GR in the infrared, e.g. $\sim \frac{1}{\Box^2} R$, are indeed non-local but have finite orders of the inverse D'Alembertian \cite{Maggiore:2013mea},\cite{Maggiore:2014sia},\cite{Conroy:2014eja},\cite{Jaccard:2013gla},\cite{Modesto:2013jea}.

\subsection*{Summary of Results}
In this section, we summarise some of relevant results, achieved within an infinite derivative gravitational framework, that are not explicitly covered in the subsequent chapters.\\\\
\emph{Newtonian Potential}\\
In \cite{Biswas:2011ar}, the Newtonian potential was studied around the weak field limit of the action \eqref{actionint}. In this case, the modified propagator was modulated by an overall factor of  $a(\Box)=e^{-\Box/M^2}$, where $M$ is the scale of modification. The exponential nature of this function was invoked in order to render the theory ghost and tachyon free, which is covered in detail in Chapter \ref{chap:GF}. For a theory with modified propagator $\Pi$, given by
\[
\Pi=\frac{1}{a(-k^2)}\Pi_{GR}
,\]
where $\Pi_{GR}$ is the physical graviton propagator and $\Box\rightarrow -k^2$ in Fourier space on a flat background, the Newtonian potential $\Phi(r)$ was found to be
\[
\label{erf}
\Phi(r)\sim\frac{m\pi}{2M_P^2 r}\mbox{Erf}(\frac{r M}{2})
.\]
Here, we observe that the potential contains the familiar $1/r$ divergence of GR, modulated by an error function Erf$(r)$. At the limit $r\rightarrow \infty$ \footnote{Alternatively, if we take $M\rightarrow\infty$, which is the limit to return IDG to a local theory, we recapture the familiar $1/r$ divergence of GR, as expected.}, $\mbox{erf}(r)/r\rightarrow 0$ returning flat space. Furthermore. at the limit $r\rightarrow 0$, the potential converges to a constant, thus ameliorating the $1/r$ drop-off of GR and displaying improved behaviour in the UV. The explicit calculation  can be found in Appendix \ref{chap:NewtPot}. The behaviour of the Newtonian potential in an IDG theory was further expanded upon in \cite{Edholm:2016hbt}, where a more general ghost-free form factor
\[
a(\Box)=e^{-\gamma(\Box/M^2)}
\]
was studied, where $\gamma$ is some entire function. In this case, identical limits were observed at $r\rightarrow \infty$. Furthermore, using laboratory data on the gravitational potential between two masses at very small distances, the lower limit $M>0.004eV$ was placed on the the scale of modification.
\begin{figure}[h]
\centering
\includegraphics[scale=0.4]{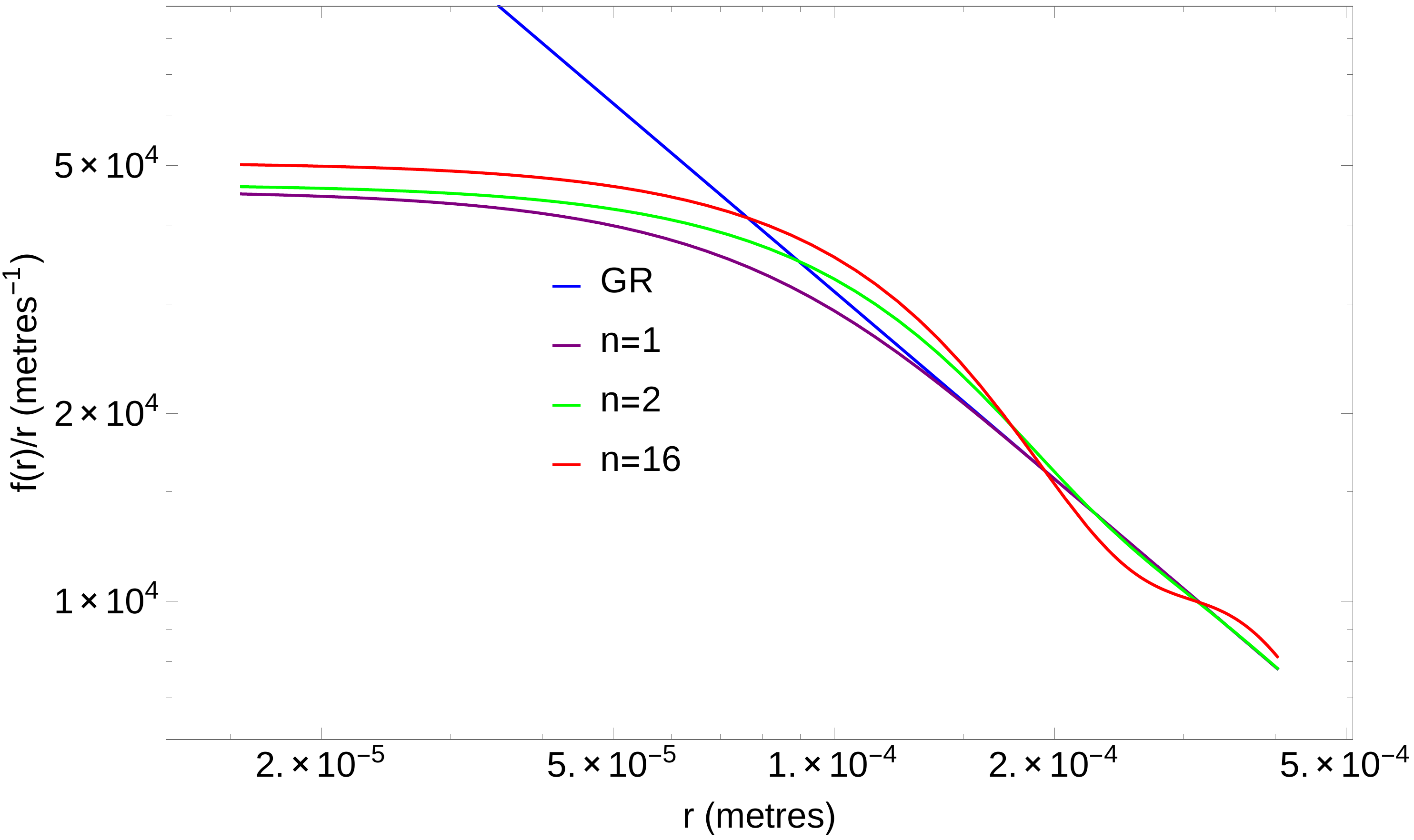}\cite{Edholm:2016hbt}
\caption{A plot of the Newtonian potential $\Phi(r)\sim f(r)/r$ vs. $r$ where $n=1$ corresponds to \eqref{erf} with $a(\Box)=e^{-\Box/M^2}$. Higher orders of $n$ are given by the exponential modification $a(\Box)=e^{-(\Box/M^2)^n}$, where $M$ has been taken to be the value of the lower bound, $M=0.004$eV for illustrative purposes} 
\end{figure}
\\\\\emph{Entropy}\\
The gravitational Wald entropy for IDG theories was investigated across two papers in \cite{Conroy:2015wfa} and \cite{Conroy:2015nva}. It was found in \cite{Conroy:2015wfa} that the gravitational entropy accounting for the UV-modified
sector vanishes around an axisymmetric black-hole metric when one requires that no additional degrees of freedom are introduced in the linear regime -- a condition which results in a ghost-free theory. The
resulting entropy was given simply by the famous area law, 
\[
S_{\text{\it Wald}}=\frac{\mbox{Area}}{4G}
.\]
In \cite{Conroy:2015nva}, the analysis was extended to consider the gravitational entropy around an (A)dS metric, where a lower bound on the leading order modification term was calculated which precludes non-physical spacetimes characterised by negative entropy. This bound was found to have cosmological significance in terms of avoiding singularities around a linearised de Sitter background. See Section \ref{sec:Entropy} for an outline of this result.
\\\\\emph{Quantum Loop Gravity}\\
Quantum aspects have been studied for  IDG theories, specifically from the point of view of a toy model, see \cite{Talaganis:2014ida}. Here,  explicit 1-loop and 2-loop computations were performed where it was found that, at 1-loop, a divergence arises. However, counter terms can be introduced to remove this divergence, in a similar fashion to loop computations in GR. Furthermore, at 2-loops the theory becomes finite. The article \cite{Talaganis:2014ida} then suggests a method for rendering arbitrary n-loops finite.
\\\\\emph{Modifications in the Infrared}\\\begin{figure}[h]
\centering
\includegraphics[scale=0.8]{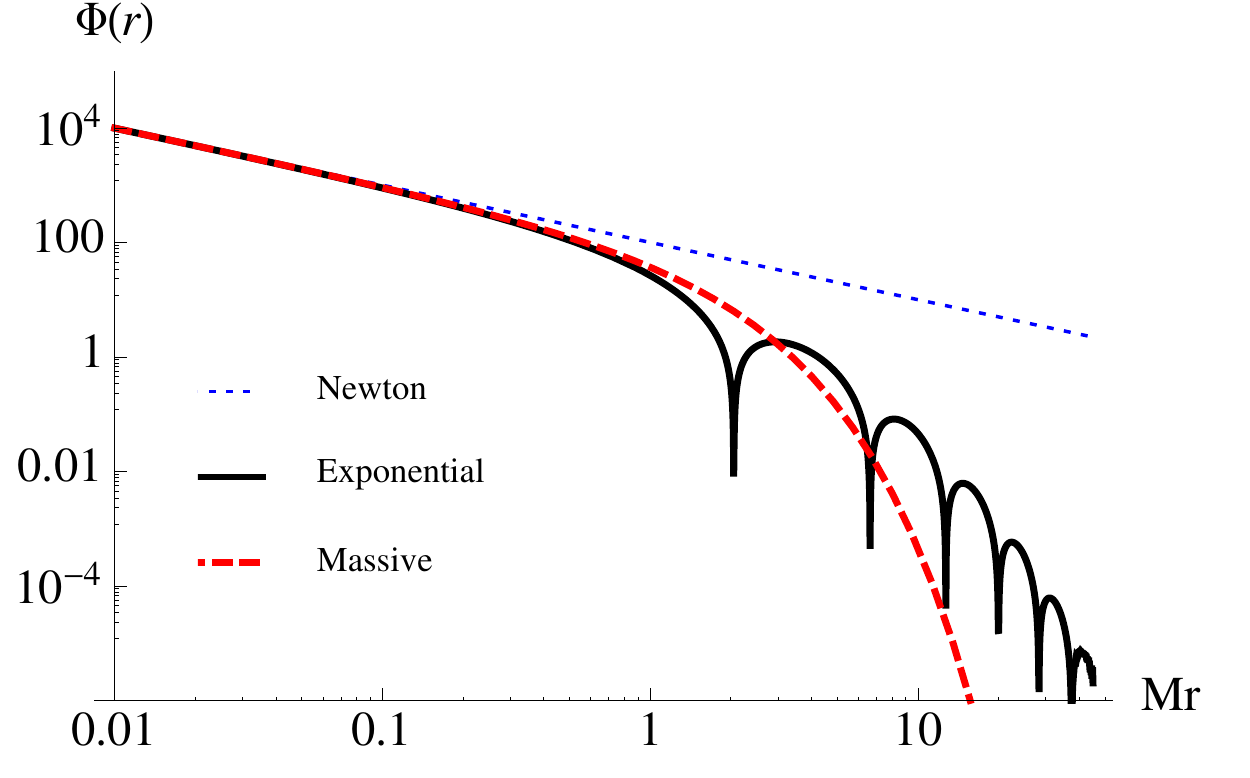}
\caption{Plot showing the suppression of the gravitational potential in the exponential model. The thick black line is the potential of the exponential model. Few initial oscillations are visible as the potential is suppressed with respect to the Newtonian 1/r behaviour depicted by the thin blue dotted line. For comparison, we also show the pure Yukawa suppression of massive gravity as the dashed red line.} 
\end{figure}The present work focuses solely on modifications to GR in the UV. Recently, however, interest has been generated in the field of non-local modifications in the infrared (IR). Such theories are characterised by the presence of inverse D'Alembertian ($1/\Box$) corrections in the gravitational action. Most notably, recent work has centred on the idea of constructing a theory of gravity which confers a non-zero mass upon the graviton, known as \emph{massive gravity}. Massive gravity theories are formulated via a $\frac{m_{gr}^2}{\Box^{2}}R$ -type extension to the Einstein-Hilbert action, where $m_{gr}$ is the mass of the massive graviton. Such theories have been explored in a number of papers as a means to explain the proliferation of dark energy in the Universe \cite{Maggiore:2013mea},\cite{Maggiore:2014sia},\cite{Modesto:2013jea},\cite{Jaccard:2013gla},\cite{Dirian:2014ara}\cite{Dirian:2014xoa}. 

In \cite{Conroy:2014eja}, the full non-linear field equations for a generalised action made up of an infinite series of inverse D'Alembertian operators was derived for the first time. The gravitational action can be formulated by replacing the form factors ${\cal F}_i(\Box)$ in \eqref{actionint} with
\[
{\cal {\bar F}}_i(\Box)=\sum_{n=1}^{\infty} f_{-n}(M^2/\Box)^n
.\]
Similar methods to \cite{Biswas:2011ar} were employed in order to derive the modified Newtonian potentials, with the added complexity that models with an additional degree of freedom in the scalar propagating mode were not excluded, i.e. $a\neq c$ in Appendix \ref{chap:NewtPot}, resulting in two Newtonian potentials. An upper bound was placed on the ratio of these potentials, known as the Eddington parameter, via the Cassini tracking experiment and various models were analysed as a means of explaining dark energy, including $Rf(R/\Box)$-models \cite{Deser:2007jk} and massive gravity. In the context of massive gravity, the massive graviton was tested and found to fall within the appropriate limits to be considered a possible dark energy candidate. Finally, a novel approach to infrared modifications was introduced, making use of all infinite inverse derivatives, which displayed a reduction in the gravitational field at large distances -- a common feature of IR extensions to GR.  
\newpage
\section{Organisation of Thesis}
The content of the thesis is organised as follows:
\begin{enumerate}
\item[Chapter 2:] This chapter begins with a derivation of the most general, generally covariant, infinite derivative action of gravity that is quadratic in curvature, before moving on to the main focus of the chapter: the highly non-trivial task of attaining the full non-linear field equations. The general methodology is outlined before moving on to the explicit calculation. Finally the linearised field equations are derived around both Minkowski and de Sitter spacetimes for later use in the context of ghost and singularity free cosmologies.

\item[Chapter 3:] The general ghost and tachyon criteria that a viable theory must conform to are established, with specific examples of pathological behaviour  given. The correction to the graviton propagator from the infinite derivative extension is attained, as are the ghost-free conditions around Minkowski space.

\item[Chapter 4:] This final chapter is the crux of the thesis, combining the field equations (Chapter 2) and the ghost-free conditions (Chapter 3) to formulate a viable singularity-free theory of gravity. The chapter begins with a discussion on the intriguing topic of defining a singularity, before moving on to an outline of the famous Hawking-Penrose Singularity Theorem. The Raychaudhuri equation (RE) is introduced and derived, with particular attention paid to the RE in a cosmological setting. A novel calculation then follows where the RE is applied to infinite derivative gravity theory and viable \emph{defocusing conditions} are derived around Minkowski and de Sitter spacetimes.
\end{enumerate}
\chapter{Infinite Derivative Gravity}
\label{chap:2}
\section{Derivation of Action}
Having introduced the concept of infinite derivative gravity theories and some of the progress made in the area, our goal here is to derive the most general, generally covariant infinite derivative action of gravity, with a view to formulating the associated equations of motion. Following this, in Chapter \ref{chap:GF}, we will use the field equations to understand the nature of the modified propagator. 

We begin by inspecting the fluctuations around a given background up to quadratic order in $h$, according to
\[
\label{d0}
g_{\mu\nu}=\eta_{\mu\nu}+h_{\mu\nu}
.\]
For presentation purposes, we have restricted the background metric to that of the Minkowski spacetime as in \cite{Biswas:2011ar}, while the derivation has been repeated in the more general framework of maximally symmetric spacetimes of constant curvature, i.e. Minkowski or (Anti) de Sitter space, in \cite{Biswas:2016egy},\cite{Biswas:2016etb}.  In principle, it should be possible to relax this restriction on the background metric further to any background metric with a well-defined Minkowski limit, with this latter point required to eliminate potentially singular non-local terms. 

As noted in \cite{Biswas:2013kla},\cite{Biswas:2011ar}, the most general, four dimensional, generally covariant metric-tensor-based gravitational action, with a well-defined Minkowski limit, may be expressed in the following generic form
\[
\label{d1}
S=\int d^4 x\sqrt{-g}\left[{\cal P}_0+\sum_i {\cal P}_i \prod_I \left( {\cal O}_{iI}{\cal Q}_{iI}\right)\right]
,\]
where ${\cal P}$ and ${\cal Q}$ are functions of Riemannian curvature and the metric tensor, while the operator ${\cal O}$ is made up, solely, of covariant operators, in accordance with general covariance.

Our goal is to inspect fluctuations around Minkowski space up to quadratic order. To this end, following closely to \cite{Biswas:2016egy},\cite{Biswas:2016etb},\cite{Biswas:2011ar},\cite{Biswas:2005qr}, we may recast \eqref{d1} into the following invariant form
\[
S=S_{EH}+S_{UV},\qquad\mbox{with}\qquad S_{UV}= \int d^4x\sqrt{-g}\left( R_{\mu_1 \nu_1 \lambda_1 \sigma_1}{\cal O}^{\mu_1 \nu_1 \lambda_1 \sigma_1}_{\mu_2 \nu_2 \lambda_2 \sigma_2}R^{\mu_2 \nu_2 \lambda_2 \sigma_2}\right)
\label{d3}
,\]
where $S_{EH}$ is the Einstein-Hilbert action and $S_{UV}$ constitutes the modification of GR in the ultraviolet (UV). The operator ${\cal O}_{\mu_{1}\nu_{1}\lambda_{1}\sigma_{1}}^{\mu_{2}\nu_{2}\lambda_{2}\sigma_{2}}$ represents a general covariant operator, such as the D'Alembertian operator $\Box=g^{\mu\nu}\nabla_{\mu}\nabla_{\nu}$; and the tensor $R_{\mu_{1}\nu_{1}\lambda_{1}\sigma_{1}}$ represents all possible forms of Riemannian curvature, such as the curvature scalar, Ricci Tensor, Riemann and Weyl tensors. It is worth noting that while the generic form \eqref{d3} includes all order of curvature via the commutation relation \eqref{comrel}, we restrict ourselves to a theory that is quadratic in curvature.

Noting that the differential operator ${\cal O}$ contains only the Minkowski metric coupled with covariant derivatives, we may expand the compact form \eqref{d3} to the following
\begin{eqnarray}
S&=&\int d^4x\frac{\sqrt{-g}}{2}\Big[M_P^2R+R F_1(\Box)R+R F_2(\Box)\nabla_{\nu}\nabla_{\mu}R^{\mu\nu}+R_{\mu\nu} F_3(\Box)R^{\mu\nu}\nonumber \\ 
&+&R_{\mu}^{\nu} F_4(\Box)\nabla_{\nu}\nabla_{\lambda}R^{\mu\lambda}+R^{\lambda\sigma} F_5(\Box)\nabla_{\mu}\nabla_{\sigma}\nabla_{\nu}\nabla_{\lambda}R^{\mu\nu}+R F_6(\Box)\nabla_{\mu}\nabla_{\nu}\nabla_{\lambda}\nabla_{\sigma}R^{\mu\nu\lambda\sigma}\nonumber \\ 
&+&R_{\mu\lambda} F_7(\Box)\nabla_{\nu}\nabla_{\sigma}R^{\mu\nu\lambda\sigma}+R_{\lambda}^{\rho} F_8(\Box)\nabla_{\mu}\nabla_{\sigma}\nabla_{\nu}\nabla_{\rho}R^{\mu\nu\lambda\sigma} \nonumber \\ 
& + &R^{\mu_1\nu_1} F_9(\Box)\nabla_{\mu_1}\nabla_{\nu_1}\nabla_{\mu}\nabla_{\nu}\nabla_{\lambda}\nabla_{\sigma}R^{\mu\nu\lambda\sigma}+ R_{\mu\nu\lambda\sigma} F_{10}(\Box)R^{\mu\nu\lambda\sigma} \nonumber \\ 
& + & R^{\rho}{ }_{\mu\nu\lambda} F_{11}(\Box)\nabla_{\rho}\nabla_{\sigma}R^{\mu\nu\lambda\sigma}+ R_{\mu\rho_1 \nu\sigma_1} F_{12}(\Box)\nabla^{\rho_1}\nabla^{\sigma_1}\nabla_{\rho}\nabla_{\sigma}R^{\mu\rho\nu\sigma}
\nonumber\\ &+&
R_{\mu}^{\;\nu_1\rho_1\sigma_1}F_{13}(\Box)\nabla_{\rho_1}\nabla_{\sigma_1}\nabla_{\nu_1}\nabla_{\nu}\nabla_{\rho}\nabla_{\sigma}R^{\mu\nu\lambda\sigma} \nonumber \\ 
& + & R^{\mu_1\nu_1\rho_1\sigma_1} F_{14}(\Box)\nabla_{\rho_1}\nabla_{\sigma_1}
\nabla_{\nu_1}\nabla_{\mu_1}\nabla_{\mu}\nabla_{\nu}\nabla_{\rho}\nabla_{\sigma}R^{\mu\nu\lambda\sigma}\Big]\,,
\label{d4}
\end{eqnarray} where we have liberally used integration by parts and the functions $F_i$ are defined by
 \[
 \label{Fintro}
 F_i(\Box)=\sum^{\infty}_{n=0}{\tilde f}_{i_n}(\Box/M^2)^n
 .\]
These functions contain all orders of the D'Alembertian operator $\Box=g^{\mu\nu}\nabla_\mu \nabla_\nu$\footnote{Up to quadratic order around Minkowski space, the D'Alembertian will appear in the action only as $\Box=\eta^{\mu\nu}\nabla_\mu \nabla_\nu$}, with each operator modulated by the scale of non-locality $M$ to ensure that these functions are dimensionless. The coefficients ${\tilde f}_{i_n}$, as yet unconstrained, ensure that these are arbitrary infinite derivative functions.

The action \eqref{d4} can be reduced upon noting the antisymmetric properties of the Riemann tensor,
\[
\label{RiemAnti}
R_{(\mu\nu)\rho\sigma}=R_{\mu\nu(\rho\sigma)}=0,
\]
along with the (second) Bianchi identity
\[
\label{Bianchi}
 \nabla_\alpha R^\mu_{\;\nu\beta\gamma}+\nabla_\beta R^\mu_{\;\nu\gamma\alpha}+\nabla_\gamma R^\mu_{\;\nu\alpha\beta}=0
.\]
\pagebreak

\noindent\rule[0.5ex]{\linewidth}{0.25pt} 
\emph{Example:}\\
Consider the terms
\[
RF_{1}(\Box)R+RF_{2}(\Box)\nabla_{\nu}\nabla_{\mu}R^{\mu\nu}+	R_{\mu}^{\nu}F_{4}(\Box)\nabla_{\nu}\nabla_{\lambda}R^{\mu\lambda}
.\]
These can be expressed as the following
\[
\label{2-9}
RF_{1}(\Box)R+\frac{1}{2}RF_{2}(\Box)\Box R+	\frac{1}{2}R_{\mu}^{\nu}F_{4}(\Box)\nabla_{\nu}\nabla^{\mu}R
,\]
by noting the identity $\nabla_{\mu}R^{\mu\nu}=\frac{1}{2}\nabla^{\nu}R$ and subsequently $\nabla_{\nu}\nabla_{\mu}R^{\mu\nu}=\frac{1}{2}\Box R$, which results from a contraction of the Bianchi identity \eqref{Bianchi}. We then perform integration by parts on the final term, to find that \eqref{2-9} develops as follows
\begin{align}
	&=RF_{1}(\Box)R+\frac{1}{2}RF_{2}(\Box)\Box R+\frac{1}{2}\nabla^{\mu}\nabla_{\nu}R_{\mu}^{\nu}F_{4}(\Box)R
	\\&=RF_{1}(\Box)R+\frac{1}{2}RF_{2}(\Box)\Box R+\frac{1}{4}RF_{4}(\Box)\Box R
	\\&\equiv RF_{1}(\Box)R.
\end{align}
In the last step, we have redefined the arbitrary function $F_{1}(\Box)$ to incorporate $F_{2}(\Box)$ and $F_{4}(\Box)$. \\
\noindent\rule[0.5ex]{\linewidth}{0.25pt} 
\\\\
Proceeding in a similar manner, we find that the action reduces to 
\begin{eqnarray} 
S&=&\int d^4x\frac{\sqrt{-g}}{2}\Big[M_P^2 R+R F_1(\Box)R + R_{\mu\nu} F_3(\Box)R^{\mu\nu}
+ R F_6(\Box)\nabla_{\mu}\nabla_{\nu}\nabla_{\lambda}\nabla_{\sigma}R^{\mu\nu\lambda\sigma} \nonumber \\
& + & R_{\mu\nu\lambda\sigma} F_{10}(\Box)R^{\mu\nu\lambda\sigma}  +
R_{\mu}^{\nu_1\rho_1\sigma_1}F_{13}(\Box)\nabla_{\rho_1}\nabla_{\sigma_1}\nabla_{\nu_1}\nabla_{\nu}\nabla_{\rho}\nabla_{\sigma}R^{\mu\nu\lambda\sigma} \nonumber \\ 
& + & R^{\mu_1\nu_1\rho_1\sigma_1} F_{14}(\Box)\nabla_{\rho_1}\nabla_{\sigma_1}
\nabla_{\nu_1}\nabla_{\mu_1}\nabla_{\mu}\nabla_{\nu}\nabla_{\rho}\nabla_{\sigma}R^{\mu\nu\lambda\sigma}\Big]\,.
\label{d5}\end{eqnarray}
A final important reduction comes when one notes that, as we are considering fluctuations around Minkowski space, the covariant derivatives commute freely.\\\\
\noindent\rule[0.5ex]{\linewidth}{0.25pt} 
\\\emph{Example:}
\\Take, for example, the $F_6(\Box)$ term in the above expression. We can decompose this in to two parts, like so
\[
RF_{6}(\Box)\nabla_{\mu}\nabla_{\nu}\nabla_{\lambda}\nabla_{\sigma}R^{\mu\nu\lambda\sigma}=\frac{1}{2}RF_{6}(\Box)\nabla_{\mu}\nabla_{\nu}\nabla_{\lambda}\nabla_{\sigma}R^{\mu\nu\lambda\sigma}+\frac{1}{2}RF_{6}(\Box)\nabla_{\mu}\nabla_{\nu}\nabla_{\lambda}\nabla_{\sigma}R^{\mu\nu\lambda\sigma}
.\]
We then commute one pair of derivatives in the first term to find
\[
RF_{6}(\Box)\nabla_{\mu}\nabla_{\nu}\nabla_{\lambda}\nabla_{\sigma}R^{\mu\nu\lambda\sigma}=\frac{1}{2}RF_{6}(\Box)\nabla_{\nu}\nabla_{\mu}\nabla_{\lambda}\nabla_{\sigma}R^{\mu\nu\lambda\sigma}+\frac{1}{2}RF_{6}(\Box)\nabla_{\mu}\nabla_{\nu}\nabla_{\lambda}\nabla_{\sigma}R^{\mu\nu\lambda\sigma}.
\]
Relabelling the indices gives
\[
RF_{6}(\Box)\nabla_{\mu}\nabla_{\nu}\nabla_{\lambda}\nabla_{\sigma}R^{\mu\nu\lambda\sigma}=RF_{6}(\Box)\nabla_{\mu}\nabla_{\nu}\nabla_{\lambda}\nabla_{\sigma}R^{(\mu\nu)\lambda\sigma}=0
,\]
which vanishes due to the antisymmetric properties of the Riemann tensor, \eqref{RiemAnti}.
\noindent\rule[0.5ex]{\linewidth}{0.25pt}
\\\\
Taking this into account, we can now  express the general form of the modified action as follows
\[ \label{f_action}
S=\int d^4x\frac{\sqrt{-g}}{2}\left[M_P^2R+R F_1(\Box)R
+R_{\mu\nu} F_3(\Box)R^{\mu\nu}
+R_{\mu\nu\lambda\sigma} F_{10}(\Box)R^{\mu\nu\lambda\sigma}\right]\,.
\]
We complete the derivation of the most general, generally covariant action of gravity that is quadratic in curvature with a little bookkeeping. First of all, it is preferable to replace the Riemann tensor in the gravitational action with the Weyl tensor, which is defined by
\[
\label{weyl}
C_{\;\alpha\nu\beta}^{\mu}\equiv R_{\;\alpha\nu\beta}^{\mu}-\frac{1}{2}(\delta_{\nu}^{
\mu}R_{\alpha\beta}-\delta_{\beta}^{\mu}R_{\alpha\nu}+R_{\nu}^{\mu}g_{
\alpha\beta}-R_{\beta}^{\mu}g_{\alpha\nu})+\frac{R}{6}(\delta_{\nu}^{\mu}g_{
\alpha\beta}-\delta_{\beta}^{\mu}g_{\alpha\nu})
.\]
This is because the Weyl tensor vanishes precisely in a conformally-flat background, making calculations less cumbersome in, for example, a cosmological Friedmann-Robertson-Walker (FRW) setting. This substitution does not represent any fundamental change to the theory as any change arising from reformulating the above action in terms of the Weyl tensor is absorbed by the arbitrary coefficient ${\tilde f}_{i_n}$ contained within the infinite derivative functions \eqref{Fintro}. In acknowledgement of this minor change, we now rename the infinite derivative functions, like so
 \[
 \label{Fcurly}
 {\cal F}_i(\Box)=\sum^{\infty}_{n=0}f_{i_n}(\Box/M^2)^n
 ,\]
while also renaming the indices for presentation purposes. Finally, we introduce a dimensionless `counting tool' $\lambda$ which offers a straightforward limit $\lambda\rightarrow 0$ to return the theory to that of GR. Taking this into account, we now arrive at the final form of the modified action
\[ \label{action}
S=\int d^4x\frac{\sqrt{-g}}{2}\left[M_P^2R+ \lambda R{\cal F}_1(\Box)R
+\lambda R_{\mu\nu} {\cal F}_2(\Box)R^{\mu\nu}
+\lambda C_{\mu\nu\lambda\sigma} {\cal F}_{3}(\Box)C^{\mu\nu\lambda\sigma}\right]\,.
\]
\section{Equations of Motion}
Having attained the most general, generally covariant, infinite derivative action of gravity that is quadratic in curvature, the next step is to compute the field equations -- a highly non-trivial task. We begin with an overview of the methods involved, largely based upon \cite{Biswas:2013cha}, before delving into the full technical derivations.
\subsection{General Methodology}
\subsubsection*{Single $\Box$}
In order to illustrate the methods involved in deriving the field equations for the action \eqref{action}, we begin with a simple example, by way of the action,
\bg
S_s = \int d^4 x \sqrt{-g} R\Box R
,\en
where $R$ denotes the curvature scalar. Varying this, gives
\[
\label{varyss}
\delta S_{s}=\int d^{4}x\sqrt{-g}\biggl(\frac{h}{2}R\Box R+\delta R\Box R+R\delta(\Box R)\biggr),
 \]
where we are considering the variation
\[
g_{\mu\nu}\rightarrow g_{\mu\nu}+\delta g_{\mu\nu}
\]
and have defined
\bg
\delta g_{\mu\nu} \equiv h_{\mu\nu}, \mbox{ such that } \delta g^{\mu\nu}=-h^{\mu\nu}
.\footnote{Note: This second identity ($\delta g^{\mu\nu}=-h^{\mu\nu}$) follows from the first ($\delta g_{\mu\nu}\equiv h_{\mu\nu}$), along with the invariance of the Kronecker delta. 
\begin{align*}
\delta_{\nu}^{\mu}	&=g^{\mu\sigma}g_{\sigma\nu}\rightarrow(g^{\mu\sigma}+\delta g^{\mu\sigma})(g_{\sigma\nu}+\delta g_{\sigma\nu})
	=\delta_{\nu}^{\mu}+g^{\mu\sigma}\delta g_{\sigma\nu}+\delta g^{\mu\sigma}g_{\sigma\nu}+{\cal O}(h^{2}),
\end{align*}
which implies that
\[
g^{\mu\sigma}h_{\sigma\nu}=-\delta g^{\mu\sigma}g_{\sigma\nu}.
\]
Act $g^{\nu\tau}$ to both sides to find
\[
h^{\mu\tau}=-\delta g^{\mu\tau}.
\]
}
\en 
One can then compute the variation of the determinant of the metric, which gives
\[
\label{deltasqrt}
\delta\sqrt{-g}=\frac{1}{2}\sqrt{-g}h
,\]
where $h\equiv h^\mu_\mu$~\cite{Veltman:1975vx}. We then note that the D'Alembertian operator $\Box$ contains within it a metric and must therefore be subjected to variation. Upon integration by parts, we may express \eqref{varyss} in the following form
\[
\label{Christoffelh2}
\delta S_{s}=\int d^{4}x\sqrt{-g}\biggl(\frac{h}{2}R\Box R+2\delta R\Box R+R\delta(\Box)R\biggr)
. \]
We will deal with the tricky final term in due course. Firstly, however, we apply the variational principle in order to compute the variation of any relevant curvatures. Upon inspection of the definition of the Christoffel symbol, \eqref{Christoffel}, we find the variation to be of the form
\bg
\label{variedChris}
\delta\Gamma_{\mu\nu}^{\lambda}=\frac{1}{2}\left(\nabla_{\mu}h_{\nu}^{\lambda}+\nabla_{\nu}h_{\mu}^{\lambda}-\nabla^{\lambda}h_{\mu\nu}\right)
. \en
Similarly, variation of the curvature terms, found in \ref{sec:AppCurv}, give
\bg
\nn
\delta R^{\lambda}{ }_{\mu\sigma\nu}=\nabla_{\sigma}\delta\Gamma_{\mu\nu}^{\lambda}-\nabla_{\nu}\delta\Gamma_{\mu\sigma}^{\lambda}
\en 
\bg
\nn
\delta R_{\mu\nu}=\nabla_{\lambda}\delta\Gamma_{\mu\nu}^{\lambda}-\nabla_{\nu}\delta\Gamma_{\mu\lambda}^{\lambda}
 \en
\bg
\delta R=-h^{\mu\nu}R_{\mu\nu}+g^{\mu\nu}\delta R_{\mu\nu}
. \en
Substitution of the above varied Christoffel symbol \eqref{Christoffelh2} allows us to represent the varied curvature in terms of the perturbed metric tensor $h_{\mu\nu}$:
\[
\delta R^{\lambda}{ }_{\mu\sigma\nu}=\frac{1}{2}\left(\nabla_{\sigma}\nabla_{\mu}h_{\;\nu}^{\lambda}-\nabla_{\sigma}\nabla^{\lambda}h_{\mu\nu}-\nabla_{\nu}\nabla_{\mu}h_{\;\sigma}^{\lambda}+\nabla_{\nu}\nabla^{\lambda}h_{\mu\sigma}\right)
\nn \]
\[
\delta R_{\mu\nu}=\frac{1}{2}\left(\nabla_{\lambda}\nabla_{\mu}h_{\nu}^{\lambda}+\nabla_{\lambda}\nabla_{\nu}h_{\mu}^{\lambda}-\Box h_{\mu\nu}-\nabla_{\nu}\nabla_{\mu}h\right)
 \nn\]
\[
\label{varyids}
\delta R=-R^{\mu\nu}h_{\mu\nu}+\nabla_{\lambda}\nabla^{\sigma}h_{\sigma}^{\lambda}-\Box h
\] 
Identities of this type are well known. What is less clear, however, is the computation of $\delta(\Box)R$, which we shall detail below.
\subsubsection*{Computing $\delta(\Box)R$}
By expanding out the components of the D'Alembertian, we have
\bg
\label{deltaboxR0}
\delta(\Box)R=\delta(g^{\mu\nu}\nabla_{\mu}\nabla_{\nu})R=-h^{\mu\nu}\nabla_{\mu}\nabla_{\nu}R+g^{\mu\nu}\delta(\nabla_{\mu})\nabla_{\nu}R+g^{\mu\nu}\nabla_{\mu}\delta(\nabla_{\nu})R
. \en
We find here some terms that involve the variation of the covariant operator, which are not so trivial. However, by varying the tensor $\nabla_\mu\nabla_\nu R$, we may equate these non-trivial terms with the variation of ordinary tensors, like so
\bg
\delta(\nabla_{\mu})\partial_{\nu}R+\nabla_{\mu}\delta(\nabla_{\nu})R=\delta(\nabla_{\mu}\nabla_{\nu}R)-\nabla_{\mu}\nabla_{\nu}\delta R.
\en
Next, we compute the terms on the right hand side by expanding out the covariant derivative so that, after some cancellation, we get an identity in terms of the Christoffel symbol
\bg
\delta(\nabla_{\mu})\partial_{\nu}R+\nabla_{\mu}\delta(\nabla_{\nu})R=-\delta\Gamma_{\mu\nu}^{\kappa}\partial_{\kappa}R
.\en
We briefly note that the simplicity of this identity is due to the scalar nature of the curvature involved. Tensors of higher order, such as the Ricci or Riemann tensor, will result in additional terms. Upon substitution into  \eqref{deltaboxR0} in conjunction with the variation of the Christoffel stated previously in \eqref{varyids}, we arrive at a vital result in computing the field equations for a non-local action of the type \eqref{action}.
\bg
\label{deltaboxR}
\delta(\Box)R=-h^{\mu\nu}\nabla_{\mu}\nabla_{\nu}R-\nabla^{\sigma}h_{\sigma}^{\kappa}\partial_{\kappa}R+\frac{1}{2}\partial^{\kappa}h\partial_{\kappa}R
. \en
\subsubsection*{Multiple $\Box$'s}
Crucially, however, in order to derive the field equations for an infinite derivative action such as \eqref{action}, the above mechanism must be generalised to encapsulate an arbitrary number of D'Alembertian operators acting on the curvature. To shed light on this, we consider the action
\bg
\label{actionboxn}
S_{m}=\int d^{4}x\sqrt{-g}R\Box^{n}R.
 \en
Varying with respect to the metric tensor gives
\bg
\delta S_{m}=\int d^{4}x\sqrt{-g}\biggl(\frac{h}{2}R\Box^{n}R+\delta R\Box^{n}R+R\delta(\Box^{n})R+R\Box^{n}\delta R\biggr),
\en 
analogous to \eqref{varyss}. We now turn our attention to the $R\delta(\Box^{n})R$ term. Repeated application of the product rule reveals the following
\begin{align}
R\delta(\Box^{n})R	&=	R\Box\delta(\Box^{n-1})R+R\delta(\Box)\Box^{n-1} R
\nn\\&	...	
\nn\\&	=\sum_{m=0}^{n-1}	R\Box^{m}\delta(\Box)\Box^{n-m-1}R.
\label{deltaboxn}
 \end{align}
It is then straightforward to substitute \eqref{deltaboxR} into this identity, along with the previously derived varied  curvature terms \eqref{varyids} to reveal the field equations for an action of the type \eqref{actionboxn}.
\subsubsection*{Arbitrary functions of $\Box$}
We may generalise further by considering actions of the type
\bg
S_{F}\sim\int d^{4}x\sqrt{-g}R{\cal F}(\Box)R,
\en 
where ${\cal F}(\Box)$
  is an arbitrary function of D'Alembertian operators of the form 
  \[
  {\cal F}(\Box)\equiv\sum_{n=0}^{\infty}f_{1_{n}}\frac{\Box^{n}}{M^{2n}}.
\]
Here, $f_{1_{n}}$ are arbitrary constants and each non-local operator is modulated by the scale of non-locality, $M$. For such an action, the analogue of \eqref{deltaboxn} is given by
\[
\label{deltaboxf}
R\delta{\cal F}_{1}(\Box)R=\sum_{n=1}^{\infty}\frac{f_{1_{n}}}{M^{2n}}\sum_{m=0}^{n-1}R\Box^{m}\delta(\Box)\Box^{n-m-1}R.
\]
Furthermore, the prescription
\bg
\label{trick}
\int d^4 x\sqrt{-g}\sum_{l}\sum_{m}\Box^{m}A\Box^{l}B=\int d^4 x\sqrt{-g}\sum_{l}\sum_{m}\Box^{l}A\Box^{m}B,
\en 
where $A$ and $B$ are arbitrary tensors, allows us to recast the identity into the following manageable form
\bg
R\delta{\cal F}_{1}(\Box)R=\sum_{n=1}^{\infty}f_{1_{n}}\sum_{m=0}^{n-1}\Box^{m}R\delta(\Box)\Box^{n-m-1}R.
\en 
A final important point is that as $R$
  is a scalar, so too is $\Box^{n-m-1}R$,
  and as such the derived identity for $\delta(\Box)R$
  remains valid for our intentions, where one simply has to substitute $R\rightarrow\Box^{n-m-1}R$
  in \eqref{deltaboxR}. We shall see in the subsequent sections, the central role these observations play in deriving the full field equations for \eqref{action}.
\subsubsection*{Full Action}  
  We are now ready to compute the variation of our action \eqref{action}. For simplicity of presentation, we proceed by decomposing the action into its constituent parts which we denote $S_{0,...,3}$. We define the the gravitational energy momentum tensor as
 \begin{equation}
 T^{\mu\nu}=-\frac{2}{\sqrt{-g}}\frac{\delta S}{\delta g_{\mu\nu}},
 \end{equation}
 where $g\equiv |\det g_{\mu\nu}|$ is the determinant of the metric tensor, and compute the contribution to $T^{\mu\nu}$ for the individual sectors of the action separately.
\subsection{$S_0$}
$S_0$ is nothing more than the Einstein-Hilbert action
\begin{equation}
S_{0}=\frac{1}{2}\int d^{4}x\sqrt{-g}\left(M_P^2R-2\Lambda\right),
  \end{equation}
where, in our formalism, we have taken the cosmological constant $\Lambda$ to be of mass dimension $4$. Varying the action and substituting the identity for the varied curvature scalar \eqref{varyids},  along with \eqref{deltasqrt}, leads to the Einstein equation
\begin{equation}
\begin{aligned} T^{\mu\nu} & =M_P^2 G^{\mu\nu}+g^{\mu\nu}\Lambda\end{aligned}
 \end{equation}
 \subsection{$S_1$}
 The next step is to compute the variation of
\begin{equation}
S_{1}=\frac{1}{2}\int d^{4}x\sqrt{-g}R{\cal F}_{1}(\Box)R\ .
\end{equation}
Varying this and substituting values for $\delta R$ and $\delta \sqrt{-g}$, given in \eqref{varyids} and \eqref{deltasqrt}, respectively, we
find 
\begin{align}
\delta S_{1}&=\frac{1}{2}\int d^{4}x\sqrt{-g}\biggl[\biggl(\frac{1}{2}g^{\mu\nu}R{\cal F}_{1}(\Box)R+2\nabla^{\mu}\nabla^{\nu}{\cal F}_{1}(\Box)R
\nn\\&
-2g^{\mu\nu}\square{\cal F}_{1}(\Box)R-2R^{\mu\nu}{\cal F}_{1}(\Box)R\biggr)h_{\mu\nu}+R\delta{\cal F}_{1}(\Box)R\biggr]
 \end{align}
where we have integrated by parts where appropriate. To calculate the final term, we employ the identity in \eqref{deltaboxf} and substitute the value for $\delta(\Box)R$ given in \eqref{deltaboxR}. We then integrate by parts in order to factor out the perturbed metric $h_{\mu\nu}$. Further terms will simplify by noting the double summation relation given by \eqref{trick}, until we arrive at the energy-momentum tensor contribution, which is given by
\begin{align}
T_{1}^{\mu\nu}&=2G^{\mu\nu}{\cal F}_{1}(\Box)R+\frac{1}{2}g^{\mu\nu}R{\cal F}_{1}(\Box)R-2\left(\nabla^{\mu}\nabla^{\nu}-g^{\mu\nu}\Box\right){\cal F}_{1}(\Box)R
\nn\\&
-\Omega_{1}^{\mu\nu}+\frac{1}{2}g^{\mu\nu}(\Omega_{1\sigma}^{\;\sigma}+\bar{\Omega}_{1})\,,
 \end{align}
where we have defined the symmetric tensors
\begin{equation}
\Omega_{1}^{\alpha\beta}=\sum_{n=1}^{\infty}f_{1_{n}}\sum_{l=0}^{n-1}\nabla^{
\alpha}R^{(l)}\nabla^{\beta}R^{(n-l-1)},\quad\bar{\Omega}_{1}=\sum_{n=1}^{\infty
}f_{1_{n}}\sum_{l=0}^{n-1}R^{(l)}R^{(n-l)},
\end{equation}
and have introduced the notation $\Box^n R\equiv R^{(n)}$.

\subsection{$S_2$}
We now focus on
\[
\label{s2}
S_{2}=\frac{1}{2}\int d^{4}x\sqrt{-g}\left(R^{\mu\nu}{\cal F}_{2}(\Box)R_{\mu\nu}\right)\,.
 \]

Varying the action, we find
\begin{align}
\delta S_{2}	&=\frac{1}{2}\int d^{4}x\biggl[\delta\sqrt{-g}\left(R^{\mu\nu}{\cal F}_{2}(\Box)R_{\mu\nu}\right)+\sqrt{-g}\delta R^{\mu\nu}{\cal F}_{2}(\Box)R_{\mu\nu}
\nn\\&+\sqrt{-g}R^{\mu\nu}{\cal F}_{2}(\Box)\delta R_{\mu\nu}
	+\sqrt{-g}R^{\mu\nu}\delta{\cal F}_{2}(\Box)R_{\mu\nu}\,\biggr].
\end{align}
Careful substitution of the identities found in \ref{sec:Appvary}, accounts for all but the final term:
\begin{align}
\delta S_{2}	&=	\frac{1}{2}\int d^{4}x\sqrt{-g}\biggl[\biggl(\frac{1}{2}g^{\mu\nu}R^{\sigma\tau}{\cal F}_{2}(\Box)R_{\sigma\tau}-2R_{\sigma}^{\nu}{\cal F}_{2}(\Box)R^{\sigma\mu}+2\nabla_{\sigma}\nabla^{\mu}{\cal F}_{2}(\Box)R^{\sigma\nu}
		\nn\\&
		-\square{\cal F}_{2}(\Box)R^{\mu\nu}-g^{\mu\nu}\nabla_{\sigma}\nabla_{\tau}{\cal F}_{2}(\Box)R^{\sigma\tau}\biggl)h_{\mu\nu}+\int d^{4}x\sqrt{-g}R^{\mu\nu}\delta{\cal F}_{2}(\Box)R_{\mu\nu}\biggr].
 \label{varys3}
\end{align}
To compute the final term, we employ the same method outlined in the general methodology section, albeit with an added degree of difficulty. Here, we reiterate the main steps. In terms of the Ricci tensor, the analogous identity of (\ref{deltaboxn}) is arrived at by identical means,
\begin{equation}
\label{deltaFRuv}
R^{\mu\nu}\delta{\cal F}_2(\Box)R_{\mu\nu}=\sum_{m=0}^{
n-1 }
\sum_{n=1}^{\infty}f_{2_n}R^{\mu\nu(m)}\delta(\square)R_{\mu\nu}^{(n-m-1)}\,.
\end{equation}
Of vital importance, however, is the form of $\delta(\Box)R_{\mu\nu}$. We expand out the components of the D'Alembertian operator, in the same manner as in the scalar case, before taking the variation
\begin{equation}
\label{deltaboxRuv1}
\delta(\Box)R_{\mu\nu}=\delta(g^{\sigma\tau}\nabla_\sigma \nabla_\tau)R_{\mu\nu}=h^{\sigma\tau}\nabla_\sigma \nabla_\tau R_{\mu\nu}+g^{\sigma\tau}\delta(\nabla_\sigma)\nabla_\tau R_{\mu\nu}+g^{\sigma\tau}\nabla_\sigma \delta(\nabla_\tau)R_{\mu\nu}.
\end{equation}
As in the scalar case, \eqref{deltaboxR}, we compare $\delta(\nabla_\tau R_{\mu\nu})$ with $\nabla_\tau \delta R_{\mu\nu}$ to find
\begin{equation}
\delta(\nabla_{\tau})R_{\mu\nu}=-\delta\Gamma_{\tau\mu}^{\kappa}R_{\kappa\nu}-\delta\Gamma_{\tau\nu}^{\kappa}R_{\mu\kappa}
 \nn 
\end{equation}
\begin{equation}
\label{deltanablaRuv}
\delta(\nabla_{\sigma})\nabla_{\tau}R_{\mu\nu}=-\delta\Gamma_{\tau\mu}^{\kappa}\nabla_{\tau}R_{\kappa\nu}-\delta\Gamma_{\tau\nu}^{\kappa}\nabla_{\tau}R_{\mu\kappa}-\delta\Gamma_{\tau\nu}^{\kappa}\nabla_{\kappa}R_{\mu\nu}.
\end{equation}
At this point, in order to keep track of the indices, it is convenient to rewrite the perturbed Christoffel symbol \eqref{variedChris}, like so
\begin{equation}
\label{Christoffelhab}
\delta\Gamma_{\mu\nu}^{\lambda}=\frac{1}{2}\left(\delta_{\nu}^{\beta}g^{\alpha\lambda}\nabla_{\mu}+\delta_{\mu}^{\beta}g^{\alpha\lambda}\nabla_{\nu}-\delta_{\mu}^{\alpha}\delta_{\nu}^{\beta}\nabla^{\lambda}\right)h_{\alpha\beta}.
 \end{equation}
 We then substitute this into \eqref{deltanablaRuv} and in turn \eqref{deltaboxRuv1} to find the relevant identity:
 \begin{align}
 \label{deltaBoxSuv}
 \delta(\square)R_{\mu\nu}	&=	\biggl[-\nabla^{\alpha}\nabla^{\beta}R_{\mu\nu}-\nabla^{\alpha}R_{\mu\nu}\nabla^{\beta}+\frac{1}{2}g^{\alpha\beta}\nabla_{\sigma}R_{\mu\nu}\nabla^{\sigma}
		\nn\\&
		-\frac{1}{2}\delta_{(\mu}^{\beta}R_{\;\nu)}^{\alpha}\Box+\frac{1}{2}\delta_{(\mu}^{\beta}R_{\tau\nu)}\nabla^{\alpha}\nabla^{\tau}-\frac{1}{2}R_{\;(\nu}^{\alpha}\nabla^{\beta}\nabla_{\mu)}
		\nn\\&
		-\nabla^{\beta}R_{\;(\nu}^{\alpha}\nabla_{\mu)}-\delta_{(\mu}^{\beta}\nabla^{\lambda}R_{\;\nu)}^{\alpha}\nabla_{\lambda}+\delta_{(\mu}^{\beta}\nabla^{\alpha}R_{\tau\nu)}\nabla^{\tau}\biggr]h_{\alpha\beta}.
 \end{align}
Having established the form of $\delta(\square)R_{\mu\nu}$, we are now in a position to tame the troublesome term \eqref{deltaFRuv} into something manageable, like so
\begin{equation}
\int d^{4}x\sqrt{-g}R^{\mu\nu}\delta{\cal F}_{2}(\Box)R_{\mu\nu}=\int d^{4}x\sqrt{-g}\left(\Omega_{2}^{\mu\nu}-\frac{1}{2}g^{\mu\nu}(\Omega_{2\sigma}^{\;\sigma}+\bar{\Omega}_{2})+2\Delta_{2}^{\mu\nu}\right)h_{\mu\nu}\,.
 \end{equation}
where we have defined the following symmetric tensors,
\begin{equation}
\nn
\Omega_{2}^{\mu\nu}=\sum_{n=1}^{\infty}f_{2_{n}}\sum_{l=0}^{n-1}\nabla^{\mu}R_{\tau}^{\sigma(l)}\nabla^{\nu}R_{\sigma}^{\tau(n-l-1)},\quad\bar{\Omega}_{2}=\sum_{n=1}^{\infty}f_{2_{n}}\sum_{l=0}^{n-1}R_{\tau}^{\sigma(l)}R_{\sigma}^{\tau(n-l)}\,,
\end{equation}
\begin{equation}
\Delta_{2}^{\mu\nu}=\sum_{n=1}^{\infty}f_{2_{n}}\sum_{l=0}^{n-1}\nabla_{\tau}\left(R_{\;\sigma}^{\tau(l)}\nabla^{\mu}R^{\nu\sigma(n-l-1)}-\nabla^{\mu}R_{\;\sigma}^{\tau(l)}R^{\nu\sigma(n-l-1)}\right),
\end{equation}
This, combined with (\ref{varys3}), gives us the contribution of the Ricci tensor terms to the energy-momentum tensor
\begin{align}
T_{2}^{\mu\nu}	&=	-\frac{1}{2}g^{\mu\nu}R^{\sigma\tau}{\cal F}_{2}(\Box)R_{\sigma\tau}+2R_{\sigma}^{\nu}{\cal F}_{2}(\Box)R^{\sigma\mu}-2\nabla_{\sigma}\nabla^{\mu}{\cal F}_{2}(\Box)R^{\sigma\nu}
		\nn\\&+\square{\cal F}_{2}(\Box)R^{\mu\nu}+g^{\mu\nu}\nabla_{\sigma}\nabla_{\tau}{\cal F}_{2}(\Box)R^{\sigma\tau}-\Omega_{2}^{\mu\nu}+\frac{1}{2}g^{\mu\nu}(\Omega_{2\sigma}^{\;\sigma}+\bar{\Omega}_{2})-2\Delta_{2}^{\mu\nu},
 \label{P2}
\end{align}
\subsection{$S_3$}
Finally we focus on the terms involving the Weyl tensors. We proceed in much the same manner as the previous case, with a further layer of complexity due to the number of indices involved. The action we wish to vary is given by
\[
S_{3}=\frac{1}{2}\int d^{4}x\sqrt{-g}C^{\mu\nu\lambda\sigma}{\cal
F}_{3}(\Box)C_{\mu\nu\lambda\sigma},
\]
where the Weyl tensor is defined by
\[
\label{weyl}
C^{\mu}{ }_{\alpha\nu\beta}\equiv R^{\mu}{ }_{\alpha\nu\beta}-\frac{1}{2}(\delta_{\nu}^{
\mu}R_{\alpha\beta}-\delta_{\beta}^{\mu}R_{\alpha\nu}+R_{\nu}^{\mu}g_{
\alpha\beta}-R_{\beta}^{\mu}g_{\alpha\nu})+\frac{R}{6}(\delta_{\nu}^{\mu}g_{
\alpha\beta}-\delta_{\beta}^{\mu}g_{\alpha\nu})
.\]
Varying the action we find
\[\delta S_{3}=\frac{1}{2}\sqrt{-g}\int d^{4}x\biggl[\frac{1}{2}g^{\alpha\beta}h_{\alpha\beta}C^{\mu\nu\lambda\sigma}{\cal F}_{3}(\Box)C_{\mu\nu\lambda\sigma}+\delta\left(C^{\mu\nu\lambda\sigma}{\cal F}_{3}(\Box)C_{\mu\nu\lambda\sigma}\right)\biggr].
 \]
Once again, we intend to arrange the expression in terms of the metric tensor $h_{\alpha\beta}$. The second term develops as follows
\begin{align}
\delta&\left(C^{\mu\nu\lambda\sigma}{\cal F}_{3}(\Box)C_{\mu\nu\lambda\sigma}\right)	=	\delta C^{\mu\nu\lambda\sigma}{\cal F}_{3}(\Box)C_{\mu\nu\lambda\sigma}+C^{\mu\nu\lambda\sigma}\delta({\cal F}_{3}(\Box)C_{\mu\nu\lambda\sigma})
	\nn\\&=	\delta(g^{\nu\rho}g^{\phi\lambda}g^{\sigma\psi}C^{\mu}{ }_{\rho\phi\psi}){\cal F}_{3}(\Box)C_{\mu\nu\lambda\sigma}+C^{\mu\nu\lambda\sigma}\delta({\cal F}_{3}(\Box)g_{\mu\rho}C^{\rho}{ }_{\nu\lambda\sigma})
	\nn\\&=	-4h^{\nu\rho}C^{\mu}{ }_{\rho\phi\psi}{\cal F}_{3}(\Box)C_{\mu\nu}{ }{ }^{\phi\psi}+\delta C^{\mu}{ }_{\rho\phi\psi}{\cal F}_{3}(\Box)C_{\mu}{ }^{\rho\phi\psi}+C_{\rho}{ }^{\nu\lambda\sigma}\delta({\cal F}_{3}(\Box)C^{\rho}{ }_{\nu\lambda\sigma})
	\nn\\&=	-4h^{\alpha\beta}C_{\beta\mu\nu\lambda}^{\;}{\cal F}_{3}(\Box)C_{\alpha}{ }^{\mu\nu\lambda}+2\delta C^{\mu}{ }_{\nu\lambda\sigma}{\cal F}_{3}(\Box)C_{\mu}{ }^{\nu\lambda\sigma}+C_{\mu}{ }^{\nu\lambda\sigma}\delta{\cal F}_{3}(\Box)C^{\mu}{ }_{\nu\lambda\sigma}
  \end{align}
Next, using the definition of the Weyl tensor (\eqref{weyl}), we note that
\[
\delta C^{\mu}{ }_{\nu\lambda\sigma}{\cal F}_{3}(\Box)C_{\mu}{ }^{\nu\lambda\sigma}=\left(\delta R^{\mu}{ }_{\nu\lambda\sigma}-\frac{1}{2}(R_{\lambda}^{\mu}h_{\nu\sigma}-R_{\sigma}^{\mu}h_{\nu\lambda})\right){\cal F}_{3}(\Box)C_{\mu}{ }^{\nu\lambda\sigma}.
\]
Here, we have used the essential property 
\[
C^{\mu}{ }_{\nu\mu\lambda}=0
,\]
which is due to the fact that the Weyl tensor is the traceless component of the Riemann tensor. We then reformulate the variation of the Riemann tensor \eqref{varyids} like so
\[
\delta R^{\lambda}{ }_{\mu\sigma\nu}=\frac{1}{2}\left(g^{\alpha\lambda}\delta_{\nu}^{\beta}\nabla_{\sigma}\nabla_{\mu}-\delta_{\mu}^{\alpha}\delta_{\nu}^{\beta}\nabla_{\sigma}\nabla^{\lambda}-g^{\alpha\lambda}\delta_{\sigma}^{\beta}\nabla_{\nu}\nabla_{\mu}+\delta_{\mu}^{\alpha}\delta_{\sigma}^{\beta}\nabla_{\nu}\nabla^{\lambda}\right)h_{\alpha\beta}
, \]
and substitute to find
\[
2\delta C^{\mu}{ }_{\nu\lambda\sigma}{\cal F}_{3}(\Box)C_{\mu}{ }^{\nu\lambda\sigma}	=	-2\left(R_{\nu}^{\mu}+2\nabla_{\nu}\nabla^{\mu}\right){\cal F}_{3}(\Box)C_{\mu}{ }^{\alpha\nu\beta}h_{\alpha\beta},
  \]
where we have integrated by parts where appropriate. The variation of the action, thus far, is then given by
\begin{align}
\delta S_{3}	&=	\frac{1}{2}\int d^{4}x\sqrt{-g}\biggl[\biggl(\frac{1}{2}g^{\alpha\beta}C^{\mu\nu\lambda\sigma}{\cal F}_{3}(\Box)C_{\mu\nu\lambda\sigma}-4C_{\;\mu\nu\lambda}^{\beta}{\cal F}_{3}(\Box)C^{\alpha\mu\nu\lambda}
\nn\\&-2\left(R_{\mu\nu}+2\nabla_{\nu}\nabla_{\mu}\right){\cal F}_{3}(\Box)C^{\mu\alpha\nu\beta}\biggr)h_{\alpha\beta}+\frac{1}{2}C_{\mu}{ }^{\nu\lambda\sigma}\delta{\cal F}_{3}(\Box)C^{\mu}{ }_{\nu\lambda\sigma}\biggr].
 \end{align}
To compute the final term, we proceed as in the previous cases to derive $\delta(\Box)$ acting upon the Weyl tensor:
\begin{align}
\delta(\Box)C_{\mu\nu\lambda\sigma}	&=	\biggl[-\nabla^{\alpha}\nabla^{\beta}C_{\mu\nu\lambda\sigma}-\nabla^{\alpha}C_{\mu\nu\lambda\sigma}\nabla^{\beta}+\frac{1}{2}g^{\alpha\beta}\nabla_{\tau}C_{\mu\nu\lambda\sigma}\nabla^{\tau}
		\nn\\&-\frac{1}{2}C^{\alpha}{ }_{\nu\lambda\sigma}\nabla^{\beta}\nabla_{\mu}+\frac{1}{2}C^{\alpha}{ }_{\mu\lambda\sigma}\nabla^{\beta}\nabla_{\nu}+\frac{1}{2}C^{\alpha}{ }_{\mu\nu\sigma}\nabla^{\beta}\nabla_{\lambda}+\frac{1}{2}C^{\alpha}{ }_{\lambda\mu\nu}\nabla^{\beta}\nabla_{\sigma}
		\nn\\&-\nabla^{\beta}C^{\alpha}{ }_{\nu\lambda\sigma}\nabla_{\mu}+\nabla^{\beta}C^{\alpha}{ }_{\mu\lambda\sigma}\nabla_{\nu}-\nabla^{\beta}C^{\alpha}{ }_{\sigma\mu\nu}\nabla_{\lambda}+\nabla^{\beta}C^{\alpha}{ }_{\lambda\mu\nu}\nabla_{\sigma}\biggr]h_{\alpha\beta}.
 \end{align}
From this, we deduce
\[
\int
d^{4}x\sqrt{-g} C^{\mu\nu\lambda\sigma}\delta{\cal
F}_{3}(\Box)C_{\mu\nu\lambda\sigma}=\int
d^{4}x\sqrt{-g}\left(\Omega_{3}^{\alpha\beta}-\frac{1}{2}g^{
\alpha\beta}(\Omega_{3\gamma}^{\;\gamma}+\bar{\Omega}_{3})+4\Delta_{3}^{
\alpha\beta}\right)h_{\alpha\beta}\,,
\]
by defining
\[
\Omega_{3}^{\alpha\beta}=\sum_{n=1}^{\infty}f_{3_{n}}\sum_{l=0}^{n-1}\nabla^{\alpha}C_{\;\nu\lambda\sigma}^{\mu(l)}\nabla^{\beta}C_{\mu}^{\;\nu\lambda\sigma(n-l-1)},\quad\bar{\Omega}_{3}=\sum_{n=1}^{\infty}f_{3_{n}}\sum_{l=0}^{n-1}C_{\;\nu\lambda\sigma}^{\mu(l)}C_{\mu}^{\;\nu\lambda\sigma(n-l)}\,,
\nn \]
\[
 \Delta_{3}^{\alpha\beta}=\sum_{n=1}^{\infty}f_{3_{n}}\sum_{l=0}^{n-1}\nabla_{\nu}\left(C_{\;\;\;\sigma\mu}^{\lambda\nu(l)}\nabla^{\alpha}C_{\lambda}^{\;\beta\sigma\mu(n-l-1)}-\nabla^{\alpha}C_{\;\;\;\sigma\mu}^{\lambda\nu\;\;(l)}C_{\lambda}^{\;\beta\sigma\mu(n-l-1)}\right).
 \]
As such, the contribution to the energy-momentum tensor is then
\begin{align}
T_{3}^{\mu\nu}	&=	-\frac{1}{2}g^{\mu\nu}C^{\sigma\tau\lambda\rho}{\cal F}_{3}(\Box)C_{\sigma\tau\lambda\rho}+4C_{\;\sigma\tau\lambda}^{\nu}{\cal F}_{3}(\Box)C^{\mu\sigma\tau\lambda}+2\left(R_{\sigma\tau}+2\nabla_{\tau}\nabla_{\sigma}\right){\cal F}_{3}(\Box)C^{\sigma\mu\tau\nu}
		\nn\\&-\Omega_{3}^{\mu\nu}+\frac{1}{2}g^{\mu\nu}(\Omega_{3\gamma}^{\;\gamma}+\bar{\Omega}_{3})-4\Delta_{3}^{\mu\nu},
   \end{align}
   where we have reintroduced the original $_\mu,_\nu$ notation by relabelling the indices.
\subsection{The Complete Field Equations}
We are now in a position to state the full equations of motion for the action $S$ in
(\ref{action}) as a combination of $S_0,\;S_1,\;S_2$ and $S_3$ derived in the previous sections.
\begin{align}
\label{eomfull}
T_{\nu}^{\mu}	&=	M_{P}^{2}G_{\nu}^{\mu}+\delta_{\nu}^{\mu}\Lambda+2\lambda G_{\nu}^{\mu}{\cal F}_{1}(\Box)R+\frac{\lambda}{2}\delta_{\nu}^{\mu}R{\cal F}_{1}(\Box)R-2\lambda\left(\nabla^{\mu}\partial_{\nu}-\delta_{\nu}^{\mu}\square\right){\cal F}_{1}(\Box)R
		\nn\\&+2\lambda R_{\sigma}^{\mu}{\cal F}_{2}(\Box)R_{\;\nu}^{\sigma}-\frac{\lambda}{2}\delta_{\nu}^{\mu}R_{\;\tau}^{\sigma}{\cal F}_{2}(\Box)R_{\;\sigma}^{\tau}-2\lambda\nabla_{\sigma}\nabla_{\nu}{\cal F}_{2}(\Box)R^{\mu\sigma}+\lambda\square{\cal F}_{2}(\Box)R_{\nu}^{\mu}
		\nn\\&+\lambda\delta_{\nu}^{\mu}\nabla_{\sigma}\nabla_{\tau}{\cal F}_{2}(\Box)R^{\sigma\tau}-\frac{\lambda}{2}\delta_{\nu}^{\mu}C^{\sigma\tau\lambda\rho}{\cal F}_{3}(\Box)C_{\sigma\tau\lambda\rho}+4\lambda C_{\;\sigma\tau\lambda}^{\mu}{\cal {\cal F}}_{3}(\square)C_{\nu}^{\;\sigma\tau\lambda}
		\nn\\&-2\lambda\left(R_{\sigma\tau}+2\nabla_{\sigma}\nabla_{\tau}\right){\cal {\cal F}}_{3}(\square)C_{\nu}^{\;\sigma\tau\mu}-\lambda\Omega_{1\nu}^{\mu}+\frac{\lambda}{2}\delta_{\nu}^{\mu}(\Omega_{1\sigma}^{\;\sigma}+\bar{\Omega}_{1})
		\nn\\&-\lambda\Omega_{2\nu}^{\mu}+\frac{\lambda}{2}\delta_{\nu}^{\mu}(\Omega_{2\sigma}^{\sigma}+\bar{\Omega}_{2})-2\lambda\Delta_{2\nu}^{\mu}-\lambda\Omega_{3\nu}^{\mu}+\frac{\lambda}{2}\delta_{\nu}^{\mu}(\Omega_{3\gamma}^{\gamma}+\bar{\Omega}_{3})-4\lambda\Delta_{3\nu}^{\mu},
   \end{align}
where $T^\mu_{\nu}$ is the stress energy tensor for the matter components of the Universe \footnote{We have lowered an index for convenience when analysing perturbations later in the text.} and we restate the symmetric tensors, we defined earlier:
\[\nn
\label{Omegas}
\Omega_{1\nu}^{\mu}=\sum_{n=1}^{\infty}f_{1_{n}}\sum_{l=0}^{n-1}\partial^{\mu}R^{(l)}\partial_{\nu}R^{(n-l-1)},\quad\bar{\Omega}_{1}=\sum_{n=1}^{\infty}f_{1_{n}}\sum_{l=0}^{n-1}R^{(l)}R^{(n-l)},
 \]
\[\nn
\Omega_{2\nu}^{\mu}=\sum_{n=1}^{\infty}f_{2_{n}}\sum_{l=0}^{n-1}\nabla^{\mu}R_{\tau}^{\sigma(l)}\nabla_{\nu}R_{\sigma}^{\tau(n-l-1)},\quad\bar{\Omega}_{2}=\sum_{n=1}^{\infty}f_{2_{n}}\sum_{l=0}^{n-1}R_{\tau}^{\sigma(l)}R_{\sigma}^{\tau(n-l)}\,,
 \]
\[\nn
\Delta_{2\nu}^{\mu}=\sum_{n=1}^{\infty}f_{2_{n}}\sum_{l=0}^{n-1}\nabla_{\tau}\left(R_{\;\sigma}^{\tau(l)}\nabla^{\mu}R^{\nu\sigma(n-l-1)}-\nabla^{\mu}R_{\;\sigma}^{\tau(l)}R^{\nu\sigma(n-l-1)}\right)\,,
 \]
\[\nn
\Omega_{3\nu}^{\mu}=\sum_{n=1}^{\infty}f_{3_{n}}\sum_{l=0}^{n-1}\nabla^{\mu}C_{\;\tau\lambda\rho}^{\sigma(l)}\nabla_{\nu}C_{\sigma}^{\;\tau\lambda\rho(n-l-1)},\;\bar{\Omega}_{3}=\sum_{n=1}^{\infty}f_{3_{n}}\sum_{l=0}^{n-1}C_{\;\tau\lambda\rho}^{\sigma(l)}C_{\sigma}^{\;\tau\lambda\rho(n-l)}\,,
 \]
\[
\Delta_{3\nu}^{\mu}=\sum_{n=1}^{\infty}f_{3_{n}}\sum_{l=0}^{n-1}\nabla_{\tau}\left(C_{\;\;\lambda\rho\sigma}^{\tau(l)}\nabla^{\mu}C_{\nu}^{\;\lambda\rho\sigma(n-l-1)}-\nabla^{\mu}C_{\;\;\lambda\rho\sigma}^{\tau(l)}C_{\nu}^{\;\lambda\rho\sigma(n-l-1)}\right).
 \]
The trace equation is often particularly useful and we provide it below for the
general action (\ref{action}):
\begin{align}
T	&=	-M_{P}^{2}R+6\lambda\square{\cal F}_{1}(\Box)R+\lambda\square{\cal F}_{2}(\Box)R-2\lambda\nabla_{\sigma}\nabla_{\tau}{\cal F}_{2}(\Box)R^{\sigma\tau}+2\lambda C^{\mu\nu\lambda\sigma}{\cal F}_{3}(\Box)C_{\mu\nu\lambda\sigma}
	\nn\\&	+\lambda\Omega_{1\sigma}^{\;\sigma}+2\lambda\bar{\Omega}_{1}+\lambda\Omega_{2\sigma}^{\;\sigma}+2\lambda\bar{\Omega}_{2}+\lambda\Omega_{3\sigma}^{\;\sigma}+2\lambda\bar{\Omega}_{3}-2\lambda\Delta_{2\sigma}^{\;\sigma}-4\lambda\Delta_{3\sigma}^{\;\sigma}
 \label{trace}
\end{align}

\section{Linearised Field Equations around Minkowski Space}
\label{sec:linkmink}
In order to make a step towards understanding the physical implications of the non-local gravitational theory described by the action \eqref{action}, we consider the linear approximation of the theory, by analysing small fluctuations around Minkowski space, according to the algorithm 
\[
\label{pertmink}
g_{\mu\nu}=\eta_{\mu\nu}+ h_{\mu\nu},\qquad g^{\mu\nu}=\eta^{\mu\nu}-h^{\mu\nu}
.\]
Here, $\eta_{\mu\nu}$ is the Minkowski metric and $h_{\mu\nu}\equiv \delta g_{\mu\nu}$ is the variation with respect to the metric tensor. From \eqref{varyids}, we can read off the relevant curvatures up to linear order
\[
\label{MinkRiem}
R^\rho_{\;\mu\sigma\nu}=\frac{1}{2}\left(\partial_{\sigma}\partial_{\mu}h^\rho_{\nu}
+\partial_{\nu}\partial^\rho h_{\mu\sigma}-\partial_{\nu}\partial_{\mu}h^\rho_{\sigma}-\partial_{\sigma}\partial^{\rho}h_{\mu\nu}\right)\,,
\nn\]
\[
\nn\label{MinkRicci}
R_{\mu\nu}=\frac{1}{2}\left(\partial_{\sigma}\partial_{\mu}h_{\nu}^{\sigma}
+\partial_{\nu}\partial_{\sigma}h_{\mu}^{\sigma}-\partial_{\nu}\partial_{\mu}
h-\Box h_{\mu\nu}\right)\,,
\\
\] 
\[
\label{MinkR}
R=\partial_{\mu}\partial_{\nu}h^{\mu\nu}-\square h\,,
\]
 where $\Box=g^{\mu\nu}\nabla_{\mu} \nabla_{\nu}=\eta^{\mu\nu}\partial_\mu \partial_\nu$.
Substitution of  the above curvatures into \eqref{eomfull} reveals the linearised equations of motion around a Minkowski background
\begin{align}
\label{eommink}
\kappa T_{\nu}^{\mu}	&=	-\frac{1}{2}\left[1+\lambda M_{P}^{-2}{\cal F}_{2}(\Box)\square+2M_{P}^{-2}\lambda{\cal F}_{3}(\Box)\square\right]\square h_{\nu}^{\mu}
		\nn\\&+\frac{1}{2}\left[1+\lambda M_{P}^{-2}{\cal F}_{2}(\Box)\Box+2M_{P}^{-2}\lambda{\cal F}_{3}(\Box)\square\right]\partial_{\sigma}(\partial^{\mu}h_{\nu}^{\sigma}+\partial_{\nu}h^{\mu\sigma})\nn
		\\&-\frac{1}{2}\left[1-4M_{P}^{-2}\lambda{\cal F}_{1}(\Box)\square-\lambda M_{P}^{-2}{\cal F}_{2}(\Box)\square+\frac{2}{3}M_{P}^{-2}\lambda{\cal F}_{3}(\Box)\square\right]\left(\partial_{\nu}\partial^{\mu}h+\delta_{\nu}^{\mu}\partial_{\sigma}\partial_{\tau}h^{\sigma\tau}\right)
		\nn\\&+\frac{1}{2}\left[1-4M_{P}^{-2}\lambda{\cal F}_{1}(\Box)\Box-\lambda M_{P}^{-2}{\cal F}_{2}(\Box)\Box+\frac{2}{3}M_{P}^{-2}\lambda{\cal F}_{3}(\Box)\square\right]\delta_{\nu}^{\mu}\Box h\nn
		\\&-\left[2\lambda M_{P}^{-2}{\cal F}_{1}(\Box)+\lambda M_{P}^{-2}{\cal F}_{2}(\Box)+\frac{2}{3}M_{P}^{-2}\lambda{\cal F}_{3}(\Box)\right]\partial^{\mu}\partial_{\nu}\partial_{\sigma}\partial_{\tau}h^{\sigma\tau}.
   \end{align}
We may represent the linearised field equations as
 \begin{align}
 \label{Tababc}
-\kappa T_{\mu\nu}	&=\frac{1}{2}\biggl[	a(\Box)\square h_{\mu\nu}+b(\Box)\partial_{\sigma}(\partial_{\mu}h_{\nu}^{\sigma}+\partial_{\nu}h_{\mu}^{\sigma})+c(\Box)\left(\partial_{\nu}\partial_{\mu}h+g_{\mu\nu}\partial_{\sigma}\partial_{\tau}h^{\sigma\tau}\right)
		\nn\\&+d(\Box)g_{\mu\nu}\Box h+f(\Box)\partial_{\mu}\partial_{\nu}\partial_{\sigma}\partial_{\tau}h^{\sigma\tau}\biggr]
,   \end{align}
by defining the following infinite derivative functions
\ba
\nn
&a(\Box)\equiv1+M_{P}^{-2}{\cal F}_{2}(\Box)\square+2M_{P}^{-2}{\cal F}_{3}(\Box)\Box=-b(\Box)
 \\
&c(\Box)\equiv1-4M_{P}^{-2}{\cal F}_{1}(\Box)\square-M_{P}^{-2}{\cal F}_{2}(\Box)\square+\frac{2}{3}M_{P}^{-2}{\cal F}_{3}(\Box)\Box=-d(\Box)
 \nn\\&
f(\Box)\equiv 4M_{P}^{-2}{\cal F}_{1}(\Box)+2M_{P}^{-2}{\cal F}_{2}(\Box)+\frac{4}{3}M_{P}^{-2}{\cal F}_{3}(\Box)
  \label{abc}.
 \ea
One can then confirm the following relations
 \begin{align}
 \label{minkconstr}
a(\Box)+b(\Box)&=0\nn
\\c(\Box)+d(\Box)&=0\nn
\\b(\Box)+c(\Box)+f(\Box)\Box&=0
. \end{align}
These identities were found by explicit evaluation of the respective terms, but can be understood, more intuitively, as a consequence of the Bianchi identities. The stress energy tensor of any minimally coupled diffeomorphism-invariant gravitational action must be conserved, i.e.
\[
\nabla_\mu T^\mu_\nu=0
.\]
This applies equally to the linearised  equations of motion \eqref{Tababc} as it does to the full non-linear field equations \eqref{eomfull}. Recall that, in this case, $\Box=g^{\mu\nu}\nabla_{\mu} \nabla_{\nu}=\eta^{\mu\nu}\partial_\mu \partial_\nu\equiv\partial^2$, so that it suffices to take the partial derivative of \eqref{Tababc} in order to test the Bianchi identity. As such, 
\begin{equation}
-\partial_{\mu}T_{\nu}^{\mu}=(a+b)\partial_{\mu}\partial^{2}h_{\nu}^{\mu}+(b+c+f\Box)\partial_{\sigma}\partial_{\mu}\partial_{\nu}h^{\mu\sigma}+(c+d)\partial_{\nu}\partial^{2}h
,
\end{equation}
where we have suppressed the argument in the infinite derivative functions, i.e $f(\Box)\Box=f\Box$, for presentation purposes. This divergence should vanish identically, and when the coefficients of each independent term is compared with (\ref{minkconstr}), it is clear that this is the case. 
Furthermore, by appealing to the form of the curvature around Minkowski space \eqref{MinkR} and the constraints \eqref{abc}, we may recast the field equations \eqref{Tababc} into the following concise form
\[
\label{eomminkred}
\kappa T_{\mu\nu}	=	a(\Box)R_{\mu\nu}-\frac{1}{2}\eta_{\mu\nu}c(\Box)R-\frac{f(\Box)}{2}\partial_{\mu}\partial_{\nu}R
.\]
In this form, it should be immediately apparent that both the tensorial and scalar sectors of the propagator have undergone corrections by the non-local operators $a(\Box)$, $c(\Box)$ and $f(\Box)$, where $f(\Box)$ is related to $a(\Box)$ and $c(\Box)$ by $f(\Box)\Box=(a(\Box)-c(\Box))$. The trace equation is given by
\[
\label{tracemink}
\kappa T=\frac{1}{2}(a(\Box)-3c(\Box))R
\]
and will play an important role in the derivation of the ghost-free condition of the IDG theory due to its correspondence to the scalar sector of the propagator.
\section{Linearised Field Equations around de Sitter Space}
\subsection*{Reformulation of Equations of Motion}

In order to make an infinite series of D'Alembertian operators acting on the Ricci tensor more tractable in spacetimes other than Minkowski, we introduce the \emph{traceless Einstein tensor} \cite{Biswas:2016etb}
\[
S_{\nu}^{\mu}\equiv R_{\nu}^{\mu}-\frac{1}{4}\delta_{\nu}^{\mu}R,
 \]
and define
\[
\label{Ftilde}
\tilde{{\cal F}}_{1}(\Box)\equiv{\cal F}_{1}(\Box)+\frac{1}{4}{\cal F}_{2}(\Box),
 \]
so that we may write the action \eqref{action} in terms of the traceless Einstein tensor
\[
S=\int d^{4}x\frac{\sqrt{-g}}{2}\left[M_{P}^{2}R+\lambda R\tilde{{\cal F}}_{1}(\Box)R+\lambda S_{\nu}^{\mu}{\cal F}_{2}(\Box)S_{\mu}^{\nu}+C_{\mu\nu\sigma\tau}{\cal F}_{3}(\Box)C^{\mu\nu\sigma\tau}-2\Lambda\right]
, \]
with the resulting equations of motion given by
\[
\label{eomS}
\begin{aligned}M_{P}^{2}G_{\nu}^{\mu} & =T_{\nu}^{\mu}-\delta_{\nu}^{\mu}\Lambda-2\lambda S_{\nu}^{\mu}{\cal \tilde{F}}_{1}(\Box)R+2\lambda\left(\nabla^{\mu}\partial_{\nu}-\delta_{\nu}^{\mu}\square\right)\tilde{{\cal F}}_{1}(\Box)R\\
 & -\frac{\lambda}{2}R{\cal F}_{2}(\Box)S_{\nu}^{\mu}-2\lambda S_{\;\sigma}^{\mu}{\cal F}_{2}(\Box)S_{\;\nu}^{\sigma}+\frac{\lambda}{2}\delta_{\nu}^{\mu}S_{\tau}^{\sigma}{\cal F}_{2}(\Box)S_{\sigma}^{\tau}\\
 & +2\lambda\nabla_{\sigma}\nabla_{\nu}{\cal F}_{2}(\Box)S^{\mu\sigma}-\lambda\square{\cal F}_{2}(\Box)S_{\;\nu}^{\mu}-\lambda\delta_{\nu}^{\mu}\nabla_{\sigma}\nabla_{\tau}{\cal F}_{2}(\Box)S^{\sigma\tau}\\
 & +\lambda\Theta_{1\nu}^{\;\mu}-\frac{\lambda}{2}\delta_{\nu}^{\mu}\left(\Theta_{1\sigma}^{\;\sigma}+\bar{\Theta}_{1}\right)+\lambda\Theta_{2\nu}^{\;\mu}-\frac{\lambda}{2}\delta_{\nu}^{\mu}\left(\Theta_{2\sigma}^{\;\sigma}+\bar{\Theta}_{2}\right)+2\lambda{\cal E}_{2\nu}^{\;\mu}+\lambda{\cal C}_{\nu}^{\mu}.
\end{aligned}
 \]
Here, we have defined the symmetric tensors
\[
\nn
\Theta_{1\nu}^{\;\mu}=\sum_{n=1}^{\infty}\tilde{f}_{1_{n}}\sum_{l=0}^{n-1}\left(\partial^{\mu}R^{(l)}\partial_{\nu}R^{(n-l-1)}\right),\quad\bar{\Theta}_{1}=\sum_{n=1}^{\infty}\tilde{f}_{1_{n}}\sum_{l=0}^{n-1}R^{(l)}R^{(n-l)}
\]
\[
 \Theta_{2\nu}^{\;\mu}=\sum_{n=1}^{\infty}f_{2_{n}}\sum_{l=0}^{n-1}\left(\nabla^{\mu}S_{\tau}^{\sigma(l)}\nabla_{\nu}S_{\sigma}^{\tau(n-l-1)}\right),\quad\bar{\Theta}_{2}=\sum_{n=1}^{\infty}f_{2_{n}}\sum_{l=0}^{n-1}S_{\tau}^{\sigma(l)}S_{\sigma}^{\tau(n-l)}
 \nn
 \]
\[
\label{symtens}
{\cal E}_{2\nu}^{\;\mu}=\sum_{n=1}^{\infty}f_{2_{n}}\sum_{l=0}^{n-1}\nabla_{\tau}\biggl(S_{\lambda}^{\tau(l)}\nabla^{\mu}S_{\nu}^{\lambda(n-l-1)}-\nabla^{\mu}S_{\lambda}^{\tau(l)}S_{\nu}^{\lambda(n-l-1)}\biggr)
 ,\]
while ${\cal C}_{\nu}^{\mu}$ represents the contribution from the Weyl tensor which vanishes on the background. The trace equation is obtained by contracting with $\delta_{\mu}^{\nu}$
 \[
 \label{barredT}
\begin{aligned}-M_{P}^{2}R & =T-4\Lambda-6\lambda\Box\tilde{{\cal F}}_{1}(\Box)R-2\lambda\nabla_{\sigma}\nabla_{\tau}{\cal F}_{2}(\Box)S^{\sigma\tau}\\
 & -\lambda\left(\Theta_{1\sigma}^{\;\sigma}+2\bar{\Theta}_{1}\right)-\lambda\left(\Theta_{2\sigma}^{\;\sigma}+2\bar{\Theta}_{2}\right)+2\lambda{\cal E}_{2\sigma}^{\;\sigma}.
\end{aligned}
\]
\subsection*{Linear Perturbations and Equations of Motion around de Sitter} 
\label{sec:pertds}
In order to understand the cosmological significance of a theory, it is often revealing to study perturbations around de Sitter (dS) space. To analyse these fluctuations, we define the algorithm
\[
g_{\mu\nu}\rightarrow\bar{g}_{\mu\nu}+h_{\mu\nu},\qquad g^{\mu\nu}\rightarrow\bar{g}^{\mu\nu}-h^{\mu\nu}
 \]
where $\bar{g}_{\mu\nu}$
  and any subsequent `barred' tensors represent the value of the tensor on the background of de Sitter space, and as such obeys
\[
\bar{R}_{\mu\nu\lambda\sigma}=H^{2}(\bar{g}_{\mu\lambda}\bar{g}_{\nu\sigma}-\bar{g}_{\mu\sigma}\bar{g}_{\nu\lambda}),
  \]
where $H$ is the Hubble parameter (constant) for dS. We then find
\[
\label{barredR}
\bar{R}_{\nu}^{\mu}=3H^{2}\delta_{\nu}^{\mu},\qquad\bar{R}=12H^{2},\qquad\bar{S}_{\nu}^{\mu}=0.
 \]
We can express the value for $\Lambda$ by substituting the above values into the `barred' trace equation. By `barred' trace equation, we mean, replacing each curvature term in \eqref{barredT} with a `barred' background curvature term. Taking these background values for the curvature  allows us to express the cosmological constant as follows
\[
\begin{aligned}\Lambda=\frac{T}{4} & +3M_{P}^{2}H^{2}\end{aligned}
 \]
and as dS is a vacuum solution, we may write
\[
\label{barredlambda}
\Lambda=3M_P^2 H^2.
 \]
As a further check, this may be substituted into the full `barred' equations of motion to find that indeed $T_{\mu\nu}=0$.   
Next, in order to linearise around the background described by these `barred' quantities, we return to the variation principle, from which we know that
\[
\delta\Gamma_{\mu\nu}^{\lambda}=\frac{1}{2}\bar{g}^{\lambda\tau}\left(\nabla_{\mu}h_{\nu\tau}+\nabla_{\nu}h_{\mu\tau}-\nabla_{\tau}h_{\mu\nu}\right)\equiv\gamma_{\mu\nu}^{\lambda}
\nn
\]
\[
 \delta R_{\;\;\sigma\nu}^{\lambda\kappa}	=	-h^{\mu\kappa}\bar{R}_{\;\mu\sigma\nu}^{\lambda}+\bar{g}^{\mu\kappa}\left(\nabla_{\sigma}\gamma_{\mu\nu}^{\lambda}-\nabla_{\nu}\gamma_{\mu\sigma}^{\lambda}\right)\equiv r_{\;\;\sigma\nu}^{\lambda\kappa}
 \nn
\]
\[
\delta R_{\nu}^{\mu}\equiv-h^{\kappa\mu}\bar{R}_{\kappa\nu}+\bar{g}^{\kappa\mu}\left(\nabla_{\lambda}\gamma_{\kappa\nu}^{\lambda}-\nabla_{\nu}\gamma_{\kappa\lambda}^{\lambda}\right)\equiv r_{\nu}^{\mu}
 \nn
 \]
\[
\delta R\equiv-h^{\mu\nu}\bar{R}_{\mu\nu}+\nabla_{\tau}\nabla^{\sigma}h_{\sigma}^{\tau}-\Box h\equiv r.
 \]
It is preferable in this instance to vary tensors with an even number of up and down indices. This makes use of the Kronecker delta $\delta^\mu_\nu\equiv g^{\mu\rho}g_{\rho\nu}$, which is invariant under variation. Formulating the variational identities in this way has the advantage that we may contract tensors in a straightforward manner. For example, $\delta_{\nu}^{\mu}\delta R_{\mu}^{\nu}=\delta(\delta_{\nu}^{\mu}R_{\mu}^{\nu})=\delta R$, whereas in general $g^{\mu\nu}\delta R_{\mu\nu}=\delta(g^{\mu\nu}R_{\mu\nu})-\delta g^{\mu\nu}R_{\mu\nu}\neq\delta R$. Substituting the background quantities and expanding the perturbed Christoffels, we find that the perturbed Ricci tensor and curvature scalar are given by
\[
r_{\nu}^{\mu}\equiv-3H^{2}h_{\nu}^{\mu}+\frac{1}{2}\left(\nabla_{\lambda}\nabla^{\mu}h_{\nu}^{\lambda}+\nabla_{\lambda}\nabla_{\nu}h^{\mu\lambda}-\Box h_{\nu}^{\mu}-\nabla_{\nu}\partial^{\mu}h\right)
 \nn
\]
\[
r\equiv-3H^{2}h+\nabla_{\tau}\nabla^{\sigma}h_{\sigma}^{\tau}-\Box h
 .\]
 Subsequently, the perturbed traceless Einstein tensor becomes
\ba
\label{suv}
\delta S_{\nu}^{\mu}&=-3H^{2}h_{\nu}^{\mu}+\frac{1}{2}\left(\nabla_{\lambda}\nabla^{\mu}h_{\nu}^{\lambda}+\nabla_{\lambda}\nabla_{\nu}h^{\mu\lambda}-\Box h_{\nu}^{\mu}-\nabla_{\nu}\partial^{\mu}h\right)
\nn\\&-\frac{1}{4}\delta_{\nu}^{\mu}\left(-3H^{2}h+\nabla_{\tau}\nabla^{\sigma}h_{\sigma}^{\tau}-\Box h\right)\equiv s_{\nu}^{\mu}
 .\ea
The general formalism for linearisation we are following is described below 
\[
\label{perturbgen}
R\rightarrow\bar{R}+r,\qquad r_{\nu}^{\mu}\rightarrow\bar{R}_{\nu}^{\mu}+r_{\nu}^{\mu},\qquad S_{\nu}^{\mu}=\bar{S}_{\nu}^{\mu}+s_{\nu}^{\mu}
,\qquad\mbox{with}\qquad \delta_{\mu}^{\nu}r_{\nu}^{\mu}=r, \]
whereas, upon substitution of the background values, we find 
\[
R\rightarrow12H^{2}+r,\qquad r_{\nu}^{\mu}\rightarrow3H^{2}\delta_{\nu}^{\mu}+r_{\nu}^{\mu},\qquad S_{\nu}^{\mu}\rightarrow s_{\nu}^{\mu}
. \]
 Taking all this into account, we simply substitute the above values along with the background quantities \eqref{barredR} into the field equation \eqref{eomS}, neglecting terms of order $h^2$, to return the linearised equations of motion around de Sitter space. The calculation proceeds as follows: First, we perturb the field equations according to \eqref{perturbgen},
\[ \begin{aligned}M_{P}^{2}&\left((\bar{R}_{\nu}^{\mu}+r_{\nu}^{\mu})-\frac{1}{2}\delta_{\nu}^{\mu}(\bar{R}+r)\right)  =T_{\nu}^{\mu}-\delta_{\nu}^{\mu}\Lambda-2\lambda s_{\nu}^{\mu}{\cal \tilde{F}}_{1}(\Box)(\bar{R}+r)\\
 & +2\lambda\left(\nabla^{\mu}\partial_{\nu}-\delta_{\nu}^{\mu}\square\right)\tilde{{\cal F}}_{1}(\Box)(\bar{R}+r)-\frac{\lambda}{2}(\bar{R}+r){\cal F}_{2}(\Box)s_{\nu}^{\mu}\\
 & -2\lambda s_{\;\sigma}^{\mu}{\cal F}_{2}(\Box)s_{\;\nu}^{\sigma}+\frac{\lambda}{2}\delta_{\nu}^{\mu}s_{\tau}^{\sigma}{\cal F}_{2}(\Box)s_{\sigma}^{\tau}+2\lambda\nabla_{\sigma}\nabla_{\nu}{\cal F}_{2}(\Box)s^{\mu\sigma}\\
 & -\lambda\square{\cal F}_{2}(\Box)s_{\;\nu}^{\mu}-\lambda\delta_{\nu}^{\mu}\nabla_{\sigma}\nabla_{\tau}{\cal F}_{2}(\Box)s^{\sigma\tau}\\
 & +\lambda\Theta_{1\nu}^{\;\mu}-\frac{\lambda}{2}\delta_{\nu}^{\mu}\left(\Theta_{1\sigma}^{\;\sigma}+\bar{\Theta}_{1}\right)+\lambda\Theta_{2\nu}^{\;\mu}-\frac{\lambda}{2}\delta_{\nu}^{\mu}\left(\Theta_{2\sigma}^{\;\sigma}+\bar{\Theta}_{2}\right)+2\lambda{\cal E}_{2\nu}^{\;\mu}+\lambda{\cal C}_{\nu}^{\mu}.
\end{aligned}
\]
Ignoring, for the moment, the symmetric tensors on the bottom line of this equation, we reduce all other terms to linear order in $h$,
\[
\begin{aligned}M_{P}^{2}&\left((\bar{R}_{\nu}^{\mu}+r_{\nu}^{\mu})-\frac{1}{2}\delta_{\nu}^{\mu}(\bar{R}+r)\right) =T_{\nu}^{\mu}-\delta_{\nu}^{\mu}\Lambda-2\lambda s_{\nu}^{\mu}{\cal \tilde{F}}_{1}(\Box)\bar{R}\\
 & +2\lambda\left(\nabla^{\mu}\partial_{\nu}-\delta_{\nu}^{\mu}\square\right)\tilde{{\cal F}}_{1}(\Box)(\bar{R}+r)-\frac{\lambda}{2}\bar{R}{\cal F}_{2}(\Box)s_{\nu}^{\mu}\\
 & +2\lambda\nabla_{\sigma}\nabla_{\nu}{\cal F}_{2}(\Box)s^{\mu\sigma}-\lambda\square{\cal F}_{2}(\Box)s_{\;\nu}^{\mu}-\lambda\delta_{\nu}^{\mu}\nabla_{\sigma}\nabla_{\tau}{\cal F}_{2}(\Box)s^{\sigma\tau}\\
 & +\lambda\Theta_{1\nu}^{\;\mu}-\frac{\lambda}{2}\delta_{\nu}^{\mu}\left(\Theta_{1\sigma}^{\;\sigma}+\bar{\Theta}_{1}\right)+\lambda\Theta_{2\nu}^{\;\mu}-\frac{\lambda}{2}\delta_{\nu}^{\mu}\left(\Theta_{2\sigma}^{\;\sigma}+\bar{\Theta}_{2}\right)+2\lambda{\cal E}_{2\nu}^{\;\mu}+\lambda{\cal C}_{\nu}^{\mu},
\end{aligned}
\] 
before noting that the background curvature, expressed as the `barred' terms given in \eqref{barredR}, are constants. Thus, any derivatives on the background curvature will vanish, so that the field equations reduce further,
\[
\begin{aligned}
M_{P}^{2}&\left((\bar{R}_{\nu}^{\mu}+r_{\nu}^{\mu})-\frac{1}{2}\delta_{\nu}^{\mu}(\bar{R}+r)\right)  =T_{\nu}^{\mu}-\delta_{\nu}^{\mu}\Lambda-2\lambda\tilde{f}_{1_{0}}s_{\nu}^{\mu}\bar{R}+2\lambda\left(\nabla^{\mu}\partial_{\nu}-\delta_{\nu}^{\mu}\square\right)\tilde{{\cal F}}_{1}(\Box)r\\
 & -\frac{\lambda}{2}\bar{R}{\cal F}_{2}(\Box)s_{\nu}^{\mu}+2\lambda\nabla_{\sigma}\nabla_{\nu}{\cal F}_{2}(\Box)s^{\mu\sigma}-\lambda\square{\cal F}_{2}(\Box)s_{\;\nu}^{\mu}-\lambda\delta_{\nu}^{\mu}\nabla_{\sigma}\nabla_{\tau}{\cal F}_{2}(\Box)s^{\sigma\tau}\\
 & +\lambda\Theta_{1\nu}^{\;\mu}-\frac{\lambda}{2}\delta_{\nu}^{\mu}\left(\Theta_{1\sigma}^{\;\sigma}+\bar{\Theta}_{1}\right)+\lambda\Theta_{2\nu}^{\;\mu}-\frac{\lambda}{2}\delta_{\nu}^{\mu}\left(\Theta_{2\sigma}^{\;\sigma}+\bar{\Theta}_{2}\right)+2\lambda{\cal E}_{2\nu}^{\;\mu}+\lambda{\cal C}_{\nu}^{\mu}.
\end{aligned}
\]
Return now to the symmetric tensors on the bottom line. Following the initial perturbation and taking into account the constant nature of the background curvature, the symmetric tensors \eqref{symtens} are given by
\[
\nn
\Theta_{1\nu}^{\;\mu}=\sum_{n=1}^{\infty}\frac{\tilde{f}_{1_{n}}}{M^{2n}}\sum_{l=0}^{n-1}\left(\partial^{\mu}r^{(l)}\partial_{\nu}r^{(n-l-1)}\right),\]\[
\nn\bar{\Theta}_{1}=\sum_{n=1}^{\infty}\frac{\tilde{f}_{1_{n}}}{M^{2n}}(\bar{R}+r)(r)^{(n)}+\sum_{n=1}^{\infty}\frac{\tilde{f}_{1_{n}}}{M^{2n}}\sum_{l=1}^{n-1}r^{(l)}r^{(n-l)}
\]
\[
\Theta_{2\nu}^{\;\mu}=\sum_{n=1}^{\infty}\frac{\tilde{f}_{2_{n}}}{M^{2n}}\sum_{l=0}^{n-1}\left(\nabla^{\mu}s_{\tau}^{\sigma(l)}\nabla_{\nu}s_{\sigma}^{\tau(n-l-1)}\right),\quad\bar{\Theta}_{2}=\sum_{n=1}^{\infty}\frac{\tilde{f}_{2_{n}}}{M^{2n}}\sum_{l=0}^{n-1}s_{\tau}^{\sigma(l)}s_{\sigma}^{\tau(n-l)}
\nn\]
\[
{\cal E}_{2\nu}^{\;\mu}=\sum_{n=1}^{\infty}\frac{\tilde{f}_{2_{n}}}{M^{2n}}\sum_{l=0}^{n-1}\nabla_{\tau}\biggl(s_{\lambda}^{\tau(l)}\nabla^{\mu}s_{\nu}^{\lambda(n-l-1)}-\nabla^{\mu}s_{\lambda}^{\tau(l)}s_{\nu}^{\lambda(n-l-1)}\biggr),
\]
whereas if we consider only up to linear order in $h$, we retain only the solitary term,
\[
\bar{\Theta}_{1}=\bar{R}\sum_{n=1}^{\infty}\frac{\tilde{f}_{1_{n}}}{M^{2n}}\Box^n r=\bar{R}\tilde{{\cal F}_{1}}(\Box)r-\tilde{f}_{1_{0}}\bar{R}r
.\]
Simple substitution of the  background curvature \eqref{barredR}, \eqref{barredlambda}, then reveals the linearised field equations around de Sitter space
\[
\label{eomdS}
\begin{aligned}&\left(M_{P}^{2}+24H^{2}\lambda\tilde{f}_{1_{0}}\right)\left(r_{\nu}^{\mu}-\frac{1}{2}\delta_{\nu}^{\mu}r\right) =T_{\nu}^{\mu}+2\lambda\left(\nabla^{\mu}\partial_{\nu}-\delta_{\nu}^{\mu}\square\right)\tilde{{\cal F}}_{1}(\Box)r-6\lambda H^{2}\delta_{\nu}^{\mu}\tilde{{\cal F}}_{1}(\Box)r\\
 & -6H^{2}\lambda{\cal F}_{2}(\Box)s_{\nu}^{\mu}+2\lambda\nabla^{\sigma}\nabla_{\nu}{\cal F}_{2}(\Box)s_{\sigma}^{\mu}-\lambda\square{\cal F}_{2}(\Box)s_{\nu}^{\mu}-\lambda\delta_{\nu}^{\mu}\nabla_{\sigma}\nabla^{\tau}{\cal F}_{2}(\Box)s_{\tau}^{\sigma} +\lambda{\cal C}_{\nu}^{\mu}.
\end{aligned}
 \]
 We will return to these field equations when discussing the defocusing conditions around de Sitter space in Section \ref{sec:defocusdS}.
%
\chapter{Ghost-free Conditions}
\label{chap:GF}
Historically speaking, higher derivative theories of gravity have been beset by the presence of \emph{ghosts} - free fields bearing negative kinetic energy. During the process of renormalization, this negative contribution may be offset by the repeated introduction of derivatives, via additional curvature terms in the gravitational action. These higher derivative terms affect the short-range behaviour of the theory rendering it renormalizable, but with a cost. If a finite number of higher derivatives are introduced to the theory, so too is a massive spin-2 `ghost' particle into the propagator, resulting in a break down of unitarity \footnote{With the exception of $f(R)$-theories, which we discuss shortly.} \cite{ellis2012relativistic}. The aim of this chapter is to provide the conditions whereby the IDG theory \eqref{action} may invoke its infinite number of derivatives in order to render the gravitational theory ghost and tachyon-free. We do this by appealing to the form of the modified graviton propagator around Minkowski space, which we also derive. We begin with some clarifications.
\\\\\emph{Ghosts}
\\As opposed to the Faddeev Popov ghosts of field theory, which are added to gauge field theories in order to absorb non-physical degrees of freedom, ghosts in relativity are physical excitations which come with a negative residue in the graviton propagator \cite{Clifton:2011jh}, \cite{VanNieuwenhuizen:1973fi}, which is defined via
\[
\Pi^{-1\sigma\tau}_{\mu\nu}h_{\sigma\tau}=\kappa T_{\mu\nu}.
\]
 Classically, the graviton propagator details how a gravitational field propagates through space when sourced by a current $J(x)$ \cite{Schwartz:2013pla}. Any modification to the gravitational action will necessarily modify the propagator. If this modification comes with the wrong sign, then implicitly the theory admits physical states of negative energy, or ``ghosts'', leading to an instability, even at the classical level. This phenomenon, known as the \emph{Ostrogradksy instability}, results in perturbations carrying both positive and negative energy modes \cite{Abramo:2009qk}. 
\\\\\emph{The Ostrogradsky Instability}\\
A major consequence of Ostrogradsky's Theorem of 1850 \cite{Ostrogradsky:1850fid} is that powerful constraints are placed on the formulation of stable, higher-derivative theories of gravity \footnote{\emph{Higher}-derivative theories refers to gravitational theories containing more than two derivatives of the metric tensor.}. We outline the results below before offering a caveat.

Consider a Lagrangian of the type
\[
L=L(q,{\dot q},{\ddot q})
,\]
which is \emph{non-degenerate} on ${\ddot q}$, i.e. $\frac{\partial^2 L}{\partial {\ddot q}^2}\neq 0$, where dots denote derivatives with respect to some parameter $\lambda$. The Euler Lagrange equation is given by
\[
\frac{\partial L}{\partial q}-\frac{d}{dt}\frac{\partial L}{\partial {\dot q}}+\frac{d^2}{dt^2}\frac{\partial L}{\partial {\ddot q}}=0
\]
Due to the non-degeneracy of the Lagrangian, solutions will depend on four pieces of initial data $q_0,{\dot q_0},{\ddot q_0},{\dddot q}_0$. We then make the following four canonical choices 
\begin{align}
Q_1&=q,\qquad P_1=\frac{\partial L}{\partial {\dot q}}-\frac{d}{dt}\frac{\partial L}{\partial {\ddot q}}
,\\
 Q_2&={\dot q},\qquad P_2=\frac{\partial L}{\partial {\ddot q}}
.
\end{align}
One can now express ${\ddot q}$ in terms of $Q_1,\;Q_2$ and $P_2$, like so ${\ddot q}=f(Q_1,Q_2,P_2)$, so that the Hamiltonian of the theory can be written as
\[
H=P_1Q_1+P_2f(Q_1,Q_2,P_2)-L(Q_1,Q_2,f)
,\]
as in \cite{Clifton:2011jh},\cite{Woodard:2015zca}. Now, as this Hamiltonian is linear only in the canonical momentum $P_1$, a system of this form cannot be stable. When interactions take place in such a system the vacuum decays into both positive and negative kinetic energy states, which we call the \emph{Ostrogradsky instability}.

It is commonly stated that a consequence of Ostragradsky's theorem is that gravitational theories containing higher-than-two derivatives in the action will suffer from this instability. This is generically true for theories of the type
\[
S=\frac{1}{2}\int d^4x \sqrt{-g} f(R,R^{\mu\nu},R^{\mu\nu\sigma\lambda})
,\]
where $f$ is some non-trivial function. An example of such a theory would be Stelle's fourth order gravity theory, discussed in the introduction. However, there is a caveat in that $f(R)$-theories, given by
\[
S=\frac{1}{2}\int d^4x \sqrt{-g} f(R),
\]
may avoid this instability and are indeed ghost-free. In terms of the Hamiltonian, this is due to the fact that one cannot express ${\ddot q}$ as ${\ddot q}=f(Q_1,Q_2,P_2)$ for each component of the metric. In terms of the modified propagator, such theories are characterised by a single additional scalar degree of freedom which contains all higher-order derivatives. We will see in Section \ref{sec:GFST} that as long as there is at most one additional pole in the scalar sector, the theory may be considered to be ghost-free, while in Chapter \ref{chap:sing}, we reveal the vital role of this additional pole in singularity avoidance. 
 \\\\\emph{Tachyons} 
\\We define tachyons as particles with imaginary mass, i.e. $m^2_{tachyon}<0$. In a classical sense, such a particle will always travel faster than the speed of light, according to the equation
\[
E=\frac{mc^2}{\sqrt{1-\frac{|v|^2}{c^2}}}
.\]
As the energy $E$ is real and observable, if a particle has imaginary mass $m$, then the denominator must also be imaginary, implying that the velocity of the particle $v$ is greater than the speed of light $c$. It is this behaviour which lends the particle its name - `tachy' ($\tau\alpha\chi\acute{\upsilon}\varsigma$) being the Greek for `rapid'.

Gerald Feinberg, who coined the term in the context of Quantum Field Theory  \cite{PhysRev.159.1089}, proposed that fields with imaginary mass would necessarily produce physical particles that propagate at speeds faster than light, however, this was later found to be untrue. Instead, fields defined in this way, technically suffer from an instability known as \emph{tachyon condensation}, where a field is tachyonic and unstable around the local maximum of its potential, but as the field reaches its local minimum, its associated quanta are not tachyonic but ordinary particles with positive mass squared, such as the Higgs boson \cite{Peskin:257493}.

In the present work, however, we will refer to particles with imaginary mass as `tachyons', as this notion has no classical interpretation and represents a pathology in the theory. 
\\\\\emph{Ghost and tachyon criteria}\\
In order to avoid the spectre of ghosts or tachyons, we require that: \begin{enumerate}
\item The propagator will contain only first order poles at $k^2+m^2$ with real mass $m^2\geq0$ so as to avoid \emph{tachyons}.
\item Such a pole will not contain any negative residues or \emph{ghosts}.
\end{enumerate}
\section{Modified Propagator around Minkowski Space}
\label{sec:prop}
Now that we have established the general ghost-free criteria that the modified propagator must ascribe to, it is time to derive the precise form of the propagator for the linearised equations of motion \eqref{Tababc} around a Minkowski background. The field equations are expressed in terms of the inverse propagator $\Pi^{-1\sigma\tau}_{\mu\nu}$, like so,
\[
\label{inversePi}
\Pi^{-1\rho\sigma}_{\mu\nu}h_{\rho\sigma}=\kappa T_{\mu\nu}
,\]
where $\kappa=8\pi G=M_P^{-2}$.
\paragraph{Derivation}
We begin by reminding the reader of the relevant identities concerning the spin projector operators in $D$-dimensional Minkowski space, see~\cite{Biswas:2013kla},\cite{VanNieuwenhuizen:1973fi},\cite{Conroy:2015nva}.
\begin{equation}
\mathcal{P}^{2}_{\mu\nu\rho\sigma}=\frac{1}{2}(\theta_{\mu \rho}\theta_{\nu \sigma}+\theta_{\mu
\sigma}\theta_{\nu \rho} ) - \frac{1}{D-1}\theta_{\mu \nu}\theta_{\rho
\sigma},\nn
\end{equation}
\begin{equation}
\nn
\mathcal{P}^{1}_{\mu\nu\rho\sigma}=\frac{1}{2}( \theta_{\mu \rho}\omega_{\nu \sigma}+\theta_{\mu
\sigma}\omega_{\nu \rho}+\theta_{\nu \rho}\omega_{\mu \sigma}+\theta_{\nu
\sigma}\omega_{\mu \rho} ),
\end{equation}
\begin{equation}
\nn
(\mathcal{P}_{s}^{0})_{\mu\nu\rho\sigma}=\frac{1}{D-1}\theta_{\mu \nu} \theta_{\rho \sigma},
\qquad
(\mathcal{P}_{w}^{0})_{\mu\nu\rho\sigma}=\omega_{\mu \nu}\omega_{\rho \sigma},
\end{equation}
\begin{equation}
(\mathcal{P}_{sw}^{0})_{\mu\nu\rho\sigma}=\frac{1}{\sqrt{D-1}}\theta_{\mu \nu}\omega_{\rho \sigma},
\qquad 
(\mathcal{P}_{ws}^{0})_{\mu\nu\rho\sigma}=\frac{1}{\sqrt{D-1}}\omega_{\mu \nu}\theta_{\rho \sigma},
\label{spin1}
\end{equation}
where
\begin{equation}
\theta_{\mu \nu}=\eta_{\mu \nu}-\frac{k_{\mu}k_{\nu}}{k^2},\;
\mbox{  and  } \;
\omega_{\mu \nu}=\frac{k_{\mu}k_{\nu}}{k^2}.
\label{spin2}
\end{equation}
Combining the final two identities gives the useful relation
\[
\eta_{\mu\nu}=\theta_{\mu\nu}+\omega_{\mu\nu}
.\]
We then compute each component making up the linearised field equations \eqref{Tababc}, while transforming into momentum space with $\partial_\mu\rightarrow ik_\mu$, such that $\Box\rightarrow -k^2$, like so
\begin{align}\nn
a(\Box)h_{\mu \nu}& \rightarrow\ a(-k^2)\left[\mathcal{P}^{2}+\mathcal{P}^{1}+\mathcal{P}_{s}^{0}+\mathcal{P}_{w}^{0}\right]_{\mu\nu}{ }{ }^{\rho\sigma}h_{\rho\sigma},
\\\nn
b(\Box)\partial_{\sigma}\partial_{(\nu}h_{\mu)}^{\sigma} &\rightarrow -b(-k^2)k^2\left[\mathcal{P}^1+2\mathcal{P}_{w}^{0}\right]_{\mu\nu}{ }{ }^{\rho\sigma}h_{\rho\sigma},
\\
c(\Box)(\eta_{\mu \nu}\partial_{\rho}\partial_{\sigma}h^{\rho \sigma}+\partial_{\mu}\partial_{\nu}h
 )&\rightarrow\  -c(-k^2)k^2 \left[2\mathcal{P}_{w}^{0}+\sqrt{D-1}\left(\mathcal{P}_{sw}^{0}+\mathcal{P}_{ws}^{0}\right)
\right]_{\mu\nu}{ }{ }^{\rho\sigma}h_{\rho\sigma},
\nn
\\
 \eta_{\mu \nu} d(\Box)h 
&\rightarrow\  d(-k^2)\left[(D-1)\mathcal{P}_{s}^{0}+\mathcal{P}_{w}^{0}+\sqrt{D-1}\left(\mathcal{P}_{sw}^{0}+\mathcal{P}_{ws}^{0}\right)
\right]_{\mu\nu}{ }{ }^{\rho\sigma}h_{\rho\sigma},
\nn\\ 
f(\Box)\partial^{\sigma}\partial^{\rho}\partial_{\mu}\partial_{\nu}h_{\rho
\sigma} &\rightarrow\ f(-k^2)k^4(\mathcal{P}_{w}^{0})_{\mu\nu}{ }{ }^{\rho\sigma}h_{\rho\sigma}.
\label{proj1}
\end{align}
On inspection of the spin projector operators \eqref{spin1},\eqref{spin2}, we note that the multiplets ${{\cal P}^2,{\cal P}^1,{\cal P}_s^0,{\cal P}_w^0}$ conform to the following:
\[
\qquad ({\cal P}^{2}+{\cal P}^{1}+{\cal P}_{s}^{0}+{\cal P}_{w}^{0})_{\mu\nu\rho\sigma}=\frac{1}{2}(\eta_{\nu\rho}\eta_{\mu\sigma}+\eta_{\nu\sigma}\eta_{\mu\rho})
.\]
Using this property along with \eqref{spin1} and \eqref{spin2} allows us to express the inverse propagator \eqref{inversePi} in terms of the projection operators, like so
\begin{align}
\label{inversepi2}
\Pi_{\mu\nu}^{-1\rho\sigma}h_{\rho\sigma}&	=\sum_{i=1}^{6}C_{i}{\cal P}_{\mu\nu}^{i}{}{}^{\rho\sigma}h_{\rho\sigma}
	\nn\\&=\kappa({\cal P}^{2}+{\cal P}^{1}+{\cal P}_{s}^{0}+{\cal P}_{w}^{0})_{\mu\nu}{}{}^{\rho\sigma}T_{\rho\sigma}
	\nn\\&=\frac{1}{2}\kappa(\delta_{\nu}^{\rho}\delta_{\mu}^{\sigma}+\delta_{\nu}^{\sigma}\delta_{\mu}^{\rho})T_{\rho\sigma}
	\nn\\&=\kappa T_{\mu\nu}
,
\end{align}
where the coefficients $C_i$ are dependent only on $k^2$ in momentum space. We then find that \eqref{proj1} reduces to
\begin{align}
ak^2\mathcal{P}^2_{\mu\nu}{}{}^{\rho\sigma}h_{\rho\sigma}&=\kappa \mathcal{P} ^2_{\mu\nu}{}{}^{\rho\sigma} T_{\rho\sigma} \nn
\\ \nn
(a+b)k^2 \mathcal{P}^{1}_{\mu\nu}{}{}^{\rho\sigma}h_{\rho\sigma}&=\kappa\mathcal{P}^{1}_{\mu\nu}{}{}^{\rho\sigma}T_{\rho\sigma} \nn\\
[(a+(D-1)d)k^2\mathcal{P}_{s}^{0}+(c+d)k^2\sqrt{D-1}\mathcal{P}_{sw}^{0}]_{\mu\nu}{}{}^{\rho\sigma}h_{\rho\sigma}&=\kappa(\mathcal{P}_{s}^{0}){ }_{\mu\nu}{}{}^{\rho\sigma}T_{\rho\sigma},
\nn\\
\bigl[(c+d)k^2\sqrt{D-1}\mathcal{P}_{ws}^{0}+(a+2b+2c+d+f)k^2\mathcal{P}_{w}^{0}\bigr]_{\mu\nu}{}{}^{\rho\sigma}h_{\rho\sigma}&=\kappa(\mathcal{P}_{w}^{0}){ }_{\mu\nu}{}{}^{\rho\sigma}T_{\rho\sigma}.
\end{align}
In turn, from \eqref{minkconstr}, each identity reduces like so
\begin{align}
\mathcal{P}_{\mu\nu}^{2}{}{}^{\rho\sigma}h_{\rho\sigma}&=\kappa\left(\frac{\mathcal{P}^{2}}{ak^{2}}\right){}_{\mu\nu}{}{}^{\rho\sigma}T_{\rho\sigma},\nn
\\ \nn
\mathcal{P}^{1}{}_{\mu\nu}{}{}^{\rho\sigma}T_{\rho\sigma} &=0,\nn\\
(\mathcal{P}_{s}^{0}){}_{\mu\nu}{}{}^{\rho\sigma}h_{\rho\sigma}&=\kappa \frac{(\mathcal{P}_{s})^{0}{}_{\mu\nu}{}{}^{\rho\sigma}}{(a-(D-1)c)k^2}T_{\rho\sigma},
\nn \\
\kappa(\mathcal{P}_{w}^{0}){}_{\mu\nu}{}{}^{\rho\sigma}T_{\rho\sigma} &=0.
\end{align}
Thus, we have succeed in inverting the field equations so that from \eqref{inversepi2}, we may read off the non-local, $D$-dimensional, propagator around Minkowski space:
\\\\
\emph{Non-local, $D$-dimensional, propagator around Minkowski} 
\begin{equation}
\label{MinkPropD}
\Pi^{(D)}(-k^{2})=\frac{\mathcal{P}^{2}}{k^{2}a(-k^{2})}+\frac{\mathcal{P}_{s}^{0}}{k^{2}(a(-k^{2})-(D-1)c(-k^{2}))}
, \end{equation}
where, hereafter we shall suppress the indices in the propagator and projection operators. The first thing to note here is that at $a(0)$ and $c(0)$, the theory returns to that of General Relativity. It is straightforward to confirm, upon reference to \eqref{minkconstr}, that indeed the physical graviton propagator is recovered at this limit
\[
\label{GRprop}
\frac{\mathcal{P}^{2}}{k^{2}a(0)}+\frac{\mathcal{P}_{s}^{0}}{k^{2}(a(0)-(D-1)c(0))}=\frac{\mathcal{P}^{2}}{k^{2}}-\frac{\mathcal{P}_{s}^{0}}{(D-2)k^{2}}=\Pi_{GR}^{(D)}.
\]
The propagator \eqref{MinkPropD} is the most general form of the D-dimensional propagator around Minkowski space for an action of the type \eqref{action}. In practice, however, we will largely be dealing within a $4$-dimensional framework, which is given by:
\\\\\emph{Non-local Propagator around Minkowski in $D=4$-spacetime dimensions}
\[
\label{MinkProp4}
\Pi(-k^{2})=\frac{\mathcal{P}^{2}}{k^{2}a(-k^{2})}+\frac{\mathcal{P}_{s}^{0}}{k^{2}\left(a(-k^{2})-3c(-k^{2})\right)}
. \]
Another interesting variant of the IDG modified propagator occurs when one requires that no additional degrees of freedom, other than the massless graviton, are allowed into the system. Upon inspection of \eqref{abc}, we note that by choosing $a=c$, we do indeed avoid the introduction of additional scalar degrees of freedom. 
\\\\ \emph{Non-local propagator with no  additional scalar degrees of freedom ($D=4$)}
\\
\begin{equation}
\label{MinkPropAC}
\Pi(-k^{2})=\frac{1}{a(-k^{2})}\biggl(\frac{\mathcal{P}^{2}}{k^2}-\frac{\mathcal{P}_{s}^{0}}{2k^2}\biggr).
 \end{equation}
 In this case, we note that the derived non-local propagator modifies the physical graviton propagator by a factor of $\sim 1/a$
 \[
 \Pi(-k^2)=\frac{1}{a(-k^{2})}\Pi^{(4)}_{GR}
. \]
\section{Examples of Pathological Behaviour}
\label{sec:patho}
\emph{The Benign Ghost of General Relativity}\\
Let us clarify what is sometimes called the benign ghost of General Relativity. This refers to the appearance of a negative residue in GR, which does not represent any instability in the theory. The propagator of the physical graviton in four-dimensional GR is given by:
\[
\Pi_{GR}=\frac{{\cal P}^2}{k^2}-\frac{{\cal P}^0_s}{2k^2},
.\]
Naively, it appears here as if there is a negative residue at the point $k^2=0$. However, upon reference to \eqref{spin1}, we find that the coefficients attached to the spin projector operators cancel exactly so that no such negative residue survives.
\\\\\emph{$f(R)$-gravity}\\
As previously discussed, a particular class of extended theory, known as the  $f(R)$ model of gravity, is defined by the action
\[
\label{fR}
S=\frac{1}{2}\int d^4 x \sqrt{-g} f(R)
.\]
This is a generalisation of Einstein's theory where the curvature scalar is replaced with an arbitrary function of $R$. The simplest example of such a theory is known as the \emph{Starobinsky Model}, defined by
\[
\label{starobinksymodel}
S=\frac{1}{2}\int d^4 x \sqrt{-g} \left(M_P^2R+f_0 R^2\right)
,\]
where $f_0$ is an arbitrary constant. A large amount of interest was generated in theories of this type from Starobinsky's initial idea that for a positive constant $f_0=\frac{1}{6 M^2}$, the model would mimic the behaviour of the cosmological constant for a sufficiently large $R$. The curvature squared term of the action leads to a period of exponential expansion, sufficient for forming the large scale structures we see in the Universe today. After this period of rapid expansion, the higher-order curvature terms become less important as the model moves away from the Planck scale, signalling the end of inflation \cite{Starobinsky:1979ty}. We have already learned that such theories avoid the Ostrogradsky instability. The Starobinsky model, however, has proved less successful in pairing cosmic inflation with a non-singular cosmology.  As we shall see in Section \ref{sec:defocusmink}, in order to avoid an initial singularity within this model, we would require the constant $f_0$ to be negative, resulting in tachyonic behaviour, outlined below. 
 \\\\ \emph{Example of Tachyons}\\
Let us consider the action
\[
\label{R-R2}
S=\frac{1}{2}\int d^4 x \sqrt{-g} \bigl(M_P^2R-R^2\bigr)
.\]
By taking the appropriate limits of \eqref{abc} and decomposing into partial fractions, we can read off the propagator of the Starobinsky model from \eqref{MinkProp4},
\[
\label{propR2int}
\Pi_{R^2}=\Pi_{GR}+\frac{1}{2}\frac{{\cal P}_s^0}{k^2+m^2}
,\]  
where $m^2=M_P^2/6f_0$. As the coefficient of the $R^2$ term is given by $f_0=-1$, 
the value of $m^2$ for the spin-0 particle now takes the form 
\[
m^2=-\frac{1}{6}M_P^2<0
,\]
which is decidedly negative. Thus, the first order pole $k^2+m^2=0$ in the scalar mode of \eqref{propR2int} contains an imaginary mass $m$, which is the very definition of a tachyonic field.
\\\\ \emph{Fourth Order Gravity}\\
A natural generalisation of $f(R)$-gravity, \eqref{fR}, is to not only allow scalar modifications to GR but tensorial modifications too. The gravitational action will then be made up of a an arbitrary function of the curvature scalar, Ricci tensor and Weyl/Riemann tensor, like so
\[
\label{fRRuvRunop}
S=\frac{1}{2}\int d^4 x \sqrt{-g} f(R,R^{\mu\nu},R^{\mu\nu\sigma\lambda})
.\]
An interesting subclass of this generalisation comes in the form of \emph{Fourth Order Gravity}, where
\[
\label{fthorder}
{\cal L}=R+f_1 R^2 +f_2 R_{\mu\nu}R^{\mu\nu}+f_3 R_{\mu\nu\sigma\lambda}R^{\mu\nu\sigma\lambda}
.\]
As you may recall from the introduction, choosing $f_1=1,\;f_2=-4,\;f_3=1$ is nothing more than the Gauss-Bonnet action of gravity, which reproduces the Einstein Equation precisely in four dimensions. More generally, in 1977,  Stelle found that fourth-order gravity theories, \eqref{fthorder}, were perturbatively renormalizable, raising hopes for a quantum field theory of gravity. As previously discussed, such theories suffer from the Ostrogradsky instability. An example of such a pathology comes in the form of the \emph{Weyl Ghost}.  
\\\\ \emph{The Weyl Ghost}\\
Consider a Lagrangian of the type
\[
{\cal L}_{W}\sim M_P^2R+C^{\mu\nu\lambda\sigma}C_{\mu\nu\lambda\sigma}
,\]
where $C_{\mu\nu\lambda\sigma}$ is the Weyl tensor, lending the name of \emph{Weyl squared gravity} to theories of this type. It is straightforward to compute that
\[
C^2=R_{\mu\nu\lambda\sigma}R^{\mu\nu\lambda\sigma}-2R_{\mu\nu}R^{\mu\nu}+\frac{1}{3}R^2
,\]
and to note that this is a particular example of a fourth-order  theory, whereas, in terms of the IDG action \eqref{action}, we simply take ${\cal F}_1={\cal F}_2=0$ and ${\cal F}_3=1$. The relevant functions, $a$ and $c$, that make up the propagator can then be read off from \eqref{abc} so that the propagator for Weyl squared gravity is given by 
\[
\Pi_{C^2}=\frac{{\cal P}^2}{(1-(2k/M_P)^2 k^2}-\frac{{\cal P}^0_s}{2k^2}=\Pi_{GR}-\frac{{\cal P}^2}{k^2+m^2}.
\]
We see here that there is an additional pole in the spin-2 portion of the propagator and, moreover, this comes with a negative sign, or residue. This is known as the \emph{Weyl Ghost} and in order to avoid such a situation, we stipulate that the operator $a(\Box)$ which modifies the tensorial sector has no roots (to avoid additional poles) and contributes positively to the propagator (ghost-free).\cite{Biswas:2013kla}
\section{Ghost-free Conditions}
We have already established that $f(R)$-theories are the only  \emph{finite}, higher-order extension of GR that are potentially stable \cite{Woodard:2015zca} \footnote{We are considering here only metric-tensor-based extensions of GR.}. What then of infinite derivative extensions of GR? Should such theories lead to a infinite number of ghosts or can their non-locality be invoked in some way to curtail these pathologies? One way to shed light on this is to study the equivalent scalar-tensor action of such a theory, first discussed in \cite{Biswas:2005qr}. As it stands, the general form of the modified propagator \eqref{MinkProp4} may contain ghosts or tachyons due to the arbitrary nature of the infinite derivative functions \eqref{abc} contained within. The purpose of this section is motivate the constraints that must be placed on these infinite derivative functions in order to describe a theory that is free from such infirmities. We will then derive the precise covariant ghost-free form around Minkowski space. Following an interesting discussion on the minimal IDG action that can be rendered singularity-free in Section \ref{chap:sing}, we will return to the ghost-free conditions around de Sitter space.
\subsection{Motivation from Scalar-Tensor Theory}
\label{sec:GFST}
In order to motivate the methodology for rendering the theory ghost-free it is instructive to discuss the scalar sector of the action \eqref{action} following closely to \cite{Biswas:2005qr}. The scalar sector is given by
\[
\label{sc1}
S_{sc}=\frac{1}{2}\int d^4 x \sqrt{-g} \biggl(M_P^2 R+R{\cal F}(\Box)R\biggr)
,\]
which accounts for the spin-0 sector of the propagator, where ${\cal F}(\Box)=\sum^{\infty}_{n=0}\frac{f_n}{M^{2n}} \Box^n$, while the equivalent scalar-tensor action takes the form
\[
\label{sc2}
S_{sc}=\frac{1}{2}\sqrt{-g}\biggl(M_P^2(\Phi R-\psi(\Phi-1))+\psi {\cal F}(\Box)\psi\biggr).
\]
The equivalence of \eqref{sc1} and \eqref{sc2} can be seen by taking a look at the equation of motion for $\Phi$. Varying the action with respect to $\Phi$ gives
\[
\frac{\delta S_{sc}}{\delta \Phi}=\frac{1}{2}\sqrt{-g}(R-\psi)
\]
so that the field equation
\[
\frac{\delta S_{sc}}{\delta \Phi}=0\qquad\mbox{implies}\qquad R=\psi
.\] 
Substituting $\psi=R$ into \eqref{sc2} then recovers the action \eqref{sc1}. The next step is to invoke the conformal transformation
\[
e_\mu^a=\Phi^{-1/2}e^{\prime a}_\mu
.\]
Further note that the D'Alembertian transforms as $\Box\psi=\Box^\prime\psi +{\cal O}(\phi^2,\phi\psi,\psi^2)$, while the metric tensor transforms as
\begin{align}
g_{\mu\nu}& =e^a_\mu g_{ab} e^b_\nu
\nn\\& =\Phi^{-1}e^{\prime\;a}_\mu g_{ab} e^{\prime\;b}_\nu
\nn\\& =\Phi^{-1}g^\prime_{\mu\nu}.
\end{align}
Subsequently, the square root of the determinant of the metric is given by
\[
\sqrt{-g}=\Phi^{-2}\sqrt{-g^\prime},
\]
so that the relevant form of the curvature scalar transforms as
\[
\sqrt{-g}\Phi R=\sqrt{-g^{\prime}}\biggl[R^{\prime}+\frac{3}{2}\phi\Box^{\prime}\phi\biggr]
, \]
where we have defined $\Phi=e^\phi$, which up to linear order gives $\Phi=1+\phi+{\cal O}(\phi^2)$. Substitution into the action \eqref{sc2} up to quadratic order then gives
\[
\label{sc3}
S_{sc}=\frac{M_P^2}{2}\sqrt{-g}\biggl( R^\prime+\frac{3}{2}\phi\Box^\prime \phi-\psi\phi+M_P^{-2}\psi {\cal F}(\Box^\prime)\psi\biggr).
\]
By varying with respect to $\phi$ and $\psi$ respectively, we can read off the field equations
\[
\psi=3\Box^\prime \phi
,\qquad
\phi=2M_P^{-2}{\cal F}(\Box^\prime)\psi.
\] 
Substitution then reveals
\[
\label{gammast}
\biggl(-1+6M_P^{-2}{\cal F}(\Box^\prime)\Box^\prime\biggr) \phi\equiv \Gamma_{sc}(\Box)\phi=0,
\]
where the scalar sector of the propagator is given by $\sim \frac{1}{\Gamma_{sc}(-p^2)}$ in momentum space with $\partial_\mu=ip_\mu$. Further note, upon reference to the trace equation \eqref{tracemink} and the  infinite derivative functions \eqref{abc}, that the left hand side of \eqref{gammast} conforms to the trace equation of the IDG action \eqref{trace} with ${\cal F}_2={\cal F}_3=0$. This is as expected, as we are considering the scalar sector of the theory, which the trace equation describes. We may then rewrite \eqref{gammast} as follows
\[
\label{a-3cst}
\frac{1}{2}\biggl(a(\Box^{\prime})-3c(\Box^{\prime})\biggr)\phi\equiv\Gamma_{sc}(\Box)\phi=0
 .\]
We must now examine the restrictions that must be placed on the infinite derivative functions $a$ and $c$ in order to avoid ghosts and tachyons. To this end, we express the scalar propagator $\Gamma_{sc}$ as a finite power series
 \[
 \Gamma_{sc}(-p^2)=(p^2+m^2_0)(p^2+m^2_1)\dots(p^2+m^2_n),
\]
where $m_i^2$ represents the mass of a spin-0 particle, which must be positive and real in order to avoid the theory becoming tachyonic. 

In this power series, each root represents an additional pole in the scalar section of the propagator. If we were to consider two distinct poles, i.e. $m_0\neq m_1$, then one of these poles will have a negative residue and  therefore be ghost-like. To show this, we assume that there are two adjacent poles with $m_0^2<m_1^2$, so that the propagator takes the form
\[
\Gamma_{sc}(-p^2)=(p^2+m_0^2)(p^2+m_1^2){\bar a}(-p^2),
\]
with the adjacent roots given by $p^2=-m_0^2$ and $p^2=-m_1^2$. As these poles are adjacent, there can be no more zeroes contained within ${\bar a}(-p^2)$  in the range $-m_1^2<p^2<-m_0^2$, with the consequence that the sign of ${\bar a}(-p^2)$ has not changed within these limits \cite{Biswas:2005qr}. 
 We can best illustrate this by decomposing the inverse propagator into partial fractions like so
\[
\Gamma_{sc}^{-1}(-p^2)=\frac{1}{{\bar a}(-p^2)}\biggl(\frac{1}{m_1^2-m_0^2}\biggr)\biggl(\frac{1}{p^2+m_0^2}-\frac{1}{p^2+m_1^2}\biggr)
.\] 
Here, we can see that two massive spin-0 particles $m_0$ and $m_1$ have been introduced to the propagator, with poles of differing sign. When the propagator is evaluated at $p^2=-m_1^2$ and $p^2=-m_0^2$, we observe a residue of differing signs so that one of these poles must necessarily be ghost-like.

The upshot is that we must restrict the scalar propagator to at most one additional pole in order for the theory to remain ghost-free. The scalar propagator is then given by
\[
\label{Gammastabar}
\Gamma_{sc}(\Box)=(\alpha\Box-m_0^2){\bar a}(\Box)
,\]
where ${\bar a}$ contains no roots and $\alpha$ is a constant. In order to ensure that the propagator is \emph{analytic} - differentiable across the whole of the complex plane - and does not contain additional poles, ${\bar a}$ must take the form of an \emph{exponent of an entire function}, i.e
\[
\Gamma_{sc}(\Box)=(\alpha\Box-m_0^2)e^{\gamma(\Box)}
,\]
where $\gamma(\Box)$ is an entire function. Finally, if we wish to restrict ourselves further to a propagator that is proportional to the physical graviton propagator - with no additional roots - we simply set $\alpha=0$:
\[
\Gamma_{sc}(\Box)=e^{\gamma(\Box)}
.\] 
\subsection{Entire functions}
In the preceding discussion, we made use of the properties of analytic functions to express ${\bar a}(\Box)$ in terms of an exponent of an entire function. At first glance, this may seem to be a \emph{construction} or at least a reduction in the possible forms of ${\bar a}(\Box)$. However, as we shall see, this is the only form that such a function may take. Formally, an analytic function is defined as \cite{Ablowitz:CV}:
\begin{definition}
A function $f(z)$, where $z\in {\mathbb C}$, is said to be \textbf{analytic} at a point $z_0$ if it is differentiable in a \emph{neighbourhood} of $z_0$. Similarly, a function $f(z)$ is said to be \textbf{analytic in a region} if it is analytic in every point in that region. 
\end{definition}
Recall that, the \textbf{neighbourhood} of $z_0$ is simply the region that contains all the points   within a radius $\epsilon$, excluding the boundary, where $\epsilon$ is a small positive number. From first principles, we know that we can express  the derivative of a function $f(z)$ like so
\[
f^{\prime}(z)=\lim_{\Delta z\rightarrow 0} \frac{f(z+\Delta z)-f(z)}{\Delta z}
.\]
For the function $f(z)=u(x,y)+i v(x,y)$, the variation is given by
\[
\delta f(z)=\frac{\partial f}{\partial z}\delta z=f^{\prime}(z)(\delta x +i \delta y)
,\]
while for any function of two variables, we may write
\[
\delta f(x,y) = \frac{\partial f}{\partial x}\delta x+\frac{\partial f}{\partial y}\delta y
.\]
As such
\[
f^{\prime}(z)=\frac{\partial f}{\partial x},\qquad\mbox{and}\qquad if^{\prime}(z)=\frac{\partial f}{\partial y}
.\]
Thus, the real and imaginary derivatives are given by
\[
\label{cauchy1}
f^\prime (z)=u_x (x,y)+i v_x(x,y)
\]
and
\[
\label{cauchy2}
i f^\prime (z)=u_y (x,y)+i v_y(x,y)
,\]
respectively, where the subscript represents partial differentiation, i.e. $u_x=\partial u /\partial x$. Requiring that \eqref{cauchy1} and \eqref{cauchy2} are equal gives the \textbf{Cauchy-Riemann Conditions}
\[
\label{cauchy3}
u_x=v_y\qquad \mbox{and}\qquad v_x=-u_y
.\]
These are the necessary and sufficient conditions that must hold if $f(z)$ is to be analytic. This leads us neatly to the the definition of an entire function, alluded to in the previous section.

\begin{definition} A function that is analytic at each point on the ``entire'' (finite) complex plane is known as an \textbf{entire function}. 
\end{definition}
Entire functions encompass a broad range of functions. For instance, the exponential function $f(z)=e^z$ is an entire function as can be seen by expressing the exponent in terms of trigonometrical functions, like so
\[
f(z)=e^z=\cos (y)(\cos (x)+\sinh (x))+i\sin (y)(\cosh (x)+\sinh (x))
\]
and differentiating to find that the Cauchy-Riemann criteria are satisfied (for all $z$), according to
\[
u_{x}=\cos(y)\left(\sinh(x)+\cosh(x)\right)=v_{y}
 ,\qquad
v_{x}=\sin(y)\left(\sinh(x)+\cosh(x)\right)=-u_{y}
. \]
Similarly, $\sin (z)$ and $\cos (z)$ are  entire functions, as indeed are all polynomials of $z$. Herein lies the problem. If in \eqref{Gammastabar}, we were to simply define ${\bar a}$ as an entire function, this would allow additional poles to be introduced into the propagator. Consider the function
\[
{\bar a}(z)=z^2-1.
\]
This is a polynomial and is indeed an entire function, conforming to \eqref{cauchy3} with
\[
u_x=2x=v_y\qquad\mbox{and}\qquad v_x=2y=-u_y
.\]
However, this function comes hand-in-hand with two additional poles at $z=1$ and $z=-1$, which must be ghost-like in terms of the propagator. The only solution then is to consider entire functions that contain no roots. One such function is the exponential function, $f(z)=e^z$, which, as we have already established, is an entire function. 
Generalising further, to incorporate all wholly analytic and rootless functions, we must consider an exponent (contains no roots) of an entire function (analytic across the entire complex plane), i.e. $f(z)=e^{\gamma(z)}$. Thus, in \eqref{Gammastabar}, ${\bar a}(\Box)$ must take the form
\[
{\bar a}(\Box)=e^{\gamma(\Box)}
,\]
where $\gamma$ is an entire function.
\subsection{Ghost-free condition around Minkowski Space}
\label{sec:GF}
Having resolved the modified form of the propagator for the IDG action \eqref{action}, it is now pertinent to derive the necessary form that the non-local functions $a(\Box)$ and $c(\Box)$ must take so as to render the propagator ghost-free. We begin by taking the trace of the linearised field equations \eqref{eomminkred}
\[
\label{tracemink2}
\kappa T=\frac{1}{2}(a(\Box)-3c(\Box))R
. \]
As alluded to in the prior discussion on scalar-tensor theory, the trace equation accounts for the scalar sector of the propagator, which can be seen by comparing the above trace equation with the modified propagator \eqref{MinkProp4}. Furthermore, we have just learned that $a(\Box)-3c(\Box)$ can contain a maximum of one pole. We therefore construct the equality
 \[
 \label{GFtrace}
T=\frac{M_P^2}{2} \left(a(\Box)-3c(\Box)\right)R=(\alpha\Box-{\bar m}^{2}){\bar a}(\Box)R
, \]
which is analogous to \eqref{Gammastabar}. Here, ${\bar m}^2$ is a Brans-Dicke scalar, $\alpha$ is a constant and ${\bar a}(\Box)$ is an exponent of an entire function with unit dimension, containing no zeros. Substituting the operator $\Box\rightarrow 0$ reveals that the Brans-Dicke scalar ${\bar m}^2$ is none other than the Planck Mass, i.e.
\[
{\bar m}^2=M_P^2
.\]
As such, we have
\[
\label{a-3cGF}
\left(a(\Box)-3c(\Box)\right)R=2(\alpha\Box M_{P}^{-2}-1){\bar{a}}(\Box)R
 \]
Furthermore, expanding to first order allows us to express the constant $\alpha$ as follows
\[
\label{alpha}
\alpha=6f_{1_{0}}+2f_{2_{0}}-\frac{M_{P}^{2}}{M^{2}}
 .\]
 This should prove useful as it contains within it the root of the Starobinsky model as $M\rightarrow \infty$, with $f_{2_0}=0$. It should also be noted that taking $\alpha=0$ imposes that the function $c(\Box)$ contains within it no roots and so there are no additional poles introduced into the propagator. The case of $a(\Box)=c(\Box)$, given by \eqref{MinkPropAC}, is one such example whereby $\alpha$ vanishes. 
 \\\\
\emph{Ghost-free form}
\\ Taking into account the value of the Brans-Dicke scalar, and reordering, allows us to express the necessary ghost-free form of the non-local function $c(\Box)$ as
\[
\label{GF}
c(\Box)=\frac{a(\Box)}{3}\left[1+2(1-\alpha M_P^{-2}\Box){\tilde a}(\Box)\right]
, \]
where we have defined a new entire function ${\tilde a}(\Box)={\bar a}(\Box)/a(\Box)$, which contains no roots.
\\\\
\emph{Ghost-free modified propagator}
\\
To display the necessary form of the ghost-free modified propagator, we substitute \eqref{a-3cGF} into \eqref{MinkProp4}, before decomposing into partial fractions.
\[
\label{MinkPropGF}
\Pi(-k^2)=\frac{1}{a(-k^2)}\biggl[\frac{{\cal P}^2}{k^2}-\frac{1}{2{\tilde a}(-k^2)}\biggl(\frac{{\cal P}_2^0}{k^2}-\frac{{\cal P}_s^0}{k^2+m^2}\biggr)\biggr]
,\]
where we have defined the spin-0 particle $m^2\equiv M_P^2/\alpha$. 
\pagebreak
\\\\\emph{Tachyon criteria}
\\
The spin-0 particle $m$ must have real mass to ensure that the correction is non-tachyonic. Subsequently, $m^2$ must be positive so that the condition whereby tachyons are prohibited from the gravitational theory is given by
\[
\label{alphatach}
\alpha\geq0
.\]
\emph{$R^2$-Gravity}
\\
Taking the limit $M\rightarrow\infty$ effectively removes all non-locality from the gravitational theory, stripping it back to fourth order gravity. Taking this limit on the functions \eqref{abc}, when ${\cal F}_2={\cal F}_3=0$, reduces the theory to the Starobinsky model, with ${\cal L}\sim R+f_{1_0} R^2$. These functions are then given by
\[
\label{R2ac}
a(\Box)=1, \qquad c(\Box)=1-4M_P^{-2}f_{1_0}\Box
\]
Recall, that each D'Alembertian contained within the non-local functions ${\cal F}_i$ is modulated by the scale of non-locality $M$. Thus
\[
\lim_{M\rightarrow\infty} {\cal F}_i(\Box)=\lim_{M\rightarrow\infty} f_{i_0}\frac{\Box^n}{M^{2n}}\rightarrow f_{i_0}
.\] 
Substituting the values \eqref{R2ac} into the general form of the propagator \eqref{MinkProp4} (with $\Box=-k^2$ in momentum space on a flat background), and performing the same method of decomposing into partial fractions, allows us to write the propagator for $R^2$-gravity, as was previously stated in \eqref{propR2int},
\[
\label{PropR2}
\Pi_{R^2}=\Pi_{GR}+\frac{1}{2}\frac{{\cal P}_s^0}{k^2+m^2}
,\]
where $m^2=M_P^2/\alpha$. Here, it is simply the scalar sector of the propagator that is modified and, we note also, in comparison to \eqref{MinkPropGF}, $a={\tilde a}\rightarrow 1$ at this limit. Further to this, we remind the reader that the constant $\alpha$ contains within it the root of the Starobinsky model. By taking the same limits as described bove, we find from equation \eqref{alpha}, that $\alpha$ is given by $\alpha=6f_{1_0}$~\footnote{Here, we have omitted the `counting tool' $\lambda$ which serves no physical purpose other than offering a simple means of returning the theory to GR at $\lambda=0$}. Thus, in order for the Starobinsky model to avoid becoming tachyonic, from the criteria described in \eqref{alphatach}, we require the coefficient attached to the $R^2$ term in the action to be positive, i.e.
\[
f_{1_0}\geq0
,\]
where $f_{1_0}=0$ returns the theory to GR.
\chapter{Singularity-free Theories of Gravity}
\label{chap:sing}
\section{What is a Singularity?}
One of Einstein's great insights was to devise a gravitational theory that is described by curvature alone. As discussed in the introduction, this insight stemmed from the universality of the tensor transformation law, which allowed Einstein to formulate a gravitational action made up of tensors, which would necessarily preserve covariance, universally. As such the physical laws of any gravitational theory made up of tensors will remain invariant under arbitrarily differentiable coordinate transformations. This is the principle of General Covariance. Unfortunately, however, entailed within this principle is a notorious difficulty in formulating a precise definition of a singularity.  We ask the question then: \emph{What is a Singularity?} \cite{Geroch:1968ut}

An intuitive answer to this question would be that \emph{a singularity is a `place' where the curvature `blows up'} \cite{Wald:GR}. This response comes with a number of difficulties, most notably, the idea of a singularity as a `place'. What separates GR from other physical theories is that it is formulated independently of the manifold or a specified metric structure of the spacetime. Without a manifold or given metric, the very idea of a `place' remains undefined. This is very different to, for instance, electrodynamics, where the manifold is clearly defined and solutions exist, such as the Coulomb solution, which render the electromagnetic field infinite. In this case, the electromagnetic field is undefined and characterises an electromagnetic singularity.

In GR, as opposed to other physical theories, the goal is to solve for the structure of spacetime itself. For example, if we consider the Schwarzschild solution of GR, which is  well known to contain an essential singularity at the point $r=0$. It is at this point that the metric gives way to pathologies. However, without a prescribed manifold, it is not possible to discuss the concept of `outside' the manifold.
%

A possible solution to this intransigence lies in considering the associated geodesic congruences of the theory. If we consider an ingoing geodesic which extends into past infinity, then one would infer that the spacetime is non-singular. This is known as geodesic completeness. Conversely, a spacetime that is defined by converging geodesics would be said to be beset by spacetime pathologies and indeed `holes' in the fabric of spacetime, through which the geodesics can not pass. This intuitive framework is appealing in its simplicity and forms the basis of the Hawking Penrose Singularity Theorems but   comes with a number of caveats. As a vector field may be timelike, spacelike or null, one would assume that if a spacetime contains a `hole' in its fabric in one of these cases, it would be true for all cases. However, this is not the case as the various forms are manifestly not equivalent and one can imagine the different possible permutations of completeness and incompleteness of the three forms, see  \cite{Geroch:1968ut} and \cite{Wald:GR} for specific examples of these potential contradictions.

Despite geodesic completeness falling short of a satisfyingly precise mathematical definition of a singularity, the fact remains that a spacetime which is null or timelike incomplete will contain some serious physical malady. In such a spacetime, a freely falling particle, will at some finite time, simply cease to exist and as such can be justifiably considered to be singular. As a result, throughout this discussion, we will consider a theory containing causal geodesic congruences, which \emph{focus} to a point in a finite time to be singular, as it is this definition which forms the basis for the Penrose-Hawking singularity theorems. 

\section{Hawking-Penrose Singularity Theorem}
\label{sec:Hawking}
We now state the Hawking-Penrose Singularity Theorem, \cite{Hawking:1973uf},\cite{Borde:1996pt}, applicable to an open or flat Universe and concerned with null geodesic congruences.
\begin{theorem}[Singularity Theorem 1]
\label{theorem:sing1}
A spacetime $\{{\cal M},g\}$ cannot be null-geodesically complete in the past direction if
\begin{enumerate}
\item $R_{\mu\nu}k^\mu k^\nu \geq 0$ for all null tangent vectors $k^\mu$;
\item There is a non-compact Cauchy surface ${\cal H}$ in ${\cal M}$;
\item There is a closed trapped surface ${\cal T}$ in ${\cal M}$.
\end{enumerate}
\end{theorem}
~\\\emph{Notes}
\begin{enumerate}
\item Here, $R_{\mu\nu}$ is the Ricci tensor and $k^\mu$ are null geodesic congruences or `rays.' This inequality is known as the \emph{null convergence condition} and describes a spacetime where null rays \emph{focus} or converge to a point in a finite `time' (affine parameter), thus describing a singular spacetime,  according to our definition above. The full significance of this convergence condition will be revealed during the subsequent discussion on the Raychaudhuri Equation, Section \ref{sec:RE}.
\item A Cauchy surface is defined as a closed, achronal set $\Sigma$ for which the full \emph{domain of independence} $D(\Sigma)={\cal M}$, where 
\[
D(\Sigma)=D^+(\Sigma)\cup D^-(\Sigma)
,\]
and the superscripts $^+$ and $^-$ refer to the future and past domain, respectively, \cite{Wald:GR}. More formally, these domains are defined as
\begin{align}
D^{+/-}(\Sigma)=\{p\in {\cal M} \;|&\; \mbox{Every past/future inextendible \emph{causal} curve}
\\&\nn \mbox{through p intersects }\Sigma\}
\end{align}
As $\Sigma$ is \emph{achronal}, we may consider such a surface to be an \emph{instant of time}. A spacetime that possesses a Cauchy Surface is called \emph{globally hyperbolic}. A globally hyperbolic spacetime is causally simple in that the entire future or past history of the Universe can be predicted from well-defined initial conditions \cite{Narita:1998pt} or, indeed, a well-defined `instant of time.'
\item A \emph{closed trapped surface} is a topological space where the congruences of null geodesics, orthogonal to the topological space, converge. This convergence is typified by  negative ingoing and outgoing expansion. In a geometrically-flat spacetime or open Universe, the convergence condition, denoted as point (1.) in Theorem \ref{theorem:sing1},  necessarily implies a closed trapped surface, point (3.)\cite{Vilenkin:2013tua},\cite{Hawking:1973uf}. However, this is not necessarily the case for a closed Universe, see Ellis \cite{Ellis:2003mb}. We return to the notion of trapped surfaces in Section \ref{sec:cosexp}.
\end{enumerate}
\section{The Raychaudhuri Equation and General Relativity}
\label{sec:RE}
Before we delve into the derivation of the Raychaudhuri Equation, it is instructive to define some of the terminology involved. Firstly, we will often refer to \emph{congruences of geodesics}, sometimes abbreviated to simply `geodesics'. The tangents to these congruences yield a vector field, which we refer to as `tangent vectors' or indeed `rays' when discussing null geodesic congruences. A \emph{congruence} of geodesics is simply a bundle or family of geodesics. More formally, this bundle resides in an open subset of a manifold ${\cal M}$, which we may consider to be our spacetime, such that each point in this spacetime passes through precisely one geodesic within this bundle. The tangents to such a congruence yield a \emph{causal} vector field within the spacetime. Causal means that these tangents can be \emph{timelike} or \emph{null}, by which we denote the vector fields $\xi^{\mu}$ and $k^{\mu}$, respectively \cite{Wald:GR},\cite{Kar:2006ms}.
\subsection{Normalization of Timelike and Null Geodesics} 
When analysing the defocusing conditions of a spacetime, we will largely restrict ourselves to analysing the null vector fields $k^{\mu}$ and their associated null geodesic congruences. This is because null rays more readily converge than their timelike counterparts and, as our intent is to describe the conditions under which a singularity-free cosmology may flourish, it is more revealing to restrict our discussion to null geodesic congruences. We will return to this point in a more formal manner later in this section. In the meantime, however, it is instructive to contrast some of the differences in behaviour of null and timelike geodesics, beginning with a discussion on normalization. In contrast to the timelike case, which can be readily normalized to unit length, there is no natural way of normalizing a tangent vector field such as $k^\mu$. In the timelike case, a tangent field $\xi^\mu$ is defined in terms of proper time $\tau$, like so
\[
\xi^\mu =\frac{dx^\mu}{d\tau}
,\]
so that the line element $ds^2=g_{\mu\nu}dx^\mu dx^\nu$, restricted to a timelike curve $ds^2\rvert_{\mbox{{\tiny timelike}}}=-d\tau^2$, gives the straightforward normalization:
\[
\label{timelikenorm}
g_{\mu\nu}\xi^\mu \xi^\nu=-1
.\]
We can then construct a  metric $t_{\mu\nu}$ which satisfies, what we shall call, the \emph{spatial condition},
\[
t_{\mu\nu}\xi^\mu=0.
\]
This is given by
\[
\label{timelikemetric}
t_{\mu\nu}=g_{\mu\nu}+\xi_\mu\xi_\nu.
\] 
Similarly, the line element $ds^2=g_{\mu\nu} dx^\mu dx^\nu$, restricted to a light-like or null curve, is given by $ds^2\rvert_{\mbox{{\tiny null}}}=0$. Here, the null tangent field $k^{\mu}$ is defined as
\[
k^\mu=\frac{dx^\mu}{d\lambda}
,\]
 where $\lambda$ is an affine parameter.
Thus, the normalization analogous to the timelike case \eqref{timelikenorm} is given by
  \[
  \label{nullnorm}
 g_{\mu\nu}k^\mu k^\nu=0
 .\]
The outstanding task for the null tangent vectors $k^\mu$ is to construct a metric $p_{\mu\nu}$, which satisfies the spatial condition $p_{\mu\nu}k^\mu=0$ along with the above normalization. Naively, one would assume a choice of $p_{\mu\nu}=g_{\mu\nu}+k_\mu k_\nu$, would suffice, as this choice worked so well in the timelike case. However, on quick inspection, we find that this does not satisfy the spatial condition $p_{\mu\nu}k^\mu=0$. For this choice of $p_{\mu\nu}$, we have
\[
p_{\mu\nu}k^\mu=g_{\mu\nu}k^\mu+k^\mu k_\mu k_\nu=k_\nu+0\neq 0.
\]
 A popular  resolution, \cite{Kar:2006ms}, of this difficulty is to introduce an additional null vector $N^\mu$, such that $N^\mu N_\mu=0$ and $N^\mu k_\mu=-1$. We can then construct the two-dimensional metric
\[
\label{nullmetric}
p_{\mu\nu}=g_{\mu\nu}+k_\mu N_\nu +k_\nu N_\mu 
,\]
satisfying the necessary conditions $p_{\mu\nu}k^\mu=p_{\mu\nu}N^\mu=0$, as well as $k^\mu k_\mu=0$. This is by no means a unique choice of metric but it is sufficient in the subsequent derivation of the Raychaudhuri Equation and does not result in any loss of generality. In practice, the precise form of the metric will be largely irrelevant for our purposes, so long as one has in mind a two-dimensional metric that conforms to the spatial condition and satisfies \eqref{nullnorm}.
 
\subsection{Derivation of Raychaudhuri Equation}
Let us now give the derivation of what will be the key instrument in our analysis of singularity-free cosmologies - the Raychaudhuri equation for null geodesic congruences. Using the aforementioned vector field $k^{\mu}$  and following closely to \cite{Wald:GR}, we define a tensor field
\[
\label{DefB}
B_{\mu\nu}=\nabla_{\nu}k_{\mu}
 \]
which satisfies the spatial condition $B_{\mu\nu}k^{\mu}=B_{\mu\nu}k^{\nu}=0$
 , due to the fact that two vector fields of the same coordinate basis will commute. We then attribute to it the positive definite, two dimensional, spatial metric $p_{\mu\nu}$ such that $p_{\mu\nu}k^{\mu}=p_{\mu\nu}k^{\nu}=0$, as discussed previously.  We now define the \emph{expansion}, \emph{shear} \cite{Sommers} and \emph{twist}, respectively, as
\[
\label{expshtw}
\theta\equiv p^{\mu\nu}B_{\mu\nu}=\nabla_{\mu}k^{\mu},\qquad\sigma_{\mu\nu}\equiv B_{(\mu\nu)}-\frac{1}{2}\theta p_{\mu\nu},\qquad\omega_{\mu\nu}\equiv B_{[\mu\nu]}
 \]
 
To develop an intuitive understanding of these geometric terms, it is perhaps best to first consider congruences of timelike geodesics. If we consider a set of test particles, making up a sphere and centred on a geodesic, the expansion is the change in volume of the sphere; the shear is the deformation of the geometry of the sphere into an ellipsoid; and the twist is simply a rotation of the geometry \cite{tHooft:2009bh}.

In order to illustrate the analogous evolution of null rays, we must first introduce the notion of \emph{screen space}. An observer's screen space is a two dimensional space orthogonal to $k^\mu$. \emph{Images} are carried by the null rays and are displayed upon the screen space. The shape and size of these images, which are independent of the observer, are what concerns us in defining the kinematic quantities of the expansion, shear and rotation. As such, the null expansion measures the change in area of the image; the shear distorts the image; and the twist rotates the image. These quantities make up the kinematic \emph{flow}, generated by the tangent vector $k^\mu$ \cite{ellis2012relativistic},\cite{Kar:2006ms}.
 
 Returning to mathematical identities for these geometric quantities \eqref{expshtw}, we may now decompose $B_{\mu\nu}$ to the following
\[
\label{Bexpand}
B_{\mu\nu}=\frac{1}{2}\theta p_{\mu\nu}+\sigma_{\mu\nu}+\omega_{\mu\nu}
 \]
Next, consider the term $k^{\lambda}\nabla_{\lambda}B_{\mu\nu}$, which from \eqref{DefB}, becomes
\begin{align}\nn
k^{\lambda}\nabla_{\lambda}B_{\mu\nu}	&=	k^{\lambda}\nabla_{\lambda}\nabla_{\nu}k_{\mu}
	\\&=	k^{\lambda}[\nabla_{\lambda},\nabla_{\nu}]k_{\mu}+k^{\lambda}\nabla_{\nu}\nabla_{\lambda}k_{\mu}
	.\end{align}
Recall that the commutator of two covariant derivatives acting upon a tensor can be expressed in terms of the Riemann tensor, like so 
\begin{align}
\label{CommRiem}
[\nabla_{\rho},\nabla_{\sigma}]X^{\mu_{1}...\mu_{k}}{ }{ }{ }{ }_{\nu_{1}...\nu_{l}}	&=	R^{\mu_{1}}{ }_{\lambda\rho\sigma}X^{\lambda\mu_{2}...\mu_{k}}{ }{ }{ }{ }_{\nu_{1}...\nu_{l}}+R^{\mu_{2}}{ }_{\lambda\rho\sigma} X^{\mu_{1}\lambda\mu_{3}...\mu_{k}}{ }{ }{ }{ }_{\nu_{1}...\nu_{l}}+...
	\nn	\\&-R^{\lambda}{ }_{\nu_{1}\rho\sigma} X^{\mu_{1}...\mu_{k}}{ }{ }{ }{ }_{\lambda...\nu_{l}}-R^{\lambda}{ }_{\nu_{2}\rho\sigma}X^{\mu_{1}...\mu_{k}}{ }{ }{ }{ }_{\nu_{1}\lambda\nu_{3}...\nu_{l}}-...\;,
		\end{align}
	so that the term $k^{\lambda}\nabla_{\lambda}B_{\mu\nu}$ develops as follows
 \begin{align}
	k^{\lambda}\nabla_{\lambda}B_{\mu\nu} &=	-k^{\lambda}R^{\kappa}{ }_{\mu\lambda\nu}k_{\kappa}+k^{\lambda}\nabla_{\nu}\nabla_{\lambda}k_{\mu}
	\nn\\&=	-k^{\lambda}R^{\kappa}{ }_{\mu\lambda\nu}k_{\kappa}+\nabla_{\nu}(k^{\lambda}\nabla_{\lambda}k_{\mu})-\nabla_{\nu}k^{\lambda}\nabla_{\lambda}k_{\mu}
	\nn\\&=	-k^{\lambda}R^{\kappa}{ }_{\mu\lambda\nu}k_{\kappa}+\nabla_{\nu}(k^{\lambda}B_{\mu\lambda})-B^{\lambda}{ }_{\nu}B_{\mu\lambda}.
 \end{align}
The middle term then vanishes as the shear and rotation tensors are purely spatial as is the metric $p_{\mu\nu}$, so that from \eqref{Bexpand}, $B_{\mu\nu}k^\mu=0$. Thus,
\[
\label{preraych}
k^{\lambda}\nabla_{\lambda}B_{\mu\nu}=-k^{\lambda}R^{\kappa}{ }_{\mu\lambda\nu}k_{\kappa}-B^{\lambda}{ }_{\nu}B_{\mu\lambda}
. \]
We then take the trace to find
\[
k^{\lambda}\nabla_{\lambda}\theta	=	-R^{\kappa}{ }_{\lambda}k_{\kappa}k^{\lambda}-B^{\lambda\mu}B_{\mu\lambda}
 \]
which upon reference to \eqref{Bexpand}, leads us to the \emph{Raychaudhuri Equation for null geodesic congruences}, which we express as follows
\[
\label{Raych}
\frac{d\theta}{d\lambda}+\frac{1}{2}\theta^{2}=-\sigma_{\mu\nu}\sigma^{\mu\nu}+\omega_{\mu\nu}\omega^{\mu\nu}-R_{\mu\nu}k^{\mu}k^{\nu}.
 \]
Here, we have noted that $k^{\lambda}\nabla_{\lambda}\theta=\frac{d\theta}{d\lambda}$, with affine length $\lambda$. Following the same approach, one may also derive the \emph{Raychaudhuri equation for timelike vectors}, which is given by
\[
\label{Raychtimelike}
\frac{d\theta}{d\tau}+\frac{1}{3}\theta^{2}=-{\bar \sigma}_{\mu\nu}{\bar \sigma}^{\mu\nu}+{\bar \omega}_{\mu\nu}{\bar \omega}^{\mu\nu}-R_{\mu\nu}\xi^{\mu}\xi^{\nu}.
 \]
One can immediately see that these two identities take a broadly similar form with some key differences. The factor of $3$ in the denominator of the timelike equation arises from the  metric \eqref{timelikemetric}, which is $3$-dimensional as opposed to the $2$-dimensional null metric \eqref{nullmetric}. This difference in metric also accounts for the `bar' placed on the shear and twist tensors in the the timelike case. Finally, the Ricci tensor is contracted with the null (timelike) vectors fields $k^\mu$ ($\xi^\mu$), such that $k^\mu k_\mu=0$ ($\xi^\mu \xi_\mu=-1$) in the null (timelike) formulation. 
\subsection{Convergence Conditions}
\label{sec:CC}
By making a number of straightforward observations about the geometric terms in the Raychaudhuri equation \eqref{Raych}, we may reduce this identity to an inequality, which when satisfied necessitates that the associated null geodesics cannot be maximally extended in the past direction. This is known as the \emph{null convergence condition} and depicts congruences that converge to meet a singularity in a finite time. As previously discussed, the shear tensor is purely spatial and therefore contributes positively to the RHS of \eqref{Raych}, whereas the twist tensor vanishes if we take the congruence of null rays to be orthogonal to a hypersurface. Applying these constraints to the RE gives us the \emph{null convergence condition} (null CC):
\[
\label{nullCC0}
\frac{d\theta}{d\lambda}+\frac{1}{2}\theta^2\leqq - R_{\mu\nu}k^\mu k^\nu
.\]
\emph{General Relativity}\\
The behaviour of null rays in GR can be discerned by referring to the perfect fluid equation in Appendix \ref{sec:introcos}. The null energy condition (NEC) requires \cite{Kar:2006ms},\cite{Vachaspati:1998dy},\cite{Borde:1996pt}
\[
R_{\mu\nu}k^\mu k^\nu=\kappa T_{\mu\nu}k^\mu k^\nu=\kappa (\rho+p)(k^0)^2\geq 0
.\]
Thus, a spacetime will not be geodesically past-complete and will be plagued by a singularity if either of the following conditions are met
\[
\label{nullCC}
\frac{d\theta}{d\lambda}+\frac{1}{2}\theta^2\leqq 0,\qquad R_{\mu\nu}k^\mu k^\nu\geq 0.
\]
Thus, General Relativity will contain a singularity as long as the associated energy condition is retained. Another question remains, however, and that concerns the geometric terms we managed to `edit' out of proceedings. Is it possible for the shear and the twist to distort the geometry in such a way that geodesic congruences may be made past-complete? We discuss this below.
\subsection{Rotation and Convergence}
We may express the null convergence condition of \eqref{Raych}, more generally, without making any refinements to the geometric tensors. In this case, null geodesic congruences will converge in accordance with \cite{Kar:2006ms}
\[
\label{convshtw}
R_{\mu\nu}k^\mu k^\nu +\sigma^2 -\omega^2 \geq 0
\]
where $\sigma^2 \equiv \sigma_{\mu\nu}\sigma^{\mu\nu}$ and $\omega^2 \equiv \omega_{\mu\nu}\omega ^{\mu\nu}$ \cite{Kar:2006ms}. Thus, the shear induces convergence, whereas the rotation inhibits it. Upon studying the inequality given in \eqref{convshtw}, one might think, naively, that as the rotation inhibits convergence, this may be enough to render the null rays past-complete, without the aid of gravity.
\\\\\emph{Shear-free Expansion of Dust}\\ To understand the role of rotation more clearly, let us consider the simplest case of a rotating and expanding Universe, which is given by the shear-free expansion of dust. In order to do this, let us first redefine the expansion as $\theta =2\frac{{\dot F}}{F}$, where ${\dot F}=\frac{dF}{d\lambda}$, and substitute this into the RE \eqref{Raych} to find \cite{Tipler:1977zzb}
\[
2\frac{{\ddot F}}{F}+\sigma^{2}-\omega^{2}+R_{\mu\nu}k^{\mu}k^{\nu}=0
.\]
In the case of shear-free expansion of dust, we have $\sigma=0$, $\theta>0$, $\omega=\Omega/F^2$ and ${\dot \Omega}=0$ \cite{ellis2012relativistic}, whereas from the Einstein field equations with vanishing pressure, we have $R_{\mu\nu}k^{\mu}k^{\nu}=\kappa\rho(k^0)^2$, where $k^0$ is taken to be constant along $k^{\mu}$. The above equation can then be rewritten as
\[
2{\ddot{F}}\dot{F}-\frac{\dot{F}}{F^{3}}\Omega^{2}+\dot{F}F\kappa\rho(k^{0})^{2}=0.
\]
Here, we have multiplied by a factor of ${\dot F}$ on both sides so that we may integrate in a straightforward manner. Integrating, we find
\[
\label{dustF}
\dot{F}^{2}+\frac{\Omega^{2}}{2F^{2}}+\frac{1}{2}\kappa\rho(k^{0})^{2}F^{2}=\mbox{constant}
.\]
This equation suggests that there exists a solution to Einstein's equations where a period of intense rotation at early times will result in sufficient centrifugal force so as to cause a bounce in the stead of the initial singularity of the Universe \cite{ellis2012relativistic}. However,  it was shown by Ellis \cite{Ellis:1966ta} that no such solution can exist, via the Dust Shear-Free Theorem:
\begin{theorem}[Dust Shear-Free Theorem]
\label{theorem:sing1}
If a dust solution of the Einstein Field Equations is shear-free in a domain $U$, it cannot both expand and rotate in $U$:
\[
\{{\dot u}^\mu=0,\sigma_{\mu\nu}=0\}\qquad \implies\qquad \omega_{\mu\nu} \theta=0.
\]
\end{theorem}
~\\
Thus, if a dust solution is shear-free and expanding, the rotation must vanish, so that we may not realise the period of intense rotation outlined above \cite{Senovilla:1997bw},\cite{Ellis:2011pi}.
\\\\\emph{G{\" o}del Universe}\\
Another exact solution of Einstein's Field Equations is the G{\" o}del solution \cite{Godel}, which can be described by the line element \cite{Hawking:1973uf}
\[
ds^2=-dt^2+dx^2-\frac{1}{2}e^{2\sqrt{2} \omega x}dy^2+dz^2-2 e^{\sqrt{2} \omega x} dt dy
,\]
where $\omega$ is the magnitude of the vorticity. G{\" o}del's solution has vanishing expansion and shear but is characterised by a non-zero rotation and rotational symmetry around every point. In this case, the Raychaudhuri equation in terms of timelike tangent vectors \eqref{Raychtimelike}, can be expressed as
\[
\label{REtimegodel}
R_{\mu\nu}\xi^\mu \xi^\nu-2\omega^2=0
,\]
where in the timelike case $R_{\mu\nu} \xi^\mu \xi^\nu=\frac{\kappa}{2}(\rho+3p)-\Lambda=\kappa \rho$. Comparing these identities in the absence of pressure reveals $\frac{\kappa}{2}\rho=-\Lambda$, while substitution into \eqref{REtimegodel} gives $\omega^2=-\Lambda$. We may then write the analogous Friedmann equation for the G{\" o}del Universe as
\[
\kappa\rho+\Lambda-\omega^2=0
.\] 
Such a Universe is indeed geodesically complete but leads to a breakdown in causality due to the prevalence of \emph{closed timelike curves}. Whereas proper time can be measured consistently along a given world line, there exists no concept of cosmic time. An observer travelling along such a closed curve will travel forward in time as measured locally by the observer but, globally, may return to an event in the past \cite{ellis2012relativistic},\cite{Barrow:2003ph},\cite{Lobo:2010sz}. While undoubtedly revealing with regards to the role of rotation in gravitational theories and geodesic-completeness, as well as offering a tantalising  glimpse at the possibility of time travel, G{\" o}del universe's violation of causality means that we cannot consider it to be a viable non-singular theory in the present text.
\subsection{Cosmological Expansion}
\label{sec:cosexp}
In this section, we will build upon the mathematical definition of the expansion given by \eqref{expshtw}, by first giving an intuitive picture of the general concept, before moving on to the details and consequences in a cosmological setting. Given a sphere of test particles, the expansion is defined by the change in volume of this sphere and can be subdivided into \emph{ingoing} and \emph{outgoing expansions}, which are delineated by the respective ingoing and outgoing (timelike) tangent vectors. The type of surface formed by these ingoing and outgoing tangent vectors can have profound consequences for the nature of the spacetime, as we shall see below.

\subsubsection*{Normal, Trapped and Antitrapped Surfaces}
\begin{figure}[h]\cite{tHooft:2009bh}
\centering
\includegraphics[scale=0.4]{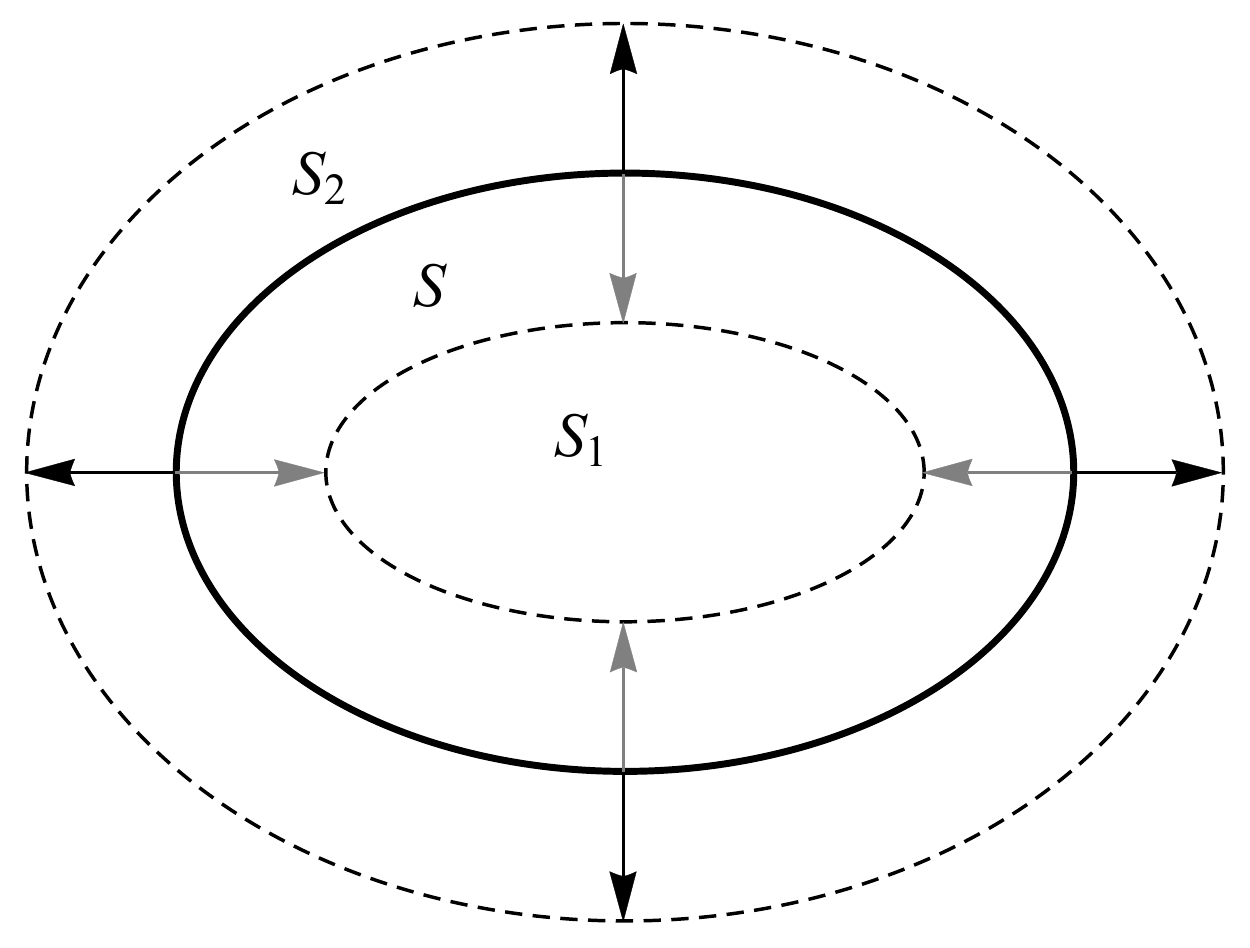}
\caption{From surface $S$, ingoing geodesics produce a smaller surface $S_1$ with area $A_1$ after a time $t=\epsilon$, while outgoing geodesics form a larger surface $S_2$ with area $A_2$. This is the behaviour of ingoing and outgoing expansions in a flat spacetime and the surfaces $S_1$ and $S_2$ are known as normal surfaces.}\label{figexp} 
\end{figure}

In order to gain a better understanding of ingoing and outgoing expansions, let us, by way of example, consider a two dimensional spatial sphere $S$ on a curved space \cite{tHooft:2009bh}. Let $A$ be the area of $S$ at cosmic time $t=0$. After a small amount of time, $t=\epsilon$, has passed, ingoing geodesics will describe a surface $S_1$ with area $A_1$, whereas outgoing rays will form a surface $S_2$ with area $A_2$. The respective expansion rates will then be given by
\[
\theta_{IN}=\frac{dA_1}{d\epsilon},\qquad \theta_{OUT}=\frac{dA_2}{d\epsilon}
.\]
Conventionally, one would expect the outgoing geodesics to describe a growing surface and the ingoing geodesics to describe a shrinking one. This is the behaviour in an asymptotically-flat
spacetime and as such the surfaces formed are known as \emph{normal surfaces}. An example of such a surface is depicted in Fig. \ref{figexp}.

However, at points close to a singularity, these expansions can behave very differently. For example, inside the event horizon of a Schwarzschild black hole, with $r<2GM$, both sets of geodesics would form a surface of smaller area, after a given time has passed. The resulting surfaces are known as \emph{trapped surfaces}, the existence of which strongly suggests and, in some case, necessitates the formation of a singularity. In broad strokes, we may say that singularities are an inevitable consequence of trapped surfaces, in a geometrically-flat or open Universe, so long as positive energy density is maintained \cite{tHooft:2009bh}. However, this is not necessarily the case for a closed Universe, see \cite{Ellis:2003mb}.

There is another type of surface called an \emph{antitrapped surface}, which is formed when the expansion is rapid enough that both in- and outgoing tangent vectors form a surface of a larger area after a given time has passed. A {minimally antitrapped surface} (MAS) has vanishing outgoing expansion and any surface greater in radius to the MAS will necessarily be antitrapped. Furthermore, an apparent horizon is found on the inner boundary of the MAS, like so
\[
\label{mashorizon}
x_{mas}=H^{-1}
,\]
where $x_{mas}$ is the physical size of the minimally antitrapped surface. We will return to this relation to the the apparent horizon shortly, but conclude here with a summary of the surfaces we have introduced:
\begin{align}
\mbox{Normal Surface:}&\qquad \theta_{IN}<0\mbox{ and } \theta_{OUT}>0
\\ \mbox{Trapped Surface:}&\qquad \theta_{IN}<0\mbox{ and } \theta_{OUT}<0
\\ \mbox{Antitrapped Surface:}&\qquad \theta_{IN}>0\mbox{ and } \theta_{OUT}>0.
\end{align}
Further note that surfaces termed \emph{marginally trapped}, are those with negative expansions as opposed to negative definite, i.e. marginally trapped surfaces include those with vanishing expansion and are defined by simply replacing the $<$ sign with $\leq$, above. Similarly for \emph{marginally antitrapped} surfaces, $>$ is replaced with $\geq$. 
\subsubsection*{Cosmological Expansion}
In order to understand more clearly the nature of these surfaces, it is pertinent to derive the exact form of the ingoing and outgoing expansions in a cosmological setting. To this end, we invoke the  spatially flat, homogenous and isotropic Friedmann-Robertson-Walker (FRW) metric
\[
\label{FRWsphere}
ds^{2}=-dt^{2}+a^{2}(t)\left(dr^{2}+r^{2}d\Omega^{2}\right)
 . \]
Recall from \eqref{expshtw}, that we may write the expansion $\theta$ as follows
\[
\theta=\partial_{\mu}k^{\mu}+\Gamma_{\mu\sigma}^{\mu}k^{\sigma}.
\]
We also note that the two geodesic equations for the time and spatial coordinate read
\[
\frac{d^{2}t}{d\lambda^{2}}+a\dot{a}\delta_{ij}\frac{dx^{i}}{d\lambda}\frac{dx^{j}}{d\lambda}=0,\qquad
\frac{d^{2}x^{i}}{d\lambda^{2}}+\frac{\dot{a}}{a}\frac{dt}{d\lambda}\frac{dx^{i}}{d\lambda}=0
, \]
respectively, where $\lambda$ is an affine parameter.
Without loss of generality, we may consider paths along the $x$-direction only \footnote{where, at this time, we are considering the isotropic form of \eqref{FRWsphere} with $dr^{2}+r^{2}d\Omega^{2}=dx^2+dy^2+dz^2$}, with $x^{\mu}(\lambda)=\{t(\lambda),x(\lambda),0,0\}$. For $ds^{2}\rvert_{\mbox{{\tiny null}}}=0$, we then have
\[
dt^{2}=a^{2}(t)dx^{2},\qquad \implies \qquad \frac{dx}{d\lambda}=\frac{1}{a}\frac{dt}{d\lambda}.
 \]
Substituting this latter identity into the geodesic equation for the time coordinate gives
\[
\frac{d^{2}t}{d\lambda^{2}}+\frac{\dot{a}}{a}\left(\frac{dt}{d\lambda}\right)^{2}=0
 \]
One can easily verify that $d\lambda=\frac{a}{N}dt$ with constant $N$ is a solution of this equation. Setting $N$ to unity we find
\[
\left(\frac{dt}{d\lambda},\frac{dx^{i}}{d\lambda}\right)=\left(\frac{1}{a},\frac{1}{a^{2}}\right)
. \]
Due to the isotropic nature of the FRW metric \eqref{FRWsphere}, the spatial components are equal and can therefore be truncated into  the index $i=\{1,2,3\}$~\footnote{Greek indices indicate spatial and temporal components, i.e. $\mu,\nu,\lambda,\ldots=\{0,1,2,3\}$, whereas Latin letters indicate the spatial components $i,j,k,\ldots=\{1,2,3\}$}. We may now express $k^{\mu}$, the tangential vector field to the congruence of null geodesics, as follows
\[
\label{nullgeo}
k^{\mu}=\left(\frac{1}{a},\pm\frac{1}{a^{2}}\right)=(k^{0},k^{i}),
 \]
where the sign attached to $k^i$ is negative for ingoing rays and positive for outgoing rays. The final step is to compute the expansion, itself. To this end, we note that the expansion can be rewritten in the following form
\[
\theta\equiv\frac{1}{\sqrt{g}}\partial_\mu(\sqrt{g} k^\mu )
\]
For the FRW metric given in \eqref{FRWsphere}, we have $g=|\det(g_{\mu\nu})|=a^6 r^4 \sin^2\psi$, allowing us to write the ingoing and outgoing expansion for the given cosmological spacetime
\[
\label{nullgeoinout}
\theta_{IN}=\frac{2}{a(t)}\left(\frac{\dot{a}}{a}-\frac{1}{ra}\right),\qquad\theta_{OUT}=\frac{2}{a(t)}\left(\frac{\dot{a}}{a}+\frac{1}{ra}\right)
. \]
\cite{Vilenkin:2014yva},\cite{Vachaspati:1998dy}.
\subsubsection*{Cosmological Apparent Horizons and Conformal Diagram}
The first thing to note from these ingoing and outgoing expansions is that the term $x\equiv ra$ denotes the physical size of the surface described by the expansion in a geometrically-flat spacetime. This can be seen by the general formula for comoving distance of a general FRW metric, which is given by \cite{Berera:2000xz}
\[
x\equiv a\int dr (1-kr^2)^{-\frac{1}{2}}=ar,\mbox{ when } k=0.
\]
As the inner boundary of an antitrapped surface is the region where the surface becomes \emph{marginal}, by definition, this region will have vanishing expansion. Thus, we may justify the assertion \eqref{mashorizon} that the minimally antitrapped surface is bounded by an apparent horizon on its inner margin, upon reference to the ingoing expansion given in \eqref{nullgeoinout}. By setting \eqref{nullgeoinout} to zero, we find that for a comoving distance $r_{mas}$, we have
\[
x_{mas}=H_{FRW}^{-1}
,\] 
where $H_{FRW}$ is the cosmological apparent horizon of the FRW background \cite{Berera:2000xz}, \cite{Vachaspati:1998dy},\cite{tHooft:2009bh},\cite{d1992approaches}.

\subsubsection*{Accelerated Expansion of the Universe} 
\begin{figure}[h]\cite{Vachaspati:1998dy},\cite{Berera:2000xz}
\centering
\includegraphics[scale=0.4]{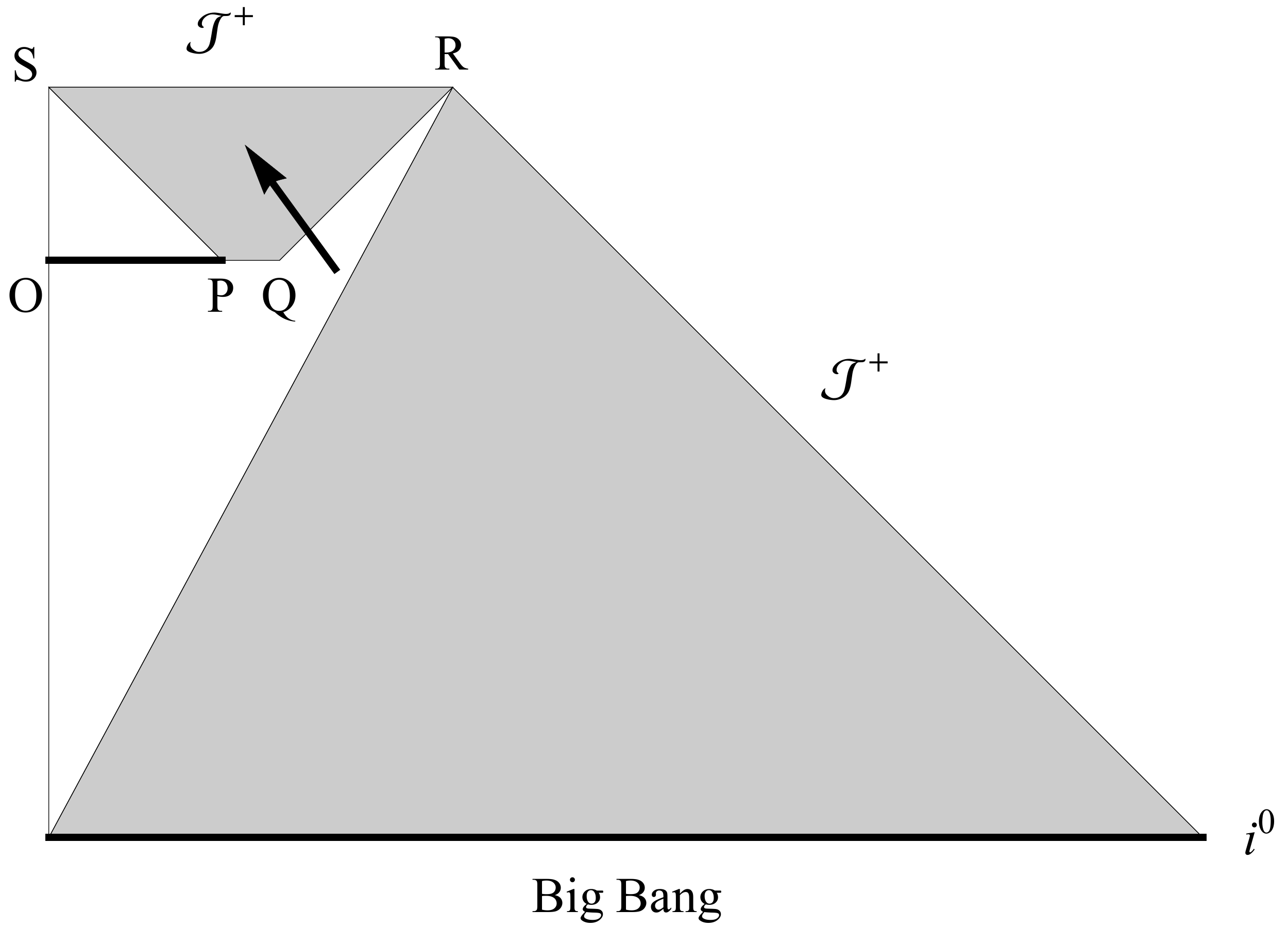}
\caption{A conformal diagram of a Big Bang cosmology with local inflation. Shaded regions are antitrapped and white regions are normal surfaces. A patch begins to inflate at cosmic time $t$ from $O$ to $Q$ with inflationary size $x_{\text {\it inf}}$, where the line $OP$ borders the apparent inflationary horizon. The arrow depicts an ingoing null ray entering an antitrapped region from a normal region, which is prohibited under the convergence condition \eqref{nullCC}. The inflationary patch $OQ$ may indeed be extended into the antitrapped region so that no such violation occurs.}\label{bigbang} 
\end{figure}

Inflation theory suggests a period of rapid expansion soon after the  Big Bang is required for the formation of the large scale structures we see today. Whereas, expansion slows after this inflationary period, observational data has found that the Universe is still undergoing accelerated expansion at present \cite{Accel}. In terms of the FRW metric \eqref{FRWsphere}, accelerated expansion of the universe is defined by
\[
\label{acc}
\ddot{a}>0,\qquad\mbox{equivalently}\qquad {\dot H}+H^2>0
.\]
On comparing this with the curvature scalar in FRW, $R=6\left(\dot{H}+2H^2+\frac{k}{a^2}\right)$, we find that in a period of accelerated expansion, the curvature scalar $R$ is always positive, at least in a geometrically flat or open Universe. Fig. \ref{bigbang} illustrates  such inflation within a Big Bang cosmology. 

\subsubsection*{Null vs. Timelike Geodesic Congruences}
We stated earlier that null geodesic congruences more readily converge than their timelike counterparts and therefore, an analysis of null rays is more illuminating in terms of defocusing and past-completeness. In this section, we will show that if a geometrically-flat spacetime is singularity-free in the context of null rays, its timelike counterpart will necessarily be singularity-free. For our purposes here, we begin with the isotropic form of the FRW metric
 \[
 ds^2=-dt^2+a^2(t)(dx^2+dy^2+dz^2)
, \]
while following closely to \cite{Albareti:2014dxa}. We then compute the null and timelike convergence conditions, which are
\[
a\ddot{a}\leq\dot{a}^{2},\qquad a\ddot{a}\leq\frac{\dot{a}^{2}}{1+\frac{3a^{2}}{2\gamma_{0}^{2}v_{0}^{2}u_{0}^{2}}}
 \]
 respectively. Here, the timelike vector field is taken to be of the form $\xi^\mu=\gamma(1,\nu^i)$, where $\gamma\equiv \frac{1}{\sqrt{1-a^2 v^2}}$; and the subscript $_0$ refers to the quantities being evaluated at $t=t_0$. Further details can be found in \cite{Albareti:2014dxa}. For the present discussion, the precise details of these quantities are not strictly relevant. We simply note that the respective convergence conditions take the form
 \[
 a\ddot{a}\leq\dot{a}^{2},\qquad a\ddot{a}\leq\frac{\dot{a}^2}{1+A^2}
\]
with $A^{2}=\frac{3a^{2}}{2\gamma_{0}^{2}v_{0}^{2}u_{0}^{2}}$, being a positive parameter. The limiting case of the null convergence condition can be found by framing the left inequality as an identity and solving the differential equation. We find this to take the form 
\[
a(t)=c_0 e^{c_1 t},
\]
for some integration constants $c_0,c_1$. Pleasingly, this conforms to the scale factor in de Sitter space,
\[
a(t)=a_0 e^{\bar{H} t}
,\]
where ${\bar H}$ is the Hubble constant \cite{Albareti:2014dxa}. Comparing this with the timelike case and we find that of the two convergence conditions, the null CC is less restrictive. In other words, timelike geodesics are more easily made past-complete by this condition and so, in a study of singularity-free cosmologies, it makes sense to study null rays over their timelike counterparts. To summarise, in a geometrically-flat cosmology, if a spacetime is non-singular for null rays, it will also be devoid of singularities for timelike geodesics.
\section{Defocusing Conditions for Infinite Derivative Gravity around Minkowski Space}
\label{sec:defocusmink}
In this section, we extend our study of geodesic congruences away from general relativity  into the novel approach of infinite derivative gravity (IDG), the groundwork of which was laid in Chapter \ref{chap:2}, where the non-linear and linearised equations of motion were derived and in Chapter \ref{chap:GF}, where the theory was rendered free of ghosts and tachyons. It is now time to return to the central question of this thesis, first raised in the Introduction, by asking the question:
\begin{quote}
\emph{Can null rays defocus in an infinite derivative theory of gravity, without introducing ghosts, tachyons or exotic matter?}
\end{quote}
In essence, we wish to show how infinite derivative extensions of gravity, in contrast to GR and finite  models, have the potential to describe a stable and singularity-free theory of gravity. 
\\\\
Recall that in Section \ref{sec:linkmink}, we derived the linearised field equations for the infinite derivative action of gravity:
\[
\label{actionRE}
S=\frac{1}{2}\int d^4x \sqrt{-g}\biggl(M_P^2 R+R{\cal F}_1(\Box)R+R^{\mu\nu}{\cal F}_2(\Box)R_{\mu\nu}+C^{\mu\nu\lambda\sigma}{\cal F}_3(\Box)C_{\mu\nu\lambda\sigma}\biggr)
,\]
where, within the form factors ${\cal F}_i(\Box)=\sum^\infty_{n=0}(\Box/M^2)^n$, each D'Alembertian operator is modulated by the scale of non-locality $M$. The resulting field equations are given by
\[
\label{eomminkRE}
\kappa T_{\mu\nu}=a(\Box)R_{\mu\nu}-\frac{1}{2}\eta_{\mu\nu} c(\Box)R-\frac{f(\Box)}{2}\partial_\mu \partial_\nu R
,\]
where the infinite derivative functions $a(\Box),c(\Box),f(\Box)$ are made up of the form factors ${\cal F}_i(\Box)$, defined in \eqref{abc}, and conform to the constraint
\[
f(\Box)\Box=a(\Box)-c(\Box)
.\]
Furthermore, through our lengthy discussion on the Raychaudhuri equation in General Relativity in Section \ref{sec:RE}, we learned of its powerful role in conveying the focusing behaviour of geodesic congruences, where the sole contribution of gravity stems from the $R_{\mu\nu}k^\mu k^\nu$ term, with $k^\mu$ representing a null tangent vector or ray. We may then find the contribution of gravity to the RE for the infinite derivative theory of gravity, described by the action \eqref{actionRE}, by contracting the linearised field equations \eqref{eomminkRE} with the tangent vectors $k^\mu$. Thus, we obtain the \emph{IDG convergence condition}
\[
\label{convergeIDG}
R_{\mu\nu} k^\mu k^\nu=a^{-1}(\Box)\biggl(\kappa T_{\mu\nu}k^\mu k^\nu +\frac{1}{2}k^\mu k^\nu f(\Box)\partial_\mu \partial_\nu R\biggr)\geq 0
.\]
If a theory satisfies this condition, the associated null rays cannot start to diverge until they reach the origin. In other words, these null rays converge towards a singularity in a finite time, as is the behaviour in GR. However, in contrast to GR, where $R_{\mu\nu}k^\mu k^\nu$ 
 must remain positive so as  not to violate the null energy condition \eqref{NEC}, we have modified the stress-energy tensor in such a way that it may indeed be possible to reverse the sign of $R_{\mu\nu}k^\mu k^\nu$ whilst retaining the NEC. We call the inequality $R_{\mu\nu}k^\mu k^\nu<0$ the \emph{defocusing condition}, as it is the condition whereby null rays may defocus, suggestive of a singularity-free theory of gravity.

\subsubsection*{Homogenous Solution.}
The linearised field equations \eqref{eomminkRE} describe the curvature of a spacetime that has been perturbed away from Minkowski space. We begin our analysis by discussing perturbations that are entirely homogenous, with all curvature dependent only on the cosmic time $t$. One could think of the time-dependent, perturbed metric $h_{ij}$ which makes up the curvature, as being closely related to the cosmological scale factor of FRW, which is useful in this context, as we are considering cosmological singularities. In the homogenous case, the D'Alembertian simply becomes $\Box=-\partial_t^2$, so that the defocusing condition $R_{\mu\nu}k^\mu k^\nu<0$ reads \cite{Conroy:2016sac}
\[
\label{divergemink1}
R_{\mu\nu} k^\mu k^\nu=a^{-1}(\Box)\biggl(\kappa T_{\mu\nu}k^\mu k^\nu -\frac{1}{2}k^\mu k^\nu f(\Box)\Box R\biggr)< 0
,\]
where in order to preserve the NEC, we have $T_{\mu\nu}k^\mu k^\nu\geq 0$. We may then say that the minimum requirement for such a theory to display the desired defocusing behaviour is given by
\[
\frac{f(\Box)\Box}{a(\Box)} R= \frac{a(\Box)-c(\Box)}{a(\Box)} R>0
\label{divergemink2}
,\]
with $T_{\mu\nu}k^\mu k^\nu$ set to zero.
Immediately, we are confronted with some important observations, which we outline below.
\subsubsection*{Observations}
\begin{enumerate}
\item[$a=c:$] If we recall the form of the modified graviton propagator from Section \ref{sec:prop}
\[
\label{propmink}
\Pi(-k^{2})=\frac{\mathcal{P}^{2}}{k^{2}a(-k^{2})}+\frac{\mathcal{P}_{s}^{0}}{k^{2}\left(a(-k^{2})-3c(-k^{2})\right)},
\]  
we find that the condition $a(\Box)=c(\Box)$ necessitates that no additional pole, other than the massless graviton, is introduced. In this case, the modified propagator is simply the physical graviton propagator modulated by an overall factor of $\sim 1/a(\Box)$, where the function $a(\Box)$ is an exponent of an entire function, containing no roots:
\[
\label{defocus0}
\Pi(-k^{2})=\frac{1}{a(-k^{2})}\biggl(\frac{\mathcal{P}^{2}}{k^2}-\frac{\mathcal{P}_{s}^{0}}{2k^2}\biggr)
.\] 
The curvature $R$ is positive as a result of accelerated expansion of the Universe, Section \ref{sec:RE}, so that from \eqref{divergemink1}, we see that the defocusing condition can only be achieved if $a(\Box)$ is negative when acting on the curvature. As should be apparent from the above form of the propagator, such a negative function would reverse the sign of the spin-2 component, leading to a negative residue and subsequently a ghost. This ghost is known as the \emph{Weyl ghost} and was discussed in Section \ref{sec:patho}. As a result, we conclude that an \emph{additional scalar degree of freedom is required} in order for null rays to display the desired defocusing behaviour.  

\item[$a\neq c:$] Having established the  need for a departure from the pure massless mode of the graviton propagator
in \eqref{propmink}, we move into the more general case of $a(\Box)\neq c(\Box)$. This condition tells us that in order for the null rays to defocus - a minimum requirement of a singularity-free theory of gravity - one requires an additional root in the spin-0 component of the graviton propagator. As such, one additional scalar degree of freedom must propagate in the spacetime besides the massless graviton, if we wish to satisfy the defocusing condition.  As $a(\Box)$ does not introduce a new pole, the spin-2 component of the graviton
propagator remains massless. 
\end{enumerate}
~\\We have already demonstrated a significant departure from general relativity, in that IDG corrections have allowed for the possibility of singularity avoidance, via the defocusing condition \eqref{defocus0}, without violating the null energy condition. 

Having established the need for an additional pole in the propagator, we must now take steps to avoid the introduction of ghosts or tachyons. In  Section \ref{sec:GF}, we derived the ghost-free condition around a Minkowski background. This condition took the form
\[
\label{GFrepeat}
c(\Box)=\frac{a(\Box)}{3}\left[1+2(1-\alpha M_P^{-2}\Box){\tilde a}(\Box)\right]
, \]
where the constant $\alpha=6f_{1_{0}}+2f_{2_{0}}-M_{P}^{2}/M^{2}$ and ${\tilde a}(\Box)$ is an exponent of an entire function, containing no roots. Substitution into \eqref{propmink} reveals the ghost-free modified propagator for an asymptotically-flat spacetime:
\[
\label{MinkPropdec}
\Pi(-k^{2})=\frac{1}{a(-k^{2})}\biggl[\frac{{\cal P}^{2}}{k^{2}}-\frac{1}{2\tilde{a}(-k^{2})}\biggl(\frac{{\cal P}_{2}^{0}}{k^{2}}-\frac{{\cal P}_{s}^{0}}{k^{2}+m^{2}}\biggr)\biggr]
, \]
where we have defined
\[
\label{Staro-mass}
m^2=M_P^2/\alpha
.\]
$m^2$ must be positive to ensure that the mass is non-tachyonic and $\alpha$ positive definite in order to retain the essential new pole, i.e. the constant $\alpha=6f_{1_{0}}+2f_{2_{0}}-M_{P}^{2}/M^{2}$ satisfies $\alpha>0$. Armed with this, we are now in a position to describe the defocusing condition which precludes the existence of ghosts. Substitution of \eqref{GFrepeat} into \eqref{divergemink2} leads to the central result
\[
\label{defocusmink}
(1-\Box/m^2){\tilde a}(\Box)R<R
.\]
\subsection{Comparison with Starobinsky Model}
Taking the limit $M \rightarrow \infty$, with  ${\cal F}_2={\cal F}_3=0$, reduces the action \eqref{action} to that of Starobinsky's model of 
inflation~\cite{Starobinsky:1979ty}. Indeed, a curious question to ask is, could Starobinsky's action avoid the cosmological singularity? At the limit $M\rightarrow \infty$,  the propagator \eqref{propmink} can be expressed as
\[
\label{propR2}
\Pi_{R^{2}}=\Pi_{GR}+\frac{1}{2}\frac{\mathcal{P}_{s}^{0}}{k^{2}+m^{2}} ,
\]
where $m$ is given by (\ref{Staro-mass}), with $\alpha=6f_{1_0}\geq 0$, and $m^2>0$, to avoid tachyonic mass.
 However, the fundamental difference can be seen by comparing the propagator for $R^2$-gravity with the IDG propagator, \eqref{MinkPropdec}.  In the local limit, $a(\Box)={\widetilde a}(\Box)\rightarrow 1$. Furthermore, as we are making comparisons with the propagator in momentum space, the D'Alembertian takes the form $\Box\rightarrow -k^2$. In this case, the defocusing inequality \eqref{divergemink1} can only be satisfied for 
\[
m^{-2} R<0
.\]
In this scenario, to avoid focusing we require $m^2< 0$, rendering the theory tachyonic. Alternatively, negative curvature would contradict the requirement of accelerated expansion of the Universe, see \eqref{acc}, which is vital to realise primordial inflation, Section \ref{sec:patho}. As such, the Starobinsky model cannot pair inflation with resolving the Big Bang Singularity.
\section{Bouncing Solution}
\label{sec:Bouncing}
\begin{figure}[h]\cite{Conroy:2014dja}
\centering
\includegraphics[scale=0.3]{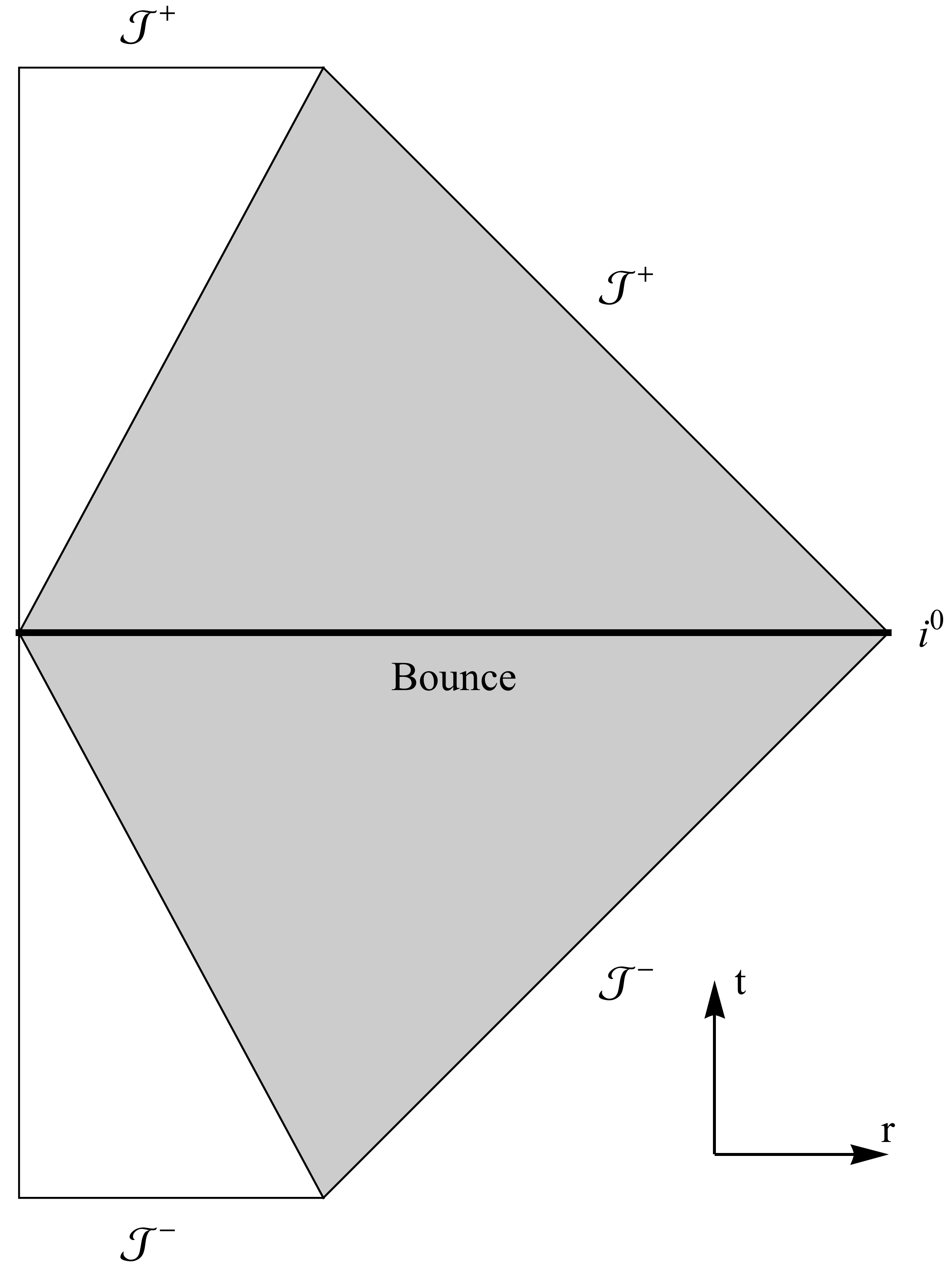}
\caption{A conformal diagram depicting a bouncing cosmology, seen as an extension of Fig. \eqref{bigbang} into past infinity. Shaded regions are antitrapped surfaces, bordered by a cosmological apparent horizon on their inner margins and white regions are normal surfaces.} \label{figbounce} 
\end{figure}
Up to this point, we have not spoken about what replaces the Big Bang singularity in a non-singular spacetime. 
This is because, as opposed to \cite{Conroy:2014dja},\cite{Koshelev:2012qn},\cite{Koshelev:2013lfm},\cite{Biswas:2011ar},\cite{Abramo:2009qk}, we have made no assumptions on the nature of the cosmological scale factor, which is closely related to the perturbed metric tensor $h_{\mu\nu}$. The term scale factor is most often associated with the function $a(t)$ in an FRW metric \eqref{FRWsphere}. However, in an homogeneous and isotropic spacetime such as the one we are considering around Minkowski space, we note that we are very closely aligned to the FRW metric, making cosmological predictions relevant. In this way, the metric can be considered to be of the form
\[
g_{\mu\nu}=\eta_{\mu\nu}+h_{\mu\nu}=\{-1,a^2(t),a^2(t),a^2(t)\}
.\]
Bouncing cosmologies replace the initial big bang singularity with that of a bounce, so that incoming geodesic congruences can be made past complete - stretching to past infinity, leading to an extension of the conformal diagram Fig. \ref{bigbang}, which is illustrated in Fig. \ref{figbounce}. Although, we do not suppose a priori that the initial singularity is replaced by a bounce, we would indeed expect cosmologies with a bouncing scale factor to satisfy the defocusing condition \eqref{defocusmink}. 

\noindent We illustrate this by way of example.
\subsection{Integral Form}
It may now be illuminating to test our defocusing condition \eqref{defocusmink}. We proceed, as in the Starobinsky case, by taking our analysis into momentum space and testing against a well-known \emph{bouncing solution}, that is, a cosmology defined by a bouncing scale factor, which is necessarily an even function. However, due to the infinite derivative nature of the function ${\tilde a}(\Box)$, this comes with an added degree of complexity. One possibility in analysing the defocusing condition \eqref{defocusmink}  lies in recasting the defocusing condition into its integral form \cite{reed1975methods}.
\newpage
\noindent\emph{Integral Form}\\
As discussed in Chapter \ref{chap:GF}, we must choose ${\tilde a}(\Box)$ to be an exponent of an entire function, the simplest case being~\footnote{We note briefly the importance of the sign in the exponent of ${\tilde a}(\Box)$. In Appendix \ref{chap:NewtPot},\cite{Edholm:2016hbt},\cite{Biswas:2011ar}, it is shown that for the correct Newtonian potential to be observed, the non-local function $a(\Box)$ must be of the form $a(\Box)=e^{-\gamma(\Box)}$, where $\gamma(\Box)$ is an entire function. As ${\tilde a}(\Box)$ is defined as ${\tilde a}(\Box)\propto 1/a(\Box)$, and both are defined as exponents of entire functions, it is reasonable to expect a difference of a minus sign in the exponent.} 
\[
{\tilde a}(\Box)=e^{\Box/M^2}
,\]
where, again, $M$ is the scale of non-locality. Thus, we wish to compute $e^{\Box/M^2}R(t)$, where $R(t)$ is the curvature scalar, solely dependent on $t$. To this end, we reformulate this expression into its integral form by first defining the Fourier transform and its inverse like so
\[
\hat{R}(k)\equiv\frac{1}{\sqrt{2\pi}}\int_{-\infty}^{\infty}e^{ikt}\tilde{R}(t)dt,\qquad R(t)\equiv\frac{1}{\sqrt{2\pi}}\int_{-\infty}^{\infty}e^{-ikt}\hat{R}(k)dk
.\]
We may then write
\[
{\tilde a}(\Box)R(t)	=	\frac{1}{\sqrt{2\pi}}\int_{-\infty}^{\infty}\exp(-k^{2}/M^{2})\exp(-ikt)\hat{R}(k)\;dk
.\]
By using the properties of the Fourier transform and defining $x\equiv k/M$, we may express this as follows
\[
{\tilde{a}}(\Box)R(t)=\frac{M}{2\pi}\int\int_{-\infty}^{\infty}\exp\left[-x^{2}+ix\left(M(\tau-t)\right)\right]\hat{R}(\tau)\;dx\;d\tau
.\]
Now, in order to compute in terms of the Gaussian integral
$
\int_{-\infty}^{\infty}\exp\left(-a(x+b)^{2}\right)dx=\sqrt{\frac{\pi}{a}}
$,
we rewrite ${\tilde a}(\Box)R(t)$ into the following form
\begin{align}{\tilde{a}}(\Box)R(t)=\frac{M}{2\pi} & \int\int_{-\infty}^{\infty}\exp\left[-\left(x-\frac{i}{2}\left(M(\tau-t)\right)\right)^{2}-\frac{1}{4}\left(M(\tau-t)\right)^{2}\right]\hat{R}(\tau)\;dx\;d\tau.\end{align}
We then compute the Gaussian integral to find
\[
{\tilde a}(\Box)R(t)=\frac{M}{2\sqrt{\pi}}\int_{-\infty}^{\infty}e^{-\frac{1}{4}M^2(\tau-t)^{2}}\hat{R}(\tau)d\tau
.\]
Similarly,
\[
-\Box/m^{2}\tilde{a}(\Box)R(t)=\frac{M^{3}}{8\sqrt{\pi}m^{2}}\int_{-\infty}^{\infty}e^{-\frac{1}{4}M^{2}(t-\tau)^{2}}\left(2-M^{2}(t-\tau)^{2}\right)\hat{R}(\tau)d\tau
.\]
The defocusing condition \eqref{defocusmink} can then be written as
\[
\label{defocusint}
\frac{M}{2\sqrt{\pi}}\biggl[\int_{-\infty}^{\infty} e^{-\frac{1}{4}M^2(\tau-t)^{2}}\biggl(1+\frac{M^{2}}{2m^{2}}-\frac{M^{4}}{4m^{2}}(t-\tau)^{2}\biggr)\hat{R}(\tau)d\tau<R(t)
\]
\\\subsubsection*{Example: $a(t)=\cosh\frac{\sigma}{2} t$}
We now turn to a particular example of a bouncing solution. In this case, we assume a scale factor of $a(t)=\cosh\frac{\sigma}{2} t$, where $\sigma$ is a  parameter of mass dimension. Solutions of this type have been studied extensively in \cite{Koshelev:2012qn},\cite{Koshelev:2013lfm},\cite{Biswas:2005qr} and found to be a solution of the field equations \eqref{eomfull}, via an Ansatz-based approach. In terms of the perturbed, t-dependent metric $h_{\mu\nu}$, this scale factor can be written as
\[
h_{\mu\nu}=\biggl\{0,\;\cosh^2(\frac{\sigma t}{2})-1,\;\cosh^2(\frac{\sigma t}{2})-1,\;\cosh^2(\frac{\sigma t}{2})-1\biggr\}.
\]
Then, from the definition of curvature around Minkowski, \eqref{MinkR}, we find the curvature to be
\[
R(t)=\frac{3}{2}\sigma^2 \cosh\sigma t
.\]
We then substitute this form of the curvature into the defocusing condition \eqref{defocusint} and compute the integral to find that, for any cosmic time $t$, defocusing may be realised according to
\[
\frac{\sigma^{2}}{m^{2}}>(1-e^{-\frac{\sigma^{2}}{M^{2}}}).
\]
This is satisfied for all real $\sigma$ such that $\sigma\neq0$,
so that we have confirmed that a known bouncing scale factor does indeed display the desired ghost-free defocusing behaviour, according to the constraint \eqref{defocusmink}.
\section{A simpler action of gravity.}
\label{sec:simpler}
From the defocusing condition \eqref{divergemink1}, we can deduce the simplest infinite derivative action that can describe a singularity-free theory of gravity. The central components for defocusing are the functions $a(\Box)$ and $c(\Box)$, which, in order to achieve freedom from ghosts, are exponents of entire functions with zero roots and one root, respectively. These functions are in turn made up of the infinite derivative form factors ${\cal F}_i(\Box)$ with $i=\{1,2,3\}$, which make up the gravitational action \eqref{action}, as given by \eqref{abc}. Consequently,  upon inspection of \eqref{abc}, it appears that we may be able to `switch off' one or two of the form factors without changing the  nature of the  functions $a(\Box)$ or $c(\Box)$. 
\\\\ \emph{Non-linear Regime}\\
For example, by setting ${\cal F}_2=0$, whilst retaining the infinite derivative form factors ${\cal F}_1$ and ${\cal F}_3$ and noting that
in a conformally-flat background, such as FRW, (A)dS or Minkowski, the Weyl tensor vanishes on the background, the action \eqref{action} reduces to the following
\[
\label{actionsimp}
S_{NL}=\frac{1}{2}\int d^{4}x\sqrt{-g}\bigl[M_{P}^{2}R+ R{\cal F}_{1}(\Box)R\bigr]\,.
\]
This reduced action would clearly prove useful in a non-linear cosmological analysis, where the contribution of the Weyl tensor would be precisely zero, even with a non-zero form factor ${\cal F}_3$, but it may also prove to be of interest in the linearised regime.
\\\\\emph{Linearised Regime}\\
On inspection of the infinite derivative functions \eqref{abc} which make up the field equations in the linearised regime, it should be clear that it is possible to `switch off' any one of the form factors ${\cal F}_i$, whilst still retaining the infinite derivative nature of the functions $a,c,f$ and thus, not adversely affecting the theory. We may extend this further to switching off two of the form factors ${\cal F}_i$. Straightforward examples, include: setting ${\cal F}_1={\cal F}_3=0$ with
\ba
&a(\Box)=1+M_{P}^{-2}{\cal F}_{2}(\Box)\square\nn\\&
a(\Box)-3c(\Box)=-2+4M_{P}^{-2}{\cal F}_{2}(\Box)\square
  \ea
and ${\cal F}_1={\cal F}_2=0$:
\ba
&a(\Box)=1+2M_{P}^{-2}{\cal F}_{3}(\Box)\Box\nn\\&
a(\Box)-3c(\Box)=-2+12M_{P}^{-2}{\cal F}_{1}(\Box)\square,
  \ea
 which retain the infinite derivative modification of both the scalar and tensorial sectors of the propagator. Slightly less clear, however, is the proposition of setting ${\cal F}_2={\cal F}_3=0$. This results in a correction of the scalar sector of the propagator, while leaving the spin-2 sector of the propagator unmodified. In this instance the relevant sectors of the propagator can be obtained from \eqref{abc} and are given by
\ba
\nn
&a(\Box)=1
\\& a(\Box)-3c(\Box)=-2+12M_{P}^{-2}{\cal F}_{1}(\Box)\square,
 \ea
One can easily check, from \eqref{defocusmink}, that this is indeed sufficient to realise the desired ghost-free defocusing behaviour. It appears then, that the modification of the spin-2 component of the propagator - which is rootless and positive so as to avoid the Weyl ghost - does not play a leading role in singularity avoidance.    Furthermore, the simplicity of this case allows us to more easily convey defocusing conditions in more complicated scenarios, such as around de Sitter space, which is our next focus.
 
 \section{Defocusing Conditions around de Sitter Space}
 \label{sec:defocusdS}
Here, we extend our discussion to include ghost-free, defocusing conditions around the de Sitter spacetime, with the reduced action
\[
\label{actionred}
S=\frac{1}{2}\int d^4x\sqrt{-g}\biggl(M_P^2 R-2\Lambda+R {\cal F}(\Box)R\biggr)
.\]
As we have seen from the previous section on a \emph{simpler action of gravity}, this is adequate for our aims of describing a ghost- and singularity-free theory.
This reduced form is equivalent to setting ${\cal F}_2={\cal F}_3=0$ in the general action \eqref{action} and dropping the remaining subscript from the function ${\cal F}_1\equiv {\cal F}$, for convenience. 

From here, we can deduce the expected behaviour of the propagator around de Sitter space by inspecting the modified propagator around Minkowski \eqref{propmink}. Upon reference to \eqref{abc}, we find that, by imposing ${\cal F}_2={\cal F}_3=0$, the spin-2 or tensorial sector of the propagator is no longer modified and all subsequent corrections take place in the scalar sector. This is as expected due to the purely scalar modification of the Einstein-Hilbert action taking place in \eqref{actionred}. An exposition of the ghost-free conditions for the full action \eqref{action} has recently been discussed in \cite{Biswas:2016egy}, where the perturbed metric $h_{\mu\nu}$ is decomposed into its 10 individual degrees of freedom via
\[
h_{\mu\nu}=h_{\mu\nu}^\perp+\nabla_\mu A_\nu^\perp+\nabla_\nu A_\mu^\perp+(\nabla_\mu\nabla_\nu-\frac{1}{4}{\bar g}_{\mu\nu}\Box)B+\frac{1}{4}{\bar g}_{\mu\nu}h
.\]
Here, the transverse and traceless, massless, spin-2 graviton $h_{\mu\nu}^\perp$, accounts for 5 degrees of freedom; the transverse vector field $A_\mu ^\perp$ contributes 3 degrees of freedom; and the two scalars $h$ and $B$ provide a further 2 degrees of freedom. In this case, \textbf{transverse} simply refers to a tensor that has vanishing divergence, i.e. $\nabla_\mu A^{\mu \nu...}=0$. In the present discussion, we invoke an alternative method, similar to the previous discussion around Minkowski space.
\newpage\noindent\emph{Field Equations}\\
The field equations of the action \eqref{actionred} can be read off from \eqref{eomdS} and are given by
\ba
&\kappa T_{\nu}^{\mu}=\left(1+24M_{P}^{-2}H^{2}\lambda f_{0}\right)\left(r_{\nu}^{\mu}-\frac{1}{2}\delta_{\nu}^{\mu}r\right)-2\lambda M_{P}^{-2}\left(\nabla^{\mu}\partial_{\nu}-\delta_{\nu}^{\mu}\square\right){\cal F}(\Box)r
\nn\\&+6\lambda M_{P}^{-2}H^{2}\delta_{\nu}^{\mu}{\cal F}(\Box)r
 . \ea
Upon reflection, we find that these field equations can be recast in precisely the same form as in the Minkowski case, \eqref{eomminkred}
\[
\label{eomdSacf}
\kappa T_{\nu}^{\mu}=ar_{\nu}^{\mu}-\frac{1}{2}\delta_{\nu}^{\mu}c(\Box)r-\frac{1}{2}\nabla^{\mu}\partial_{\nu}f(\Box)r
, \]
according to the following definitions
\begin{eqnarray}
\label{dSfuncs}
&a=1+24M_{P}^{-2}H^{2}\lambda f_{0}
 \nn\\&
c(\Box)=1+24M_{P}^{-2}H^{2}\lambda f_{0}-4\lambda M_{P}^{-2}\left(\Box+4H^{2}\right){\cal F}(\Box)
 \nn\\&
\nabla^{\mu}\partial_{\nu}f(\Box)=4M_{P}^{-2}\lambda\left(\nabla^{\mu}\partial_{\nu}+\delta_{\nu}^{\mu}H^{2}\right){\cal F}(\Box) ,
\end{eqnarray}
which return \eqref{abc} at the limit $H\rightarrow 0$\footnote{provided that ${\cal F_2}={\cal F}_3=0$ due to the reduced action \eqref{actionred}}. In order to be consistent with the Minkowski case, we next note that these infinite derivative functions must conform to the same constraints given by \eqref{minkconstr}, namely that
\[
\Box f(\Box)=a-c(\Box)
.\]
By taking the trace of the final equation in \eqref{dSfuncs}, we find that this is indeed the case. 
\\\\\emph{Ghost-free Conditions}\\
Having established established consistency with the Minkowski case, in that the field equations take the same form and obey the same generic conditions, the propagator will be modified in a similar manner, according to
\[
\Pi_{ds}(\Box)=\frac{{\cal P}_{GR}^{2}}{a}+\frac{({\cal P}_{s}^{0})_{GR}}{a-3c(\Box)}.
\]
Here, the subscript $_{GR}$ denotes the physical (GR) graviton propagators around de Sitter space and contain the GR roots of the propagator via \cite{Biswas:2016egy},\cite{DHoker:1999bve},\cite{Mora:2012zi},\cite{PhysRevD.34.3670}.
\[
{\cal P}_{GR}^2=\frac{{\cal P}^{2}}{-\Box+2H^{2}},\qquad ({\cal P}_{s}^{0})_{GR}=-\frac{{\cal P}_{s}^{0}}{\Box+4H^{2}},
\]
which reduce to the familiar root $k^2=0$ at the Minkowski limit $H\rightarrow 0$ \footnote{Note that in de Sitter space the D'Alembertian operator acting on a scalar is given by 
$\Box S	=g^{\mu\nu}\nabla_{\mu}\nabla_{\nu}S
	=g^{\mu\nu}\partial_{\mu}\partial_{\nu}S-g^{\mu\nu}\Gamma_{\mu\nu}^{\kappa}\partial_{\kappa}S
		=(-\partial_{t}^{2}-3H\partial_t )S.$
	In momentum space, we can write this as
	$
	\Box S\rightarrow (-(k^{0}k_{0})-3Hik_{0})S.
$}
%
. We note here that the spin-2 sector is modulated by the constant $a=1+24M_{P}^{-2}H^{2}\lambda f_{0}$. From our discussion on pathologies of the propagator in Section \ref{sec:patho}, we know that in order to avoid the Weyl ghost, this constant must be positive definite. In truth, the positive nature of this constant is determined by fundamental physical constraints. In Appendix \ref{sec:Entropy}, we discuss the role of this constant in the gravitational entropy of such an infinite derivative action around de Sitter space, see also \cite{Conroy:2015nva}. The upshot is that the point $a=0$ coincides with a physical system defined by vanishing entropy, while $a<0$ describes non-physical spacetimes with negative entropy. Thus, $a>0$ and as a result, the tensorial structure of the propagator can not be said to be modified in any meaningful manner, as the positive constant
 \[
 \label{dSa}
 a=1+24M_{P}^{-2}H^{2}\lambda f_{0}>0
 ,\]
could be normalized to unity, if so desired. This is as expected, as the modification that is taking place is within a purely scalar modification of GR.
\\\\\emph{Ghost-free Conditions}\\
In order to avoid negative residues in the spin-0 component of the propagator, we proceed in much the same manner as in the Minkowski case, by relating the trace equation to an exponent of an entire function that has been furnished with an additional root, like so 
\[
\label{propdS}
\kappa T=\frac{1}{2}(a-3c(\Box))r=(\alpha \Box {\bar m}^{-2}-1){\bar a}(\Box)r,
\] 
where the trace equation is given by
  \[
\kappa T=-\left(1+24H^{2}M_{p}^{-2}\lambda f_{0}\right)r+6M_{p}^{-2}\lambda\left(\Box+4H^{2}\right){\cal F}(\Box)r
.  \]
As before, the substitution $\Box\rightarrow 0$ reveals that the Brans-Dicke Scalar ${\bar m}^2=M_P^2$, whereas expanding to first order reveals the constant $\alpha$ to be now given by
\[
\alpha=6\lambda f_{0}-M_{P}^{2}/M^{2}+24\lambda H^{2}M^{-2}f_{1}.
\]
Again, we check the limit as $H\rightarrow0$ returns \eqref{alpha}, with $f_{2_0}=0$. 
\\\\\emph{Tachyon Criteria}\\
Now, by decomposing the propagator \eqref{propdS} into partial fractions, we find the modified propagator in dS to be
\[
\Pi_{dS}(\Box)=\frac{1}{a}\biggl[\frac{{\cal P}^{2}}{-\Box+2H^{2}}+\frac{1}{2\tilde{a}(\Box)}\left(\frac{m^{2}}{m^{2}+4H^{2}}\right)\left(\frac{{\cal P}_{s}^{0}}{\Box+4H^{2}}-\frac{{\cal P}_{s}^{0}}{\Box-m^{2}}\right)\biggr]
,\]
where  $m^{2}\equiv M_{P}^{2}/\alpha$ and ${\tilde a}(\Box)={\bar a}(\Box)/a$. This form of the modified propagator reduces to the previously derived propagator around Minkowski space \eqref{MinkPropGF} at the limit $H\rightarrow 0$, which implies $\Box\rightarrow -k^2$. Furthermore, the constant $\alpha$ must be positive definite in order to avoid tachyons and to retain the additional scalar pole. 
  \\\\\emph{Defocusing Conditions}\\
  We are now in a position to describe the minimum conditions whereby a spacetime, linearised around de Sitter, may indeed be considered to be non-singular, in that it avoids converging null geodesic congruences. We find the contribution of gravity to the Raychaudhuri equation by contracting the field equations \eqref{eomdSacf} with the tangent vectors $k^\mu$, like so
  \[
  r_{\nu}^{\mu}k^{\nu}k_{\mu}=\frac{1}{a}\left(\kappa T_{\nu}^{\mu}k_{\mu}k^{\nu}+\frac{1}{2}k^{\nu}k_{\mu}\nabla^{\mu}\partial_{\nu}f(\Box)r\right),
 \]
 so that the minimum condition for these null rays to defocus is given by
 \[
 r_{\nu}^{\mu}k^{\nu}k_{\mu}=\frac{1}{2a}k^{\nu}k_{\mu}\nabla^{\mu}\partial_{\nu}f(\Box)r<0
 .\]
 Expanding out the covariant derivatives , we may express the defocusing condition in the following manner
 \[
 \frac{(k^{0})^{2}}{2}\frac{(a-c(\Box))r}{a}>-\frac{2H(k^{0})^{2}}{a}\partial_{t}f(\Box)r
, \]
which can, in turn, be rewritten as
\[
\label{defocusdS}
\left(1+4H\partial_{t}\Box^{-1}\right)\left[1-(1-\Box/m^{2})\tilde{a}(\Box)\right]r>0.
\]
Here, we see that at the limit $H\rightarrow 0$, the defocusing condition around Minkowski \eqref{defocusmink} is recovered. Thus, we have then succeeded in our aim of deriving the ghost-free, defocusing condition around de Sitter space, comparable to Section \ref{sec:defocusmink}.

\chapter{Conclusion}
The stated objective of this thesis was to present a viable extension of general relativity, which is free from singularities, where `viable', in this case means devoid of ghosts, tachyons or exotic matter. With this in mind, we outline the results of the present work.
\subsubsection*{Outline of Results}
We began in Chapter \ref{chap:2}, with a lengthy computation of the non-linear field equations for the IDG action \eqref{action}. First, we outlined the general methodology and introduced a number of novel techniques, which proved useful in the explicit calculation that followed. Having attained the non-linear field equations, we perturbed around both Minkowski space and de Sitter space  up to linear order, for later use in deriving the modified propagator, ghost-free conditions and defocusing condition, required to construct a viable non-singular cosmology.

Chapter \ref{chap:GF}, focused on the notion of ghosts or tachyons that may appear at the level of the propagator. First, these manifestations were defined and illustrated with a number of examples from finite extensions of GR. The form of the modified propagator was also derived for the IDG action \eqref{action}. Next, the elimination of ghosts by way of an exponential correction in the scalar propagating mode was motivated, before being put to use in a linearised regime around Minkowski space. Typically, such an exponential function in the propagator weakens the classical and quantum effects of gravity in the UV. 
With the precise form of the modified graviton propagator in hand, the requisite tachyon criteria were also established by a decomposition into partial fractions.

Having established the foundation for presenting a viable infinite derivative extension to GR in the preceding chapters, we then turned our attention to the main crux of the present work -- the avoidance of singularities -- in Chapter \ref{chap:sing}. The chapter began with a discussion on the nature of singularities and indeed, the difficulty in defining such phenomenon, before introducing the Raychaudhuri equation and analysing its import in the context of GR. Having established the focusing behaviour of null rays via the convergence condition and Hawking-Penrose theorems, we turned our attention to spacetimes that do not conform to this convergent behaviour. By reversing the inequality, we were able to examine the behaviour of null rays as they diverge or \emph{defocus}, within a geometrically flat framework. We dubbed this reversal the \emph{defocusing condition} and null rays conforming to this condition will not converge to a point in a finite time and are said to be null geodesically complete -- stretching to past infinity. This is in direct contrast to converging rays, where a photon travelling along a geodesic of this type will cease to exist in a finite time. We illustrated this defocusing behaviour with a known example of a \emph{bouncing} cosmology.

From the behaviour around Minkowski space, we were able to deduce a simpler form of IDG action, which was purely scalar in its modification, with which one could realise the desired ghost-free, defocusing behaviour. With this action in hand, it proved quite straightforward to rearrange the linearised field equations around de Sitter in precisely the same form as in the Minkowski case. From this vantage point, the defocusing conditions around de Sitter space, were also derived and found to conform to the Minkowski case at the limit $H\rightarrow 0$, as expected. 

This thesis presents a number of novel results by the author. Firstly, the calculation of non-linear field equations for the most general, infinite derivative action of gravity that is quadratic in curvature, \eqref{action},  had never been fully captured before the work that the chapter is based on, \cite{Biswas:2013cha}, was published. Previous work, such as \cite{Schmidt:1990dh},\cite{Quandt:1990gc},\cite{Decanini:2007zz} has centred on finite orders of the D'Alembertian acting on the curvature scalar.

The form of the IDG-modified propagator around Minkowski space was established in \cite{Biswas:2010zk}, along with the associated ghost-free condition. The present work reaffirms these results, while also extending them in to de Sitter space, in a novel approach, by way of a simplification of the gravitational action - a reduced action that still exhibits the required defocusing behaviour. This allowed for the extension of the recent article, \cite{Conroy:2016sac}, which detailed the defocusing conditions around Minkowski space to include defocusing conditions also around de Sitter space. Comparisons were made with finite derivative extensions of gravity, where it was found that non-locality plays an integral role in realising the desired defocusing behaviour.

The methodology used in deriving the defocusing conditions is in stark contrast to previous work on bouncing solutions in infinite derivative theories of gravity. In \cite{Biswas:2005qr}, \cite{Koshelev:2012qn},\cite{Koshelev:2013lfm}, an Ansatz was invoked as a solution to the field equations, admitting bouncing solutions, with scale factor $\propto \cosh(\frac{\sigma}{2} t)$. In the present work, we make no assumption on the nature of the scale factor a priori, except that it must conform to the requirement of accelerated expansion of the Universe  within a homogeneous framework. Having acquired the generic ghost-free defocusing conditions, we do indeed check the bouncing solution $a(t)=\cosh(\frac{\sigma}{2} t)$ for consistency and, as expected, it did display the desired behaviour.
%
\subsubsection*{Future Work}
\emph{Homogeneous Solutions}\\
 Within the context of a homogeneous framework, the defocusing condition \eqref{defocusmink} could perhaps be analysed for specific restrictions on the curvature. For example, we analysed a bouncing solution in Section \ref{sec:Bouncing} and it is perhaps straightforward to generalise this analysis with a generic bouncing scale factor, using the same integral form method. Such a scale factor would result in the curvature being given by an even function, i.e. $R(t)=r_0+r_2 t^2+...$. This would have some similarities to the analysis in \cite{Conroy:2014dja} where, similarly, a generic bouncing scale factor was analysed but through the prism of the diffusion equation method \cite{Calcagni:2007ru}.

More illuminating still would be to solve the inequality for all forms of curvature that may satisfy the defocusing conditions \eqref{defocusmink} and \eqref{defocusdS} -- to see, explicitly, whether there curvature must conform to a bouncing scale factor or whether other solutions do exist. In this way, we could conceivably build up a precise form of non-singular metric, which would always satisfy the desired defocusing behaviour.  

Another quite straightforward approach in the homogenous setup would be to extend the methodology to non-linear FRW. Whereas the generic defocusing conditions can be derived without difficulty, some issues remain in terms of the ghost-free conditions. Recall that the ghost-free conditions and modified propagator were derived within a background of constant curvature. As the curvature in an FRW background is a time-dependent function rather than a constant, these conditions must be generalised to make revealing predictions. One possible method would be to proceed with an analysis that stipulates slowly varying curvature.

Furthermore, an extension of the progress made in \cite{Vachaspati:1998dy} to include bouncing cosmologies could be particularly illuminating. Vachaspati and Trodden found that the convergence condition \eqref{nullCC} restricted trajectories passing from normal regions to antitrapped regions, detailed in Fig. \ref{bigbang}. It would be interesting to see, geometrically, if a relaxation of the convergence condition allows such behaviour. One could also trace the trajectories of rays starting out at past infinity in Fig. \ref{figbounce} to shed light on the behaviour at times in and around the bounce. 
\\\\\emph{Other Solutions}\\
A further avenue of exploration involves extending our defocusing analysis to include \emph{inhomogenous} solutions, with spatial as well as temporal dependencies. This was briefly covered in \cite{Conroy:2016sac}, where the inhomogeneous generic defocusing condition was given by
\[
\frac{f(\bar\Box)}{a(\bar\Box)}\left(\partial_{t}^{2}+\partial_r^2\right)R^{(L)}<0
.\]
As before, we required $T_{\mu\nu}k^\mu k^\nu \geq 0$ so as not to violate the NEC. Note also that $\partial_r^2=\partial_i\partial^i$ is the Laplace operator. Although, the defocusing condition can be attained in quite a straightforward manner, a full analysis remains incomplete, in that the spatial dependencies must be made tractable. Similarly, we may also wish to consider anistropic spacetimes, conforming to a general metric of the type 
\[
ds^2=-dt^2+a^2(t)\sum_i e^{2\theta_i (t)}\sigma^i \sigma^i
,\]
as in \cite{Cai:2013vm}, where $t$ is cosmic time and $\sigma^i$ are linearly independent at all point in the spacetime. This is an example of an anisotropic but homogeneous metric in four dimensions, but could conceivably be generalised further to include spatial dependencies.
\\\\
This thesis presents a concrete methodology in describing a viable non-singular theory of gravity within the framework of an homogenous cosmology. Through the analysis it is clear that non-locality, arising from IDG, plays a pivotal role, as does an additional degree of freedom in the scalar propagating sector. Such a methodology can be extended into more complex pictures of IDG, such as those described above. As we extend our study, we will understand more about the relationship between the geometry of spacetime and gravity in a non-singular spacetime. We may even broaden our analysis into the study of the blackhole singularity problem, with the overall aim of, perhaps one day, presenting a definitive picture of a non-singular theory of gravity.


\begin{appendices}

\chapter{Useful Identities and Notations}
\section{Curvature}
\label{sec:AppCurv}
Metric signature
\[
g_{\mu\nu}=(-,+,+,+)
.\]
Christoffel Symbol
  \[
  \label{Christoffel}
  \Gamma^\lambda_{\mu\nu}=\frac{1}{2}g^{\lambda\tau}(\partial_\mu g_{\nu\tau}+\partial_\nu g_{\mu\tau}-\partial_\tau g_{\mu\nu})
  ,\]
Riemann Tensor
\[
R^{\lambda}{ }_{\mu\sigma\nu}=\partial_{\sigma}\Gamma_{\mu\nu}^{\lambda}-\partial_{\nu}\Gamma_{\mu\sigma}^{\lambda}+\Gamma_{\sigma\rho}^{\lambda}\Gamma_{\nu\mu}^{\rho}-\Gamma_{\nu\rho}^{\lambda}\Gamma_{\sigma\mu}^{\rho}
,\]
\[
R_{\mu\nu\lambda\sigma}=-R_{\nu\mu\lambda\sigma}=-R_{\mu\nu\sigma\lambda}=R_{\lambda\sigma\mu\nu}
,\]
\[
R_{\mu\nu\lambda\sigma}+R_{\mu\lambda\sigma\nu}+R_{\mu\sigma\nu\lambda}=0
.\]
Ricci Tensor
\[
R_{\mu\nu}=R^{\lambda}{ }_{\mu\lambda\nu}=\partial_{\lambda}\Gamma_{\mu\nu}^{\lambda}-\partial_{\nu}\Gamma_{\mu\lambda}^{\lambda}+\Gamma_{\lambda\rho}^{\lambda}\Gamma_{\nu\mu}^{\rho}-\Gamma_{\nu\rho}^{\lambda}\Gamma_{\lambda\mu}^{\rho}
,\]
\[
R_{\mu\nu}=R_{\nu\mu}
.\]
Curvature scalar
\[
R=g^{\mu\nu}R_{\mu\nu}=g^{\mu\nu}\partial_{\lambda}\Gamma_{\mu\nu}^{\lambda}-\partial^{\mu}\Gamma_{\mu\lambda}^{\lambda}+g^{\mu\nu}\Gamma_{\lambda\rho}^{\lambda}\Gamma_{\nu\mu}^{\rho}-g^{\mu\nu}\Gamma_{\nu\rho}^{\lambda}\Gamma_{\lambda\mu}^{\rho}
,\]
Weyl Tensor
\[
C^{\mu}{ }_{\alpha\nu\beta}\equiv R^{\mu}{ }{\alpha\nu\beta}-\frac{1}{2}(\delta_{\nu}^{
\mu}R_{\alpha\beta}-\delta_{\beta}^{\mu}R_{\alpha\nu}+R_{\nu}^{\mu}g_{
\alpha\beta}-R_{\beta}^{\mu}g_{\alpha\nu})+\frac{R}{6}(\delta_{\nu}^{\mu}g_{
\alpha\beta}-\delta_{\beta}^{\mu}g_{\alpha\nu})
.\]
\[
C^\lambda{ }_{\mu\lambda\nu}=0
\]
Einstein Tensor
\[
G_{\mu\nu}=R_{\mu\nu}-\frac{1}{2}g_{\mu\nu}R
.\]
Traceless Einstein Tensor
\[
S_{\mu\nu}=R_{\mu\nu}-\frac{1}{4}g_{\mu\nu}R
\]
General formula for commuting covariant derivatives
\begin{align}
\label{comrel}
[\nabla_{\rho},\nabla_{\sigma}]X^{\mu_{1}...\mu_{k}}{ }{ }{ }{ }_{\nu_{1}...\nu_{l}}	&=	R^{\mu_{1}}{ }_{\lambda\rho\sigma}X^{\lambda\mu_{2}...\mu_{k}}{ }{ }{ }{ }_{\nu_{1}...\nu_{l}}+R^{\mu_{2}}{ }_{\lambda\rho\sigma} X^{\mu_{1}\lambda\mu_{3}...\mu_{k}}{ }{ }{ }{ }_{\nu_{1}...\nu_{l}}+...
	\nn	\\&-R^{\lambda}{ }_{\nu_{1}\rho\sigma} X^{\mu_{1}...\mu_{k}}{ }{ }{ }{ }_{\lambda...\nu_{l}}-R^{\lambda}{ }_{\nu_{2}\rho\sigma}X^{\mu_{1}...\mu_{k}}{ }{ }{ }{ }_{\nu_{1}\lambda\nu_{3}...\nu_{l}}-...\;.
		\end{align}
\section{Bianchi Identities}
The Bianchi identity is given by
\[
\nabla_{\kappa}R_{\mu\nu\lambda\sigma}+\nabla_{\sigma}R_{\mu\nu\kappa\lambda}+\nabla_{\lambda}R_{\mu\nu\sigma\kappa}=0
.\]
Contracting with $g^{\mu\lambda}$ gives the contracted Bianchi identity,
\[
\nabla_{\kappa}R_{\nu\sigma}-\nabla_{\sigma}R_{\nu\kappa}+\nabla^{\lambda}R_{\lambda\nu\sigma\kappa}=0
.\]
Contracting further with $g^{\nu\kappa}$ implies
\[
\nabla_{\kappa}R_{\sigma}^{\kappa}=\frac{1}{2}\nabla_{\sigma}R
,\]
which similarly implies
\[
\nabla^{\sigma}\nabla_{\kappa}R_{\sigma}^{\kappa}=\frac{1}{2}\Box R
\]
and
\[
\nabla_\mu G^\mu_\nu=0
.\]
\section{Variation of Curvature}
\label{sec:Appvary}
We have from the definitions of the Riemann and Ricci tensor
\[
\delta
R^{\lambda}{ }_{\mu\sigma\nu}=(\delta\Gamma_{\mu\nu}^{\lambda})_{;\sigma}
-(\delta\Gamma_{\mu\sigma}^{\lambda})_{;\nu}\nn
\]
\[
\nn
 \delta
R_{\mu\nu}=\nabla_{\lambda}\delta\Gamma_{\mu\nu}^{\lambda}-\nabla_
{\nu}\delta\Gamma_{\mu\lambda}^{\lambda}
\]
\[
\delta\Gamma{}_{\mu\nu}^{\lambda}=\frac{1}{2}(h_{\;\nu;\mu}^{\lambda}+h_{
\;\mu;\nu}^{\lambda}-h_{\mu\nu}^{\;\;;\lambda}).
\]
Substitution of the varied Christoffel symbol reveals,
\[
\nn
\delta
R^{\lambda}{ }_{\mu\sigma\nu}=\frac{1}{2}(h_{\;\nu;\mu;\sigma}^{\lambda}-h_{
\mu\nu\;;\sigma}^{\;\;;\lambda}-h_{\;\sigma;\mu;\nu}^{\lambda}+h_{
\mu\sigma\;;\nu}^{\;\;;\lambda})
\]
\[
\delta
R_{\mu\nu}=\frac{1}{2}(h_{\;\nu;\mu;\lambda}^{\lambda}+h_{\mu\lambda\;;\nu}^{
\;\;;\lambda}-\square h_{\mu\nu}-h_{;\mu;\nu}).
\]
For simplicity, it is often preferable to arrange these identities in terms of the metric variation
$h_{\alpha\beta}$, like so,
\[
\nn
\fl \delta
R_{\mu\nu\lambda\sigma}=\frac{1}{2}[\delta_{\lambda}^{\alpha}\delta_{\nu}^{\beta
}(h_{\alpha\beta})_{;\sigma;\mu}-\delta_{\lambda}^{\alpha}\delta_{\mu}^{\beta}
(h_{\alpha\beta})_{;\sigma;\nu}+\delta_{\mu}^{\alpha}\delta_{\sigma}^{\beta}(h_{
\alpha\beta})_{;\nu;\lambda}-\delta_{\sigma}^{\alpha}\delta_{\nu}^{\beta}(h_{
\alpha\beta})_{;\mu;\lambda}]
\]
\[
\fl \delta
R_{\mu\nu}=\frac{1}{2}[\delta_{\nu}^{\beta}(h_{\alpha\beta})_{;\mu}{ }^{;\alpha}
+\delta_{\mu}^{\alpha}(h_{\alpha\beta})^{;\beta}{ }_{;\nu}-\delta_{\mu}^{\alpha}
\delta_{\nu}^{\beta}\square(h_{\alpha\beta})-g^{\alpha\beta}(h_{\alpha\beta})_{
;\mu;\nu}]
.\]
We can then find the variation of the curvature scalar, $\delta R$,
\begin{align}
\delta R&=\delta(g^{\mu\nu}R_{\mu\nu})
\nn\\&=\delta g^{\mu\nu}R_{\mu\nu}+g^{\mu\nu}\delta R_{\mu\nu}
\nn\\&=-h_{\alpha\beta}R^{\alpha\beta}+g^{\mu\nu}\delta R_{\mu\nu}
\nn\\&
=-h_{\alpha\beta}R^{\alpha\beta}+(h_{\alpha\beta})^{;\alpha;\beta}-g^{
\alpha\beta}\square(h_{\alpha\beta})
,\end{align}
where we have used the following notations
\[
h_{\mu\nu}=-h^{\alpha\beta}g_{\alpha\mu}g_{\beta\nu}
,~h=g^{\mu\nu}h_{\mu\nu}
,~h_{\mu\nu}=\delta g_{\mu\nu}
,~\nabla_{\mu}R=R_{;\mu}
.\]
In summary, we have
\[
\fl\delta
R_{\mu\nu\lambda\sigma}=\frac{1}{2}[\delta_{\lambda}^{\alpha}\delta_{\nu}^{\beta
}(h_{\alpha\beta})_{;\sigma;\mu}-\delta_{\lambda}^{\alpha}\delta_{\mu}^{\beta}
(h_{\alpha\beta})_{;\sigma;\nu}+\delta_{\mu}^{\alpha}\delta_{\sigma}^{\beta}(h_{
\alpha\beta})_{;\nu;\lambda}-\delta_{\sigma}^{\alpha}\delta_{\nu}^{\beta}(h_{
\alpha\beta})_{;\mu;\lambda}]
\nn\]
\[
\delta
R_{\mu\nu}=\frac{1}{2}[\delta_{\nu}^{\beta}(h_{\alpha\beta})_{;\mu}^{;\alpha}
+\delta_{\mu}^{\beta}(h_{\alpha\beta})_{;\nu}^{;\alpha}-\delta_{\mu}^{\alpha}
\delta_{\nu}^{\beta}\square(h_{\alpha\beta})-g^{\alpha\beta}(h_{\alpha\beta})_{
;\mu;\nu}]
\nn\]
\[
\delta
R=-h_{\alpha\beta}R^{\alpha\beta}+(h_{\alpha\beta})^{;\alpha;\beta}-g^{
\alpha\beta}\square(h_{\alpha\beta})
\nn\]
\[
\delta\Gamma{}_{\mu\nu}^{\lambda}=\frac{1}{2}(g^{\lambda\alpha}\delta_{\nu}^{
\beta}h_{\alpha\beta;\mu}+g^{\lambda\alpha}\delta_{\mu}^{\beta}h_{
\alpha\beta;\nu}-\delta_{\mu}^{\alpha}\delta_{\nu}^{\beta}h_{\alpha\beta}^{
\;\;;\lambda})
 \]
\subsection{$\delta(\Box)R$}
Recall
\[
\Box=g^{\mu\nu}\nabla_{\mu}\nabla_{\nu}
\]
Then we have
\begin{align}
\nn\delta(\Box)R&=\delta
g^{\mu\nu}R_{;\mu;\nu}+g^{\mu\nu}\delta(\nabla_{\mu})R_{;\nu}+g^{\mu\nu}[
\delta(\nabla_{\nu})R]_{;\mu}
\\&
=-h_{\alpha\beta}R^{;\alpha;\beta}+g^{\mu\nu}\delta(\nabla_{\mu})R_{;\nu}+g^{
\mu\nu}[\delta(\nabla_{\nu})R]_{;\mu}
\end{align}
From the general definition of the covariant derivative of a tensor we deduce
the following
\[
\fl g^{\mu\nu}\delta(\nabla_{\mu})R_{;\nu}=-g^{\mu\nu}\delta\Gamma_{\mu\nu}^{\lambda
}R_{;\lambda}
\]
\[
\fl g^{\mu\nu}[\delta(\nabla_{\nu})R]_{;\mu}=0
\]
The last term vanishes in this case as $R$ is a scalar. This will not be true
for $\delta(\Box)R_{\mu\nu}$ and $\delta(\Box)R_{\mu\nu\lambda\sigma}$. We then
integrate by parts to find
\[
\delta(\Box)R=-h_{\alpha\beta}R^{;\alpha;\beta}+\frac{1}{2}g^{\alpha\beta}R_{
;\lambda}(h_{\alpha\beta})^{;\lambda}-R^{;\alpha}(h_{\alpha\beta})^{;\beta}
\]
with
\[
\delta\Gamma_{\mu\nu}^{\lambda}=\frac{1}{2}[g^{\alpha\lambda}\delta_{\mu}^{\beta
}(h_{\alpha\beta})_{;\nu}+g^{\alpha\lambda}\delta_{\nu}^{\beta}(h_{\alpha\beta}
)_{;\mu;}-\delta_{\mu}^{\alpha}\delta_{\nu}^{\beta}(h_{\alpha\beta})^{;\lambda}]
\]
\subsection{$\delta(\Box)R_{\mu\nu}$}
\label{sec:dS2}
\[
\nn
\delta(\square)R_{\mu\nu}=\delta
g^{\lambda\sigma}R_{\mu\nu;\lambda;\sigma}+g^{\lambda\sigma}\delta(\nabla_{
\lambda})R_{\mu\nu;\sigma}+g^{\lambda\sigma}[\delta(\nabla_{\sigma})R_{\mu\nu}]_
{;\lambda}
\]
\[
=-h_{\alpha\beta}R_{\mu\nu}^{;\alpha;\beta}+g^{\lambda\sigma}\delta(\nabla_{
\lambda})R_{\mu\nu;\sigma}+g^{\lambda\sigma}[\delta(\nabla_{\sigma})R_{\mu\nu}]_
{;\lambda}
\]
From the general definition of the covariant derivative of a tensor we have
\[\fl
g^{\lambda\sigma}\delta(\nabla_{\lambda})R_{\mu\nu;\sigma}=-\delta\Gamma_{
\lambda\mu}^{\tau}R_{\tau\nu}^{;\lambda}-\delta\Gamma_{\lambda\nu}^{\tau}R_{
\mu\tau}^{;\lambda}-g^{\lambda\sigma}\delta\Gamma_{\lambda\sigma}^{\tau}R_{
\mu\nu;\tau}
\]
\[\fl
g^{\lambda\sigma}\nabla_{\lambda}\delta(\nabla_{\sigma})R_{\mu\nu}
=-(\delta\Gamma_{\lambda\mu}^{\tau})^{;\lambda}R_{\tau\nu}-\delta\Gamma_{
\lambda\mu}^{\tau}R_{\tau\nu}^{;\lambda}-(\delta\Gamma_{\lambda\nu}^{\tau})^{
;\lambda}R_{\mu\tau}-\delta\Gamma_{\lambda\nu}^{\tau}R_{\mu\tau}^{;\lambda}
\]
So that
\[
\delta(\square)R_{\mu\nu}=-h_{\alpha\beta}R_{\mu\nu}^{;\alpha;\beta}-g^{
\lambda\sigma}\delta\Gamma_{\lambda\sigma}^{\tau}R_{\mu\nu;\tau}-(\delta\Gamma_{
\lambda(\mu}^{\tau})^{;\lambda}R_{\tau\nu)}-2\delta\Gamma_{\lambda(\mu}^{\tau}R_
{\tau\nu)}^{;\lambda}
\]
Expanding using
$\delta\Gamma_{\mu\nu}^{\lambda}=\frac{1}{2}[g^{\alpha\lambda}\delta_{\mu}^{
\beta}(h_{\alpha\beta})_{;\nu}+g^{\alpha\lambda}\delta_{\nu}^{\beta}(h_{
\alpha\beta})_{;\mu;}-\delta_{\mu}^{\alpha}\delta_{\nu}^{\beta}(h_{\alpha\beta}
)^{;\lambda}]$, we have
\begin{align}\nn
\delta(\square)R_{\mu\nu}&=-h_{\alpha\beta}R_{\mu\nu}^{;\alpha;\beta}-(h_{
\alpha\beta})^{;\beta}R_{\mu\nu}^{;\alpha}+\frac{1}{2}g^{\alpha\beta}(h_{
\alpha\beta})^{;\sigma}R_{\mu\nu;\sigma}
\\&\nn
-\frac{1}{2}\left[\square(h_{\alpha\beta})\delta_{(\mu}^{\beta}R_{\;\nu)}^{
\alpha}-(h_{\alpha\beta})^{;\tau;\alpha}\delta_{(\mu}^{\beta}R_{\tau\nu)}+(h_{
\alpha\beta})_{;(\mu}^{\;;\beta}R_{\;\nu)}^{\alpha}\right]
\\&
-R_{\;(\nu}^{\alpha;\beta}h_{\alpha\beta;\mu)}-\delta_{(\mu}^{\beta}R_{\;\nu)}^{
\alpha;\lambda}h_{\alpha\beta;\lambda}+\delta_{(\mu}^{\beta}R_{\tau\nu)}^{
;\alpha}h_{\alpha\beta}^{\;\;;\tau})
\end{align}
\subsection{$\delta(\Box)R_{\mu\nu\lambda\sigma}$}
From the definition of the D'Alembertian operator
$\Box=g^{\mu\nu}\nabla_{\mu}\nabla_{\nu}$, we have
\begin{align}\nn
\delta(\Box)R_{\mu\nu\lambda\sigma}&=\delta
g^{\kappa\tau}R_{\mu\nu\lambda\sigma;\kappa;\tau}+g^{\kappa\tau}\delta(\nabla_{
\kappa})R_{\mu\nu\lambda\sigma;\tau}+g^{\kappa\tau}[\delta(\nabla_{\tau})R_{
\mu\nu\lambda\sigma}]_{;\kappa}
\\&
=-h_{\alpha\beta}R_{\mu\nu\lambda\sigma}^{;\alpha;\beta}+g^{\kappa\tau}
\delta(\nabla_{\kappa})R_{\mu\nu\lambda\sigma;\tau}+g^{\kappa\tau}[
\delta(\nabla_{\tau})R_{\mu\nu\lambda\sigma}]_{;\kappa}
\end{align}
and from the general definition of the covariant derivative of a tensor and
treating $R_{\mu\nu\lambda\sigma;\tau}$ as a $(0,5)$-tensor, we have
\begin{eqnarray}
&&\fl g^{\kappa\tau}\delta(\nabla_{\kappa})R_{\mu\nu\lambda\sigma;\tau}=-\delta\Gamma_
{\kappa\mu}^{\rho}R_{\rho\nu\lambda\sigma}^{;\kappa}-\delta\Gamma_{\kappa\nu}^{
\rho}R_{\mu\rho\lambda\sigma}^{;\kappa}-\delta\Gamma_{\kappa\lambda}^{\rho}R_{
\mu\nu\rho\sigma}^{;\kappa}-\delta\Gamma_{\kappa\sigma}^{\rho}R_{
\mu\nu\lambda\rho}^{;\kappa}
\nonumber\\&&
-g^{\kappa\tau}\delta\Gamma_{\kappa\tau}^{\rho}R_{
\mu\nu\lambda\sigma;\rho}
\end{eqnarray}
and
\begin{align}\nn
&g^{\kappa\tau}[\delta(\nabla_{\tau})R_{\mu\nu\lambda\sigma}]_{;\kappa}=\left[
-\delta\Gamma_{\kappa\mu}^{\rho}R_{\rho\nu\lambda\sigma}-\delta\Gamma_{\kappa\nu
}^{\rho}R_{\mu\rho\lambda\sigma}-\delta\Gamma_{\kappa\lambda}^{\rho}R_{
\mu\nu\rho\sigma}-\delta\Gamma_{\kappa\sigma}^{\rho}R_{\mu\nu\lambda\rho}\right]
^{;\kappa}
\\&=-(\delta\Gamma_{\kappa\mu}^{\rho})^{;\kappa}R_{\rho\nu\lambda\sigma}
-\delta\Gamma_{\kappa\mu}^{\rho}R_{\rho\nu\lambda\sigma}^{;\kappa}
-(\delta\Gamma_{\kappa\nu}^{\rho})^{;\kappa}R_{\mu\rho\lambda\sigma}
-\delta\Gamma_{\kappa\nu}^{\rho}R_{\mu\rho\lambda\sigma}^{;\kappa}
\nonumber\\&
-(\delta\Gamma_{\kappa\lambda}^{\rho})^{;\kappa}R_{\mu\nu\rho\sigma}
-\delta\Gamma_{\kappa\lambda}^{\rho}R_{\mu\nu\rho\sigma}^{;\kappa}
-(\delta\Gamma_{\kappa\sigma}^{\rho})^{;\kappa}R_{\mu\nu\lambda\rho}
-\delta\Gamma_{\kappa\sigma}^{\rho}R_{\mu\nu\lambda\rho}^{;\kappa}
\end{align}
So that
\begin{eqnarray}
&&\delta(\Box)R_{\mu\nu\lambda\sigma}=-h_{\alpha\beta}R_{\mu\nu\lambda\sigma}^{
;\alpha;\beta}-g^{\kappa\tau}\delta\Gamma_{\kappa\tau}^{\rho}R_{
\mu\nu\lambda\sigma;\rho}
\nonumber\\&&
-\left[(\delta\Gamma_{\kappa\mu}^{\rho})^{;\kappa}R_{\rho\nu\lambda\sigma}
+(\delta\Gamma_{\kappa\nu}^{\rho})^{;\kappa}R_{\mu\rho\lambda\sigma}
+(\delta\Gamma_{\kappa\lambda}^{\rho})^{;\kappa}R_{\mu\nu\rho\sigma}
+(\delta\Gamma_{\kappa\sigma}^{\rho})^{;\kappa}R_{\mu\nu\lambda\rho}\right]
\nonumber\\&&
-2\left[\delta\Gamma_{\kappa\mu}^{\rho}R_{\rho\nu\lambda\sigma}^{;\kappa}
+\delta\Gamma_{\kappa\nu}^{\rho}R_{\mu\rho\lambda\sigma}^{;\kappa}+\delta\Gamma_
{\kappa\lambda}^{\rho}R_{\mu\nu\rho\sigma}^{;\kappa}+\delta\Gamma_{\kappa\sigma}
^{\rho}R_{\mu\nu\lambda\rho}^{;\kappa}\right]
\end{eqnarray}
Then, using
$\delta\Gamma_{\mu\nu}^{\lambda}=\frac{1}{2}[g^{\alpha\lambda}\delta_{\mu}^{
\beta}(h_{\alpha\beta})_{;\nu}+g^{\alpha\lambda}\delta_{\nu}^{\beta}(h_{
\alpha\beta})_{;\mu;}-\delta_{\mu}^{\alpha}\delta_{\nu}^{\beta}(h_{\alpha\beta}
)^{;\lambda}]$, and the Bianchi identities, we find
\begin{eqnarray}
&&\fl\delta(\Box)R_{\mu\nu\lambda\sigma}=-h_{\alpha\beta}R_{\mu\nu\lambda\sigma}^{
;\alpha;\beta}-(h_{\alpha\beta})^{;\beta}R_{\mu\nu\lambda\sigma}^{;\alpha}+\frac
{1}{2}h{}^{;\tau}R_{\mu\nu\lambda\sigma;\tau}
\\&&\fl
-\frac{1}{2}[g^{\alpha\tau}(h_{\alpha\beta})_{;\mu}^{;\beta}R_{
\tau\nu\lambda\sigma}+g^{\alpha\tau}(h_{\alpha\beta})_{;\nu}^{;\beta}R_{
\mu\tau\lambda\sigma}+g^{\alpha\tau}(h_{\alpha\beta})_{;\lambda}^{;\beta}R_{
\mu\nu\tau\sigma}+g^{\alpha\tau}(h_{\alpha\beta})_{;\sigma}^{;\beta}R_{
\mu\nu\lambda\tau}]
\nonumber\\&&\fl
-\left[g^{\alpha\tau}(h_{\alpha\beta})_{;\mu}R_{\tau\nu\lambda\sigma}^{;\beta}
+g^{\alpha\tau}(h_{\alpha\beta})_{;\nu}S_{\mu\tau\lambda\sigma}^{;\beta}+g^{
\alpha\tau}(h_{\alpha\beta})_{;\lambda}R_{\mu\nu\tau\sigma}^{;\beta}+g^{
\alpha\tau}(h_{\alpha\beta})_{;\sigma}R_{\mu\nu\lambda\tau}^{;\beta}\right]\nn
\end{eqnarray}
\chapter{Friedmann-Lema\^{\i}tre-Robertson-Walker Framework}
\label{sec:introcos}
The Friedmann-Lema\^{\i}tre-Robertson-Walker (FRW) metric forms an exact solution of Einstein's field equations and can be expressed in terms of the following isotropic and homogenous metric
\[
\label{FRWmetricgamma}
ds^2=-dt^2 +a^2(t)\gamma_{ij}dx^i dx^j
,\]
where $\gamma_{ij}$ is a 3-dimensional  maximally symmetric metric of Gaussian curvature $k$ and the scale factor $a(t)$ is a time-dependent function of unit dimension which parametrizes the relative expansion of the Universe. To understand the geometric curvature of the spacetime more readily, it is perhaps preferable to reformulate the FRW metric in a spherically symmetric coordinate system, like so
\[
ds^2=-dt^2+a^2(t)\biggl(\frac{dr^2}{1-kr^2}+r^2 d\Omega^2\biggr)
,\]
where the spherical coordinates are contained within $d\Omega^2\equiv d\theta^2+\sin^2\theta d\varphi^2$. The spatial curvature, in terms of a hypersurface of cosmic time $t$, is given by the real constant $k$, such that
\\
  \begin{equation}
    k=
    \begin{cases}
    -1 & \text{Negatively curved hypersurface  (Closed Universe)} \\
      ~0 & \text{Flat hypersurface} \\
      ~1 & \text{Positively curved hypersurface (Open Universe)}
    \end{cases}
  \end{equation}
\\\emph{Exact Solution}\\  
As the present work is largely cosmological in focus, we will now go into some detail to verify that the FRW metric is indeed an exact solution to the Einstein field equations \eqref{EinsteinEq}. In order to do this, we must derive all the relevant components that make up the metric \eqref{FRWmetricgamma}, beginning with the components of the metric tensor
  \[
  g_{00}=-1=g^{00},\qquad g_{ij}=a^2(t)\gamma_{ij}, \qquad g^{ij}=a^{-2}(t)\gamma^{ij}
  .\]
  Next, we move on to the Christoffel symbols
  \[
  \Gamma^\lambda_{\mu\nu}=\frac{1}{2}g^{\lambda\tau}(\partial_\mu g_{\nu\tau}+\partial_\nu g_{\mu\tau}-\partial_\tau g_{\mu\nu})
  ,\]
  of which the non-vanishing components are given by
  \[
  \Gamma^i_{0j}=\Gamma^i_{j0}=\frac{\dot{a}}{a}\delta^i_j,\qquad \Gamma^0_{ij}= a \dot{a} \gamma_{ij},
  \]
  where the superscript $\cdot$ denotes a derivative with respect to cosmic time $t$. We may then use the remaining Christoffel symbols to derive the relevant forms of curvature that make up the Einstein equation, from the general definitions given in Appendix \ref{sec:AppCurv}.
  \\\\ \emph{Ricci Tensor}\\
  \[
  R_{00}=-3\left(\dot{H}+H^2\right),\qquad R_{ij}=g_{ij}\left(\dot{H}+3H^2+\frac{2k}{a^2}\right)
  \]
\emph{Curvature Scalar}\\
\[
\label{FLRWscalar}
R=6\left(\dot{H}+2H^2+\frac{k}{a^2}\right)
.\]
\emph{Einstein Tensor}\\
The Einstein tensor $G_{\mu\nu}=R_{\mu\nu}-\frac{1}{2}g_{\mu\nu}R$ is then given by
\[
G_{00}=3\left(H^2+\frac{k}{a^2}\right),\qquad G_{ij}=-g_{ij}\left(2\dot{H}+3H^2+\frac{k}{a^2}\right),\qquad G_{0i}=0
.\] 
Comparing this with the Einstein Equation \eqref{EinsteinEq}, we deduce that the energy-momentum tensor must take the form
\[
T_{00}=\rho(t),\qquad T_{ij}=p(t)g_{ij},\qquad T_{0i}=0
,\]
where $\rho$ denotes the energy density and $p$ denotes the pressure. Thus, the FRW is an exact (fluid) solution of Einstein's General Relativity   \cite{Carroll:2004st, Clifton:2011jh, Wald:GR, Blau}.
\\\\ \emph{Perfect Fluid}\\
Furthermore, this form of the energy-momentum tensor describes a \emph{perfect fluid}. A perfect fluid is one where a comoving observer views the fluid around him as isotropic \cite{Weinberg:100595}. 
 In terms of the energy-momentum tensor $T_{\mu\nu}$, isotropic spacetimes must have vanishing $T_{0i}$-components in order to remain rotationally invariant \cite{Carroll:2004st}. The remaining components are given as above, which we can express in a covariant form as follows
\[
T_{\mu\nu}=(\rho+p)u_\mu u_\nu +p g_{\mu\nu}
.\]
Here, $u_\mu$ is the fluid four-velocity, i.e. $u_{\mu}=\{1,0,0,0\}$, such that $g^{\mu\nu}u_\mu u_\nu=-1$ and $(u_\mu k^\mu)^2=(k^0)^2$. We then perform the operations of (1) contracting with this fluid four velocity, (2) contracting with the \emph{null} geodesic congruence $k^\mu$ and (3) taking the trace to express three distinct identities:
\[
\label{Tu}
T_{\mu\nu}u^\mu u^\nu=\rho 
\]
\[
\label{Tk}
T_{\mu\nu}k^\mu k^\nu=(\rho+p)(k^0)^2
\]
\[
\label{Ttrace}
T=-\rho+3p
.\]
These identities allow us to write the relevant energy conditions for the present text. The \emph{Weak Energy Condition} (WEC) states that the energy density will be positive for an observer along a timelike tangent vector $\xi^\mu$. The null energy condition (NEC), given by $T_{\mu\nu}k^\mu k^\nu$, is a special case of the WEC, where the timelike tangent vector is replaced by a null ray. In this case, the energy density may conceivably be negative so long as this is balanced by sufficiently positive pressure. In terms of the perfect fluid, these are given by 
\begin{itemize}
\item {\bf Null Energy Condition:}
 $T_{\mu\nu}k^\mu k^\nu\geq 0$ implies $\rho+p\geq0$ \label{NEC}
\item {\bf Weak Energy Condition:} $T_{\mu\nu}\xi^\mu \xi^\nu\geq 0$ implies $\rho+p\geq0$ and $\rho\geq0$ \cite{Carroll:2004st}.
\end{itemize}
The NEC, in particular, will play an important role in the later discussion on singularity-free theories of gravity in Chapter \ref{chap:sing}. 

\chapter{Newtonian Potential}
\label{chap:NewtPot}
In order to compute the Newtonian potential, we must consider the Newtonian weak field limit of the field equations \eqref{eomminkred}. In a non-relativistic system, the energy density is the only significant element of the energy-momentum tensor \cite{Carroll:2004st},\cite{Wald:GR},\cite{Blau}. As such, we have $\rho=T_{00}\gg |T_{ij}|$, where the energy density $\rho$ is static. Recall, from the discussion on the perfect fluid in Section \ref{sec:introcos}, that the trace equation is given by $ T=-\rho+3p\approx-\rho$, while the 00-component is simply $T_{00}=\rho$. Furthermore, the perturbed metric for a static, Newtonian point source is given by the static line element \cite{Quandt:1990gc},\cite{Schwartz:2013pla}
\[
\label{metricspheric}
ds^2=-(1+2\Phi(r))dt^2+(1-2\Psi(r))(dx^2+dy^2+dz^2)
.\]
We turn then to the IDG field equations around Minkowski space \eqref{eomminkred}, from which we can then read off the trace and 00-component of such a metric
\begin{align}
-\kappa\rho&=\frac{1}{2}(a(\Box)-3c(\Box))R
\nn
\\
\label{rhonewt}
\kappa\rho&=a(\Box)R_{00}+\frac{1}{2}c(\Box)R.
 \end{align}
With the line element in hand \eqref{metricspheric}, we first compute the metric $h_{\mu\nu}$, using the algorithm \eqref{pertmink}
\[
h_{00}=-2\Phi(r),\qquad h_{ij}=-2\Psi(r)\eta_{ij},
\]
before substituting these values into \eqref{MinkR} to find the pertinent values for the curvature:
\[
R=2(2\triangle\Psi-\triangle\Phi),\qquad R_{00}=\triangle \Phi
.\]
Recall that at the linearised limit $\Box=\eta^{\mu\nu}\partial_\mu \partial_\nu$ which, for a static source, reduces to $\Box=\triangle$, where $\triangle\equiv\nabla^2=\partial_i \partial^i$ is the Laplace operator. Thus, we find the energy density \eqref{rhonewt} for the given metric \eqref{metricspheric} to be
\begin{align}
-\kappa\rho&=(a(\Box)-3c(\Box))(2\triangle\Psi-\triangle\Phi)\nn
\\
\kappa\rho&=(a(\Box)-c(\Box))\triangle\Phi+2c(\Box)\triangle\Psi
.
\end{align}
By comparing these two expressions for the energy density, we find that the two Newtonian potentials relate to each like so
\[
\label{PhiPsirelate}
\triangle\Phi=-\frac{a(\Box)-2c(\Box)}{c(\Box)}\triangle\Psi
.\]
Using this identity, we find
\[
\kappa\rho=\frac{a(\Box)\left(a(\Box)-3c(\Box)\right)}{a(\Box)-2c(\Box)}\triangle\Phi=\kappa m\delta^{3}(\vec{r})
, \]
where in the weak-field limit, the energy density is simply the point source, i.e. $\rho=m\delta^3({\vec r})$ and $\delta^3$ refers to the $3$-dimensional Dirac delta-function, while $m$ is the mass of the test particle. We proceed in a manner familiar to that of the Coulomb potential, \cite{Schwartz:2013pla},\cite{Kiefer:2012boa}, by performing a Fourier transform in order to express the Newtonian potential $\Phi(r)$. Recall that the Fourier transform of the Dirac delta-function is given by
\[
\delta^3({\vec r})=\int\frac{d^3 k}{(2\pi)^3}e^{ik \vec{r}}
.\]
Thus, with $\Box\rightarrow -k^2$ in Fourier space on a flat background, we solve for $\Phi(r)$,
\[
\label{PhiPotent}
\Phi(r)	=	-\frac{\kappa m}{(2\pi)^{3}}\int_{-\infty}^{\infty}d^{3}k\frac{a-2c}{a(a-3c)}\frac{e^{ik\vec{r}}}{k^{2}}=-\frac{\kappa m}{2\pi^{2}r}\int_{0}^{\infty}dk\frac{(a-2c)}{a(a-3c)}\frac{\sin(kr)}{k},
 \]
 where we have abbreviated the functions $a=a(-k^2)$ and $c=c(-k^2)$ for convenience. It is then straightforward to compute the other Newtonian potential $\Psi(r)$, using \eqref{PhiPsirelate}
 \[
 \label{PsiPotent}
 \Psi(r)=\frac{\kappa m}{2\pi^{2}r}\int_{0}^{\infty}dk\frac{c}{a(a-3c)}\frac{\sin(kr)}{k}
. \]\\
\emph{$a=c$: No additional degrees of freedom in the scalar propagating sector}\\
Recall that, for the particular case when $a=c$ no additional poles are introduced to the scalar sector of the propagator and we retain the original degrees of freedom of the massless graviton. In this instance, one would expect the two distinct Newtonian potentials to converge to a single potential. By substituting $a=c$ into \eqref{PsiPotent} and \eqref{PhiPotent}, one can quickly verify that this is the true, with the potential then given by
\[
\label{acPotent}
\Phi(r)=\Psi(r)=-\frac{\kappa m}{(2\pi)^2 r}\int_{0}^{\infty}dk\frac{\sin(kr)}{a(-k^2)k}.
 \]
 We may then test a particular ghost-free choice of the function $a$ to see whether it exhibits the expected behaviour of a Newtonian potential. In Section \ref{sec:GF}, we found that in order for the spacetime to be ghost-free, ${a}$ must be an entire function containing no roots. The simplest choice is then, 
 \[
 a(\Box)=e^{-\Box/M^2}
. \]
Thus, we find the Newtonian potential to be \cite{Biswas:2011ar},
\[
\Phi(r)=-\frac{\kappa m}{(2\pi)^2 r}\int_{0}^{\infty}\frac{dk}{k}e^{-k^{2}/M^{2}}\sin(kr)=-\frac{\kappa m \mbox{ erf}(M r/2)}{8\pi r}.
\]
 Observe now that at the limit, $r\rightarrow \infty$ \footnote{Alternatively, if we take $M\rightarrow\infty$, which is the familiar limit to return IDG to a local theory, we recapture the familiar $1/r$ divergence of GR, as expected.}, $\mbox{erf}(r)/r\rightarrow 0$ and the metric \eqref{metricspheric} is returned to flat space. On the other hand, taking the limit $r\rightarrow 0$, results in the Newtonian potential converging to a constant
 \[
 \lim_{r\rightarrow 0} \Phi(r)=\frac{\kappa m M}{8\pi^{3/2}}
 .\]
 We see here that the Newtonian potentials remain finite with $\Phi(r)\sim m M/M_P^2$ and, as such, the linear approximation is bounded all the way to $r\rightarrow 0$.

\chapter{A Note on the Gravitational Entropy}
\label{sec:Entropy}
In this section, we give a brief outline of the connection between Wald's gravitational entropy and the defocusing conditions around de Sitter space, derived in Section \ref{sec:defocusdS}. In a recent work \cite{Conroy:2015nva}, Wald's gravitational entropy~\cite{Wald:1993nt},\cite{Myers}, was computed for a non-local action of the type
\[
\label{actionent}
I=\frac{1}{16\pi G_{4}}\int d^{4}x\sqrt{-g}\left(R-2M_{P}^{-2}\Lambda+\alpha R{\cal F}(\Box)R\right)
. \] 
This was found to take the form
 \[
 \label{dsent}
 S_{I}=\frac{A_{H}^{dS}}{4G_{4}}\left(1+8f_{1_0}\alpha M_{P}^{-2}\Lambda\right)
 ,\]
 where $\alpha$ is a constant of dimension inverse mass squared. The primary thing to note here is that a non-physical, \emph{negative entropy state} is realised if the following inequality holds:
 \[
 \label{negent}
M_{P}^{2}+8\alpha\Lambda f_{1_{0}}<0.
   \]
The action \eqref{actionent} is a simple reformulation of \eqref{action}, where ${\cal F}_2(\Box)$ and ${\cal F}_3(\Box)$ have been set to zero and we have taken the dimensionless parameter $\lambda$ to be $\lambda=\alpha M_{P}^{2}$. 

If we were to now turn our attention to the defocusing condition around de Sitter space, these may be obtained directly from the linearised field equations \eqref{eomdS} by contracting with the tangent vector $k^\mu$. Hence, we find that in order for the associated null rays to diverge, we require,
 \[
 \label{defocusent}
 \begin{aligned}r_{\nu}^{\mu}k^{\nu}k_{\mu} & =\frac{1}{M_{P}^{2}\left(1+24\alpha  H^2 f_{1_{0}}\right)}(k^{0})^{2}\biggl[(\rho+p)+2\alpha M_{P}^{2}\left(\partial_{t}^{2}-H\partial_{t}\right){\cal F}_{1}(\Box)r\biggr]<0
. \end{aligned}
\]
Here, we have used the fact that in de Sitter space, $\Lambda=3M_P^2 H^2$, see \eqref{barredlambda}. Then, it is straightforward to read off the central conditions for null rays to defocus
\[
\begin{aligned}M_{P}^{2}\left(1+24\alpha  H^2 f_{1_{0}}\right)\gtrless0
 ,\qquad(\rho+p)+2\alpha M_{P}^{2}\left(\partial_{t}^{2}-H\partial_{t}\right){\cal F}_{1}(\Box)r & \lessgtr0.\end{aligned}
  \]
From \eqref{negent}, we find that the lower signs describe a non-physical spacetime defined by negative entropy and can therefore me omitted. Thus, the central constraints are simply
\[
\begin{aligned}M_{P}^{2}+24\lambda H^2 f_{1_{0}}>0
  ,\qquad(\rho+p)+2\lambda\left(\partial_{t}^{2}-H\partial_{t}\right){\cal F}_{1}(\Box)r & <0,\end{aligned}
  \]
where we have reintroduced the counting tool $\lambda\equiv \alpha M_P^2$, in accordance with the general formalism of this work.

Now, if we turn our attention to the defocusing calculation around de Sitter space given in \ref{sec:dS2}, we find that, by \eqref{dSa}, the left-most inequality is simply the constant $a$. In general, the function $a(\Box)$ is responsible for modifying the tensorial structure of the propagator but as the action considered is scalar in its modification, \eqref{actionent}, the function $a$ reduces to the constant,
\[
a=1+24\lambda M_P^{-2} H^2 f_{1_{0}}  
.\]
We have already established that this tensorial modification must be positive in order to avoid negative residues and the Weyl ghost, see Section \ref{sec:patho} but the entropy calculation gives an interesting  insight into the physical consequences of introducing ghosts into a theory. In this case, such an addition would result in a non-physical spacetime, defined by negative entropy. 

A further, intriguing property of the gravitational entropy described by \eqref{dsent}, is the possibility of realising a zero entropy state by taking $\alpha=\frac{M_{P}^{2}}{8\Lambda}$. Taking this value \emph{saturates} the defocusing condition \eqref{defocusent}, meaning little can be inferred from this vantage point. It would be interesting to pursue this line of enquiry in order to understand if this zero entropy state is indeed physical; at what cosmic time in a bouncing cosmology such a state could be realised; and whether there are any potential implications for the laws of thermodynamics prior to the bounce.

\end{appendices}

\backmatter

\nocite{} 
\renewcommand{\bibname}{References}

\bibliographystyle{unsrt}  
\bibliography{thesis}        

\end{document}